\documentclass[12pt]{article}
\usepackage[top=30truemm,bottom=30truemm,left=25truemm,right=25truemm]{geometry}
\usepackage{amsfonts,amsmath,amssymb,epsf}
\usepackage{amsmath,braket}
\usepackage{here}
\usepackage{physics}
\usepackage{graphicx,hyperref,color}
\usepackage[dvipsnames]{xcolor}
\usepackage{cite}
\usepackage{cancel}
\usepackage{url}
\usepackage{comment}
\usepackage{appendix}
\usepackage{chngcntr}
\usepackage{etoolbox}
\usepackage{lipsum}
\usepackage{tikz}
\usepackage{amsmath}
\usetikzlibrary{arrows.meta, decorations.pathreplacing}
\AtBeginEnvironment{subappendices}{%
	\addcontentsline{toc}{section}{Appendices}
	\counterwithin{figure}{subsection}
	\counterwithin{table}{subsection}
}
\hypersetup{
	unicode=false,          
	pdftoolbar=true,        
	pdfmenubar=true,        
	linktocpage=true,       
	pdffitwindow=false,     
	pdfstartview={FitH},    
	pdfnewwindow=true,      
	colorlinks=true,       
	linkcolor=blue,          
	citecolor=blue,        
	filecolor=blue,      
	urlcolor=blue           
}

\numberwithin{equation}{section}									

\newcommand{\be}{\begin{equation}}
	\newcommand{\ba}{\begin{eqnarray}}
		\newcommand{\ea}{\end{eqnarray}}
	\newcommand{\ee}{\end{equation}}

\newcommand{\no}{\nonumber \\}

\newcommand{\bea}{\begin{eqnarray}}
	\newcommand{\eea}{\end{eqnarray}}
\newcommand{\bes}{\begin{equation*}}
	\newcommand{\beas}{\begin{eqnarray*}}
		\newcommand{\eeas}{\end{eqnarray*}}
	\newcommand{\bas}{\begin{array*}}
		\newcommand{\eas}{\end{array*}}
	\newcommand{\ees}{\end{equation*}}

\def\CC{\mathbb{C}}

\newcommand{\Annulus}{\mathcal{A}}

\usepackage{tikz}
\usetikzlibrary{arrows,arrows.meta,intersections, calc,positioning,decorations.pathreplacing,decorations.pathmorphing,shapes}
\usetikzlibrary{patterns}
\usetikzlibrary{decorations.markings}
\usetikzlibrary{knots}

\usepackage{subcaption}

\begin{document}

	\begin{titlepage}
		\thispagestyle{empty}
		
		\vspace*{-2cm}
		\begin{flushright}
			YITP-25-157
			\\
		\end{flushright}
		
		\bigskip
		
		\begin{center}
			\noindent{\bf \large {Hayden–Preskill Model via Local Quenches }}
			\vspace{1.2cm}
			
			Weibo Mao$^a$,
			Tadashi Takayanagi$^{b,c}$
			\vspace{1cm}\\
			
			{\it $^a$ Kavli Institute for Theoretical Sciences, \\
				University of Chinese Academy of Sciences, \\
				Beijing 100190, China}\\
			\vspace{1.5mm}
			{\it $^b$Center for Gravitational Physics and Quantum Information,\\
					Yukawa Institute for Theoretical Physics, Kyoto University, \\
					Kitashirakawa Oiwakecho, Sakyo-ku, Kyoto 606-8502, Japan}\\
			\vspace{1.5mm}
			{\it $^c$Inamori Research Institute for Science,\\
				620 Suiginya-cho, Shimogyo-ku,
				Kyoto 600-8411, Japan}\\
			\bigskip \bigskip
					\vskip 1em
		\end{center}

\begin{abstract}
	We model the Hayden--Preskill (HP) information recovery protocol in 2d CFTs  via local joining quenches. Euclidean path integrals with slits prepare the HP subsystems: the message $M$, its reference $N$, the Page-time black hole $B$, the early radiation $E$, and the late radiation $R$; the remaining black hole after emitting $R$ is denoted as $B'$. The single-slit geometry provides an analytically tractable toy model, while the bounded-slit geometry more closely captures the HP setup. In the free Dirac fermion 2d CFT, the mutual information $I(N\!:\!B')$ shows quasi-particle dynamics with partial or full revivals, whereas that in holographic 2d CFTs, which are expected to be maximally chaotic, exhibit sharp transitions: in the bounded-slit case, when the size of the late radiation becomes comparable to that of the reference $N$, $I(N\!:\!B')$ does vanish at late time, otherwise it remains finite. This contrast between free CFTs and holographic CFTs gives a clear characterization of the HP recovery threshold.
\end{abstract}

	\end{titlepage}
	\newpage
	\tableofcontents
	\newpage
	\section{Introduction}\label{sec:introduction}
 Black hole evaporation, as originally derived by Hawking, implies that an initially pure state evolves into seemingly thermal radiation, leading to the black hole information paradox~\cite{Hawking:1975vcx}.  To reconcile unitarity, Page argued that the entanglement entropy between a black hole and its radiation should follow the “Page curve”: it increases as radiation is emitted, peaks when roughly half the black hole entropy has been radiated, then decreases as information leaks out~\cite{Page:1993df,Page:1993wv}.  Recent semiclassical gravitational computations confirm this behaviour via quantum extremal surfaces and the island formula, wherein the dominant bulk surface undergoes a phase transition at the Page time and includes an interior “island” that purifies the radiation~\cite{Penington:2019npb,Almheiri:2019psf}.  These developments also clarify why Hawking’s original calculation—treating the radiation as strictly thermal until the end of evaporation—did not respect the correct structure of Hilbert space in quantum gravity, which differs from a naive effective field theory picture \cite{Penington:2019npb}.

\begin{figure}[htbp]
		\centering		
			\includegraphics[width=11cm]{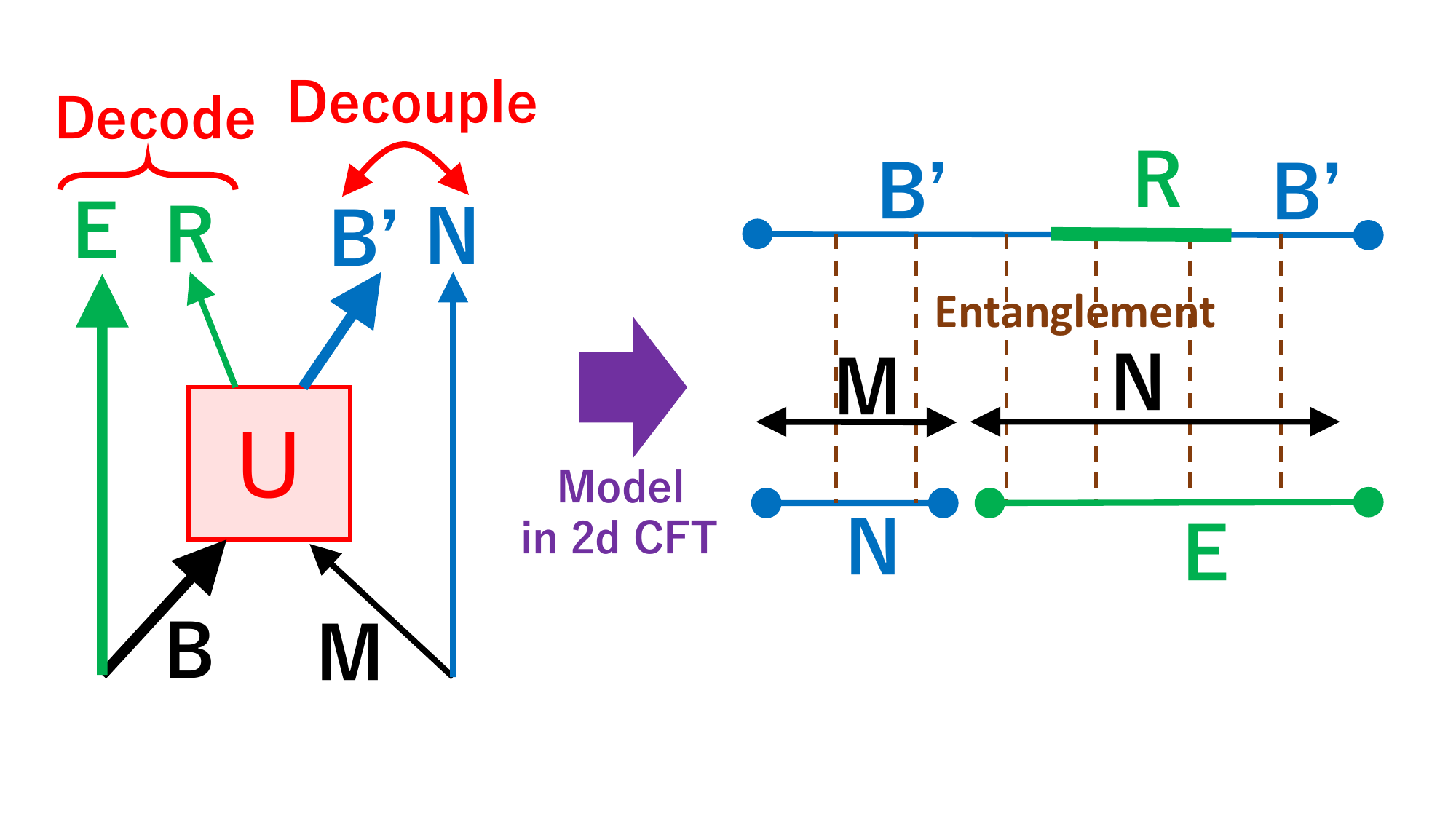}
			\caption{Setup of the Hayden-Preskill protocol (left) and its modeling in terms of a two dimensional CFT (right). 
            In the left picture, we initially decompose the full Hilbert space into $E$ (early radiation), $B$ (black hole), $M$ (message) and $N$ (the reference of the message), where $E$ and $M$ are maximally entangled with  $B$ and $N$, respectively.  After the time evolution of $B$ and $M$ described by the unitary, the new radiation $R$ is collected. The remaining subsystem is called $B'$.            
            In the right picture, the two CFT states are realized on the upper and lower horizontal line, which are entangled with each other as a thermofield double state.}
            \label{fig:HPsetup}
		\end{figure}

A characteristic phenomenon which occurs at the Page time is the fast information recovery known as the Hayden–Preskill (HP) protocol, which treats black hole evaporation as a quantum communication channel.  In this model, if a $k$-qubit message $M$, entangled with a reference $N$ is thrown into an old (post-Page) black hole $B$, then after the black hole’s scrambling time only about $k$ additional Hawking qubits $R$ are needed to decode the message, provided the internal dynamics are sufficiently chaotic~\cite{Hayden:2007cs} (refer to the left part of Fig.\ref{fig:HPsetup}). 
This “information mirror” behaviour links information recovery to the phenomena called fast scrambling expected to quantum systems dual to classical gravity in e.g. black holes or AdS backgrounds 
\cite{Sekino:2008he,Lashkari:2011yi,Shenker:2013pqa,Maldacena:2015waa}.

Two-dimensional conformal field theory (2d CFT) provides a continuum laboratory to test these ideas.  The replica trick and boundary CFT (BCFT) techniques allow analytic control over the time evolutions of entanglement entropy and mutual information after global quenches \cite{Calabrese:2004eu,Calabrese:2009qy} and local ones \cite{Calabrese:2007mtj,Nozaki:2013wia,Nozaki:2014hna}.  In rational CFT with small central charge, entanglement growth is well captured by a quasi-particle picture: entangled pairs propagate ballistically \cite{He:2014mwa} and cause oscillatory spikes in mutual information at late times~\cite{Asplund:2013zba}.  In contrast, a CFT with a large central charge (including holographic CFT) display a suppression of these free propagating effects; at strong coupling the quasi-particle description breaks down and entanglement saturates without late‑time revivals~\cite{Asplund:2015eha,Kudler-Flam:2020xqu}.  Related studies of entanglement production by local quench also reflect this dichotomy~\cite{Caputa:2014eta, Caputa:2015waa}.

On the gravitational side, the holographic  entanglement entropy formula relates the entanglement entropy in the CFT to the area of extremal surface in the bulk \cite{Ryu:2006bv,Ryu:2006ef}.  This framework has been extended to the boundary conformal field theory (BCFT) via the AdS/BCFT, where end‑of‑the‑world branes model boundary degrees of freedom~\cite{Takayanagi:2011zk,Fujita:2011fp}.  In time-dependent settings, the holographic construction describes geometrically the entanglement growth of thermofield-double states ~\cite{Hartman:2013qma}.

Most research on the Hayden--Preskill protocol in physics has so far been based on quantum circuits or random matrix models~\cite{Yoshida:2017non, Yoshida:2018vly, Piroli:2020dlx}. In this work, we propose a two-dimensional CFT model that realizes the Hayden--Preskill protocol in a continuum setting. Together, these advances motivate revisiting the information recovery and entanglement dynamics in 2d CFT, comparing the quasi-particle picture with the holographic one. In the following, we adopt up-to-date citation labels and build upon these developments to analyze the Hayden--Preskill protocol within a continuum field-theoretic framework.  In general quantum many-body systems, the relation between the fast scrambling behavior and the chaotic property is known to be very subtle and complicated \cite{Nakata:2023hwg}.
Our analysis for two dimensional (2d) CFTs in this paper, will show that they are tightly related to each other in a relatively simple way.

In this paper, we study two setups of joining quench geometries as continuum models of the HP  protocol in two dimensional CFTs, based on the conventional Euclidean path-integral formalism. One of them is the single-slit cylinder, which is  conformally mapped to the upper half-plane. This allows our analytical control and serves as a simple toy model of scrambling. 
The other model is the bounded-slit cylinder, which is mapped into an annulus. This more faithfully realizes the HP setup by preparing a four-party entangled state of the black hole $B$, early radiation $E$, the message sent to black hole $M$ and its reference system $N$ (refer to the right picture of Fig.\ref{fig:HPsetup}). We measure the subspace $R$ inside $B$ as the late radiation. The unmeasured region $(M\cup B)\setminus R$ is denoted by $N'$.  If the mutual information 
$I(N:B')$ gets vanishing, we find that the information of $M$ can be recovered from the radiation $R$. This is the condition of information recovery in the HP model \cite{Hayden:2007cs}. By evolving these states in the Lorentzian time and computing $I(N\!:\!B')$, we identify quasi-particle revivals in the free Dirac fermion CFT, while in holographic CFT the bounded-slit geometry shows the complete vanishing of $I(N\!:\!B')$, indicating that the black hole has radiated away the information, in direct analogy with the HP protocol.

The remainder of this paper is organized as follows.
Section~\ref{sec:setup} introduces the Euclidean slit geometries and their conformal maps.
Section~\ref{sec:rational-mutual-information} analyzes the dynamics of mutual information in the free-fermion 2d CFT, while Section~\ref{sec:holographic-mutual-information} turns to holographic CFT and identifies the recovery transition.
Section~\ref{sec:conclusions} summarizes our findings and discusses their implications.
Appendix~\ref{sec:method-of-maps} details the circular-slit mapping method, Appendix~\ref{sec:energy-density} derives the post-quench energy profile from the stress tensor, and Appendix~\ref{sec:detail-of-calculation} provides further computational details for the free-fermion analysis.

\section{Two dimensional CFT setups}\label{sec:setup}
To model the Hayden–Preskill protocol in a 2d CFT, we construct a Page-time black hole together with a maximally entangled reference pair $M$ and $N$ (i.e. Alice and Charlie). 
Operationally, this corresponds to joining one side of two thermofield-double states and evolving that side in real time (see the right panel of Fig.\ref{fig:HPsetup}).

Consider two decoupled two-dimensional conformal field theories, \(\text{CFT}_L\) and \(\text{CFT}_R\),
with isomorphic Hilbert spaces \(\mathcal{H}_L\) and \(\mathcal{H}_R\).
The TFD state in the doubled Hilbert space \(\mathcal{H}_L\otimes\mathcal{H}_R\) is defined as
\begin{equation}
	\label{eq:TFD-state}
	|\Psi_\beta\rangle
	= 
	\frac{1}{\sqrt{Z(\beta)}}
	\sum_{n} e^{-\frac{\beta}{2} E_n}
	\,|n\rangle_L \otimes |n\rangle_R,
	\qquad
	Z(\beta) = \sum_{n} e^{-\beta E_n},
\end{equation}
where \(|n\rangle_{L,R}\) are energy eigenstates with eigenvalue \(E_n\).

The corresponding pure-state density matrix is
\begin{equation}
	\label{eq:TFD-density}
	\rho_{\text{TFD}}
	= |\Psi_\beta\rangle\langle\Psi_\beta|
	= \frac{1}{Z(\beta)} \sum_{m,n}
	e^{-\frac{\beta}{2}(E_m+E_n)}
	\,|m\rangle_L |m\rangle_R \langle n|_L \langle n|_R.
\end{equation}

The TFD state can be prepared via a Euclidean path integral of length \(\beta/2\) on each copy: performing Euclidean evolution by \(\beta/2\) on each CFT and entangling them along their boundaries corresponds to the half cylinder construction. Only after tracing over one copy does the path-integral representation become the full thermal cylinder of circumference \(\beta\) as in \cite{Hartman:2013qma}.

For the real-time dynamics, we let the Hamiltonian act only on Alice’s subsystem $M$ and the black hole $B$, while keeping the early radiation $E$ and Charlie’s reference system $N$ static. The total Hamiltonian is then chosen as
\begin{equation}
	\label{eq:Htot-sum}
	H = H_L \otimes I .
\end{equation}
Accordingly, the time-evolved TFD state takes the form
\begin{equation}
	\label{eq:TFD-time-evolved}
	|\Psi_\beta(t_L)\rangle
	= e^{-\tau_L H_L}\, |\Psi_\beta\rangle .
\end{equation}
By setting \(\tau_L = i t\) , this is analytically continued to
\begin{equation}
	|\Psi_\beta(t)\rangle
	= e^{- i t H_L}\,|\Psi_\beta\rangle .
\end{equation}
The corresponding density matrix then evolves as
\begin{equation}
	\rho_{\text{TFD}}(t)
	= e^{-i t H_L}\,\rho_{\text{TFD}}\,e^{i t H_L}.
\end{equation}
Denote by $\rho_A=\operatorname{Tr}_{\bar{A}}\!\left[\rho_{\text{TFD}}(t)\right]$ the reduced density matrix on a finite interval $A$.  
The entanglement entropy $S_A$ can be computed using the replica trick~\cite{Calabrese:2009qy,Calabrese:2004eu}, starting from the Rényi entropies
\begin{equation}
	S_A^{(n)}(t) = \frac{1}{1-n}\,\log \operatorname{Tr}\!\left[\rho_A^n(t)\right],
\end{equation}
To evaluate the entanglement entropies introduced above, we employ the twist operator formalism in 2d CFTs. In this approach, the \(n\)-th Rényi entropy of an interval is expressed as a correlation function of twist and anti-twist operators inserted at the interval endpoints. For a single interval, one has
\begin{equation}
	S_A^{(n)}(t)
	= \frac{1}{1-n}\,\log
	\left\langle
	\sigma_n\bigl(w_{A_l},\bar w_{A_l}\bigr)\,
	\bar \sigma_n\bigl(w_{A_r},\bar w_{A_r}\bigr)
	\right\rangle_{\mathbb C},
\end{equation}
where \(w_{A_l}\) and \(w_{A_r}\) denote the locations of the interval endpoints on the thermal cylinder, and \(\sigma_n\) (\(\bar\sigma_n\)) are twist (anti-twist) operators with the conformal dimension
\begin{equation}
	\Delta_n=\frac{c}{12}\left(n-\frac{1}{n}\right).
\end{equation}
Correlation functions on the cylinder can be mapped to those on the plane or annulus via a conformal transformation, thereby reducing the computation to standard CFT correlators. Taking the limit \(n \to 1\) then yields the von Neumann entropy.
\begin{equation}
	S_A = \lim_{n \to 1} S_A^{(n)}(t).
\end{equation}

\subsection{Construction of models}
	In this section we introduce the slit-quench geometries that serve as our basic
	framework for diagnosing the Hayden--Preskill protocol in two-dimensional conformal
	field theories. The essential goal is to probe the transfer of quantum information
	by tracking the mutual information between the reference system $N$ and the black hole
	after a portion of radiation $R$ has been emitted $B'=(B\cup M)\setminus R$, with the amount of radiation controlled
	by the length of the chosen radiation interval.
	
	We will consider two complementary constructions:
	
	\begin{itemize}
		\item \textbf{Single slit:} A minimal geometry in which the Euclidean cylinder is
		cut open by a single vertical slit. The resulting state can be analyzed explicitly
		in both free fermion and holographic CFT. In the free fermion CFT, the quasi-particle picture applies and the mutual information can be computed in a closed
		form using the twist operators via the bosonization. In holographic CFTs, the same geometry can be treated
		analytically through the holographic entanglement entropy, providing a transparent and geometrical
		description of the dynamics of mutual information.
		
		\item \textbf{Bounded slit:} A more elaborate construction where the slit is inserted
		on a semi-infinite thermal cylinder, yielding a geometry that more faithfully realizes
		the Hayden--Preskill scenario. In this case the Euclidean path integral naturally
		prepares two connected thermofield-double states, leading to a four-party entangled
		structure $(N, R, B', E)$ as in the right panel of 
        Fig.\ref{fig:HPsetup}. The associated conformal mapping involves elliptic
		theta-functions, thus closed-form expressions are not generally available and the
		mutual information must instead be evaluated numerically.
	\end{itemize}

		Taken together, the two setups provide complementary advantages. The single slit
		offers an analytical control and explicit formulae in both RCFT and holographic CFT,
		while the bounded slit captures the full structure of the Hayden--Preskill protocol
		at the cost of requiring numerical analysis.
    
\subsubsection{Single slit}\label{subsec:setup1}

	Now we consider the single slit case, where the spatial domain extends over the entire real line, but we only introduce one vertical slit at $x=X_1$ in $\tau$ direction, we may use SS for short notation of single slit model. The Euclidean geometry is the full thermal cylinder
	\be
	\mathcal{C}=\{\,w=x+i\tau \in \mathbb{C} \mid x\in(-\infty,\infty),~\tau\in[0,\beta)\,\}\;\cong\;\mathbb{R}\times S^1_\beta,
	\ee
	with the compact Euclidean time direction $\tau\sim\tau+\beta$. We excise the vertical slit
	\be
	\Sigma_1=\Bigl\{\,w=X_1+i\tau ~\big|~ 	\tau\in[0,\tfrac{\beta}{2}-\alpha]\;\cup\;[\tfrac{\beta}{2}+\alpha,\beta]\,\Bigr\},
	\ee
	so that the resulting geometry is $\mathcal{C}_{\text{slit}}=\mathcal{C}\setminus\Sigma_1$.

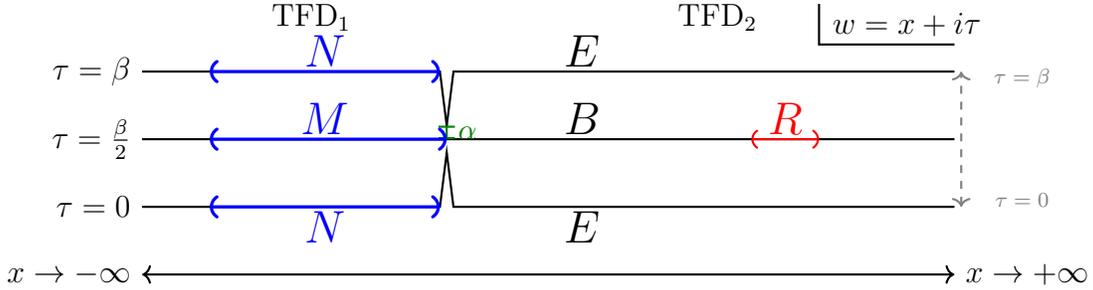
\begin{figure}[htbp]
	\centering
	\begin{tikzpicture}[scale=0.9, thick]
		\draw[<->] (-1,-1) node[left] {$x \rightarrow -\infty$} -- (11,-1) node[right] {$x \rightarrow +\infty$} ;
		
		\draw (-1,1) -- (11,1);
		
		\draw (-1,2) -- (3.4,2) -- (3.5,1.2) -- (3.6,2) -- (11,2);
		
		\draw (-1,0) -- (3.4,0) -- (3.5,0.8) -- (3.6,0) -- (11,0);
		
		
		
		\node[left] at (-1,0) {$\tau=0$};
		\node[left] at (-1,1) {$\tau=\frac{\beta}{2}$};
		\node[left] at (-1,2) {$\tau=\beta$};
		
		\node[blue] at (1.7,1.3) {\Large $M$};
		\node at (5.5,1.3) {\Large $B$};
		\node[blue] at (1.7,2.3) {\Large $N$};
		\node at (5.5,2.3) {\Large $E$};
		\node[blue] at (1.7,-0.3) {\Large $N$};
		\node at (5.5,-0.3) {\Large $E$};
		\node[red] at (8.5,1.3) {\Large $R$};
		\draw[green!50!black, thick, |-|] (3.5,1.2) -- (3.5,1) node[midway, right] {\small $\alpha$};

		\draw[blue, very thick, (-)] (0,1) -- (3.5,1); 
		\draw[blue, very thick, (-)] (0,2) -- (3.4,2); 
		\draw[blue, very thick, (-)] (0,0) -- (3.4,0); 
		
		\draw[red, thick, (-)] (8,1.0) -- (9,1.0);
		
		\node at (1.5,2.8) {TFD$_1$};
		\node at (7.5,2.8) { TFD$_2 $};
		\draw[thick] (9,3) -- (9,2.4) -- (11,2.4);
		\node at (10.3,2.7) {$w = x + i\tau$};
		
		\draw[gray, <->, dashed] (11.1,0) -- (11.1,2);
		\node[gray] at (12,0.1) {\scriptsize $\tau = 0$};
		\node[gray] at (12,1.9) {\scriptsize $\tau = \beta$};
		
	\end{tikzpicture}
	\caption{Single-slit setup: two Euclidean thermal cylinders are glued at $\tau = \beta/2$, with joining point $\alpha$ and identified boundaries $\tau = 0 \sim \beta$.}
	\label{fig:SS-model}
	\end{figure}

	The subsystem assignment is taken to be 
	\begin{equation}
		\begin{aligned}
			\textbf{Region M (Alice)}: \quad & x\in[0,X_1], \quad \tau=\tfrac{\beta}{2},\\
			\textbf{Region B (Black Hole)}: \quad & x\in(X_1,\infty), \quad \tau=\tfrac{\beta}{2},\\
			\textbf{Region N (Charlie)}: \quad & x\in[0,X_1], \quad \tau=0,\\
			\textbf{Region E (Early Radiation)}: \quad & x\in(X_1,\infty), \quad \tau=0,\\
			\textbf{Region R (Late Radiation)}: \quad & x\in[X_2,X_3], \quad \tau=i t+\tfrac{\beta}{2},\\
			\textbf{Region B$'$ (Black Hole after emission)}: \quad & x\in[0,X_2)\cup(X_3,\infty), \quad \tau=i t+\tfrac{\beta}{2}.
		\end{aligned}
	\end{equation}
	Fig.~\ref{fig:SS-model} provides an intuitive illustration of this setup.	
	
\begin{figure}[htbp]
		\centering
		
		\begin{subfigure}[b]{0.4\textwidth}
			\includegraphics[width=\linewidth]{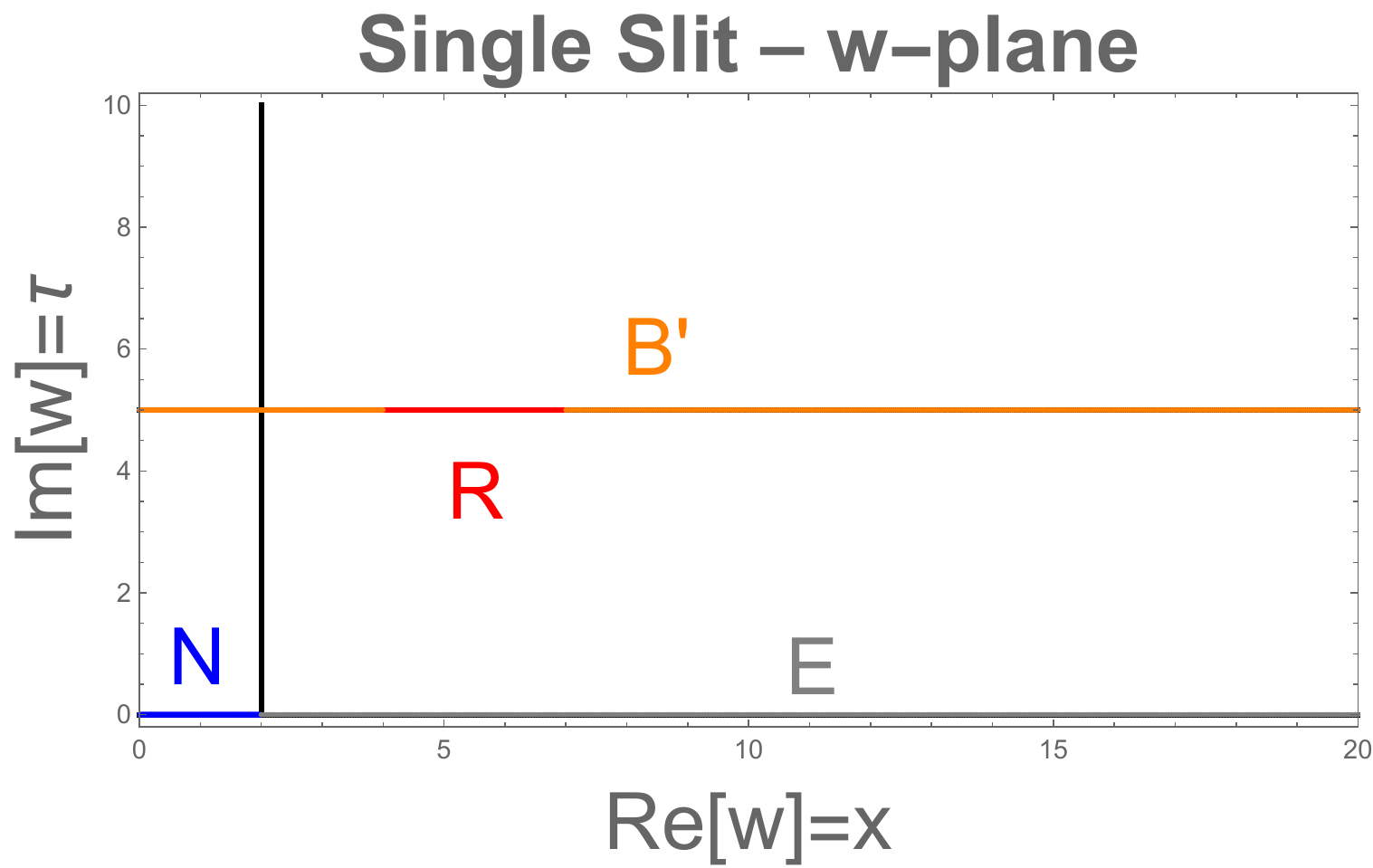}
		\end{subfigure}
		\hfill
		\begin{subfigure}[b]{0.4\textwidth}
			\includegraphics[width=\linewidth]{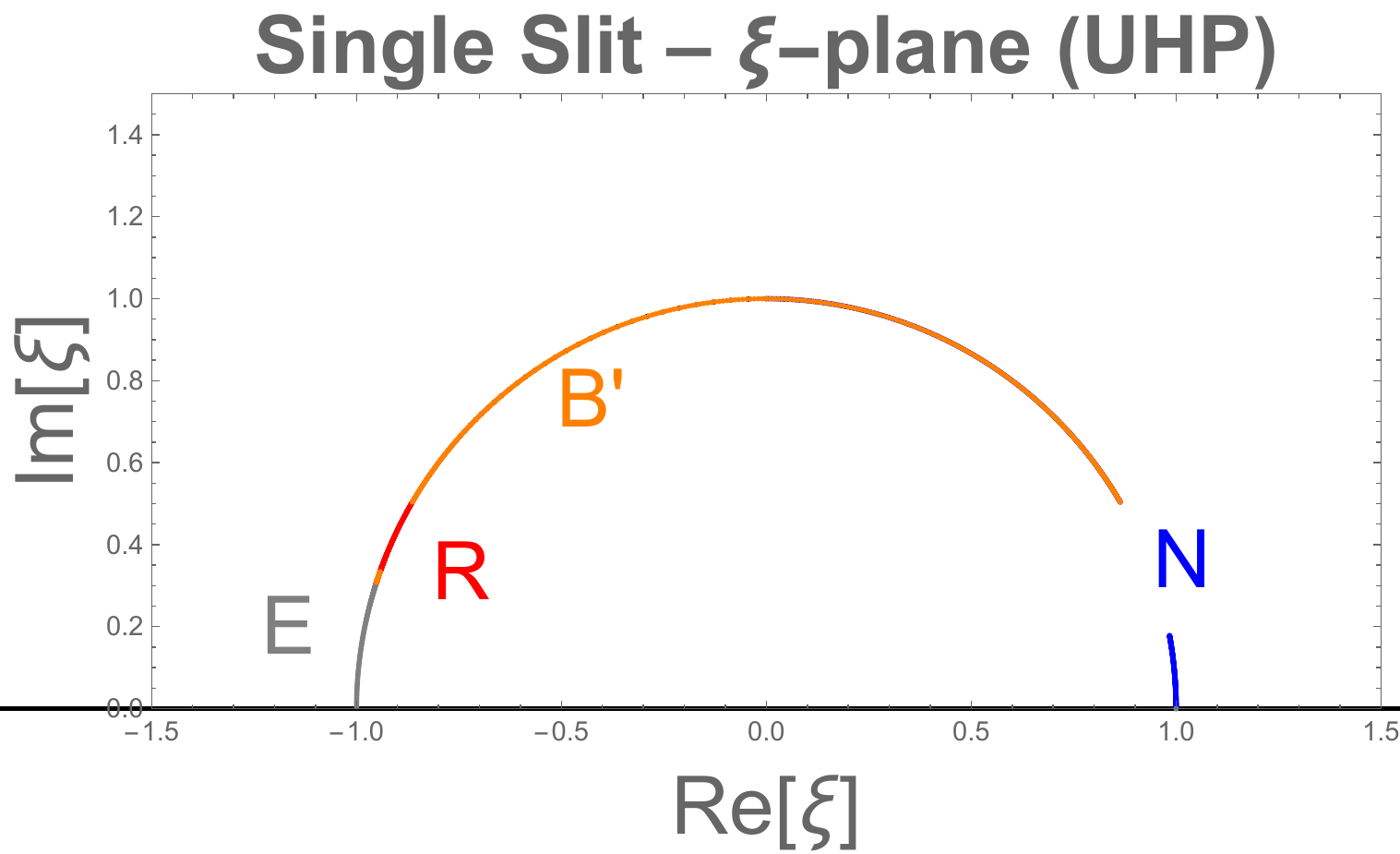}
		\end{subfigure}\\

		\caption{%
			The single-slit quench model in the $w$ and $\xi$ coordinates, with the slit explicitly shown in black. 
			The blue region denotes the $N$ subsystem, the orange double interval corresponds to $B'$, the gray interval corresponds to $E$
			and the red single interval represents $R$. 
			The left panel is plotted in the $w$ coordinate, while the right panel is in the $\xi$ coordinate.
		}
		\label{fig:single-slit-map}
	\end{figure}

	By utilizing the conformal map and its conjugate, generalized from the approach in \cite{Calabrese:2007mtj,Rajabpour:2015uqa,Rajabpour:2015nja,Rajabpour:2015xkj},
	
	\begin{equation}
		\begin{split}
			\xi(w)=&\sqrt{\frac{\sinh \left(\frac{ \pi}{\beta}\left(w-X_1\right)-i \frac{\pi (\beta/2 +\alpha)}{\beta}\right)}{\sinh \left(\frac{ \pi}{\beta}\left(w-X_1\right)-i \frac{\pi (\beta/2 -\alpha)}{\beta}\right)}},\\=&\sqrt{\frac{\cosh \left(\frac{ \pi}{\beta}\left(w-X_1\right)-i \frac{\pi \alpha}{\beta}\right)}{\cosh \left(\frac{ \pi}{\beta}\left(w-X_1\right)+i \frac{\pi \alpha}{\beta}\right)}}.
		\end{split}
	\end{equation}
	and also its conjugate $ \bar{\xi}(\bar w)= \sqrt{\frac{\cosh \left(\frac{ \pi}{\beta}\left(\bar w-X_1\right)+i \frac{\pi \alpha}{\beta}\right)}{\cosh \left(\frac{ \pi}{\beta}\left(\bar w-X_1\right)-i \frac{\pi \alpha}{\beta}\right)}}.$
	
	and the differential is given by
	\begin{equation}
		\frac{d \xi}{d w} = -\frac{i \pi  \sin \left(\frac{2 \pi  \alpha }{\beta }\right) \sech^2\left(\frac{\pi  (i \alpha +w-X_1)}{\beta }\right)}{2 \beta  \sqrt{\cosh \left(\frac{\pi  (-i \alpha +w-X_1)}{\beta }\right) \sech \left(\frac{\pi  (i \alpha +w-X_1)}{\beta }\right)}}.
	\end{equation}
	The single slit cylinder can be mapped to the upper half-plane with coordinate $(\xi, \bar{\xi})$. One can easily check that the slit itself is indeed mapped to the real axis, by requiring the branch cut of square root, while one can see the other part mapping to upper half-plane as Fig.\ref{fig:single-slit-map} shows.

\subsubsection{Bounded slit}\label{subsec:setup2}
		Although the single-slit setup can reveal certain aspects of the information-mirror phenomenon in 2d CFT, it is not sufficient to qualify as a genuine Hayden-Preskill realization. The main issue is that subsystems $N$ and $M$ are not maximally entangled at the beginning of the Lorentzian time evolution. To address this, we must introduce a slightly more elaborate construction-namely, the bounded slit setup, we may use BS for short notation.
		
	The Euclidean geometry  is given by the semi-infinite thermal cylinder
	
	\be
	\mathcal{C}=\{w=x+i \tau \in \mathbb{C} \mid x \in[0, \infty), \tau \in[0,\beta)\} \cong \mathbb{R} \times S^1_{\beta},
	\ee
	where $\tau \sim \tau+\beta$ denotes the compact Euclidean time direction.
	We introduce a slit geometry by removing a collection of vertical line segments from $\mathcal{C}$, i.e., we define the domains
	\be
	\mathcal{C}_{\text {slit }}=\mathcal{C} \backslash\left(\bigcup_j \Sigma_j\right),
	\ee
	where each slit $\Sigma_j$ is a line segment of the form
	\be
	\Sigma_j=\left\{w=x_j+i \tau \mid \tau \in\left[0, \tfrac{\beta}{2} - \alpha \right] \cup \left[\tfrac{\beta}{2} + \alpha, \beta \right]\right\},
	\ee
	for a fixed spatial coordinate $x_j \in(-\infty, \infty)$.

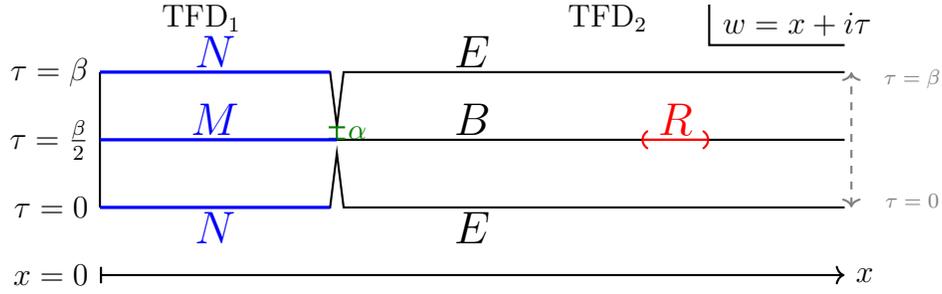
\begin{figure}[htbp] 
		\centering
		\begin{tikzpicture}[scale=0.9, thick]
			\draw[|->] (0,-1) node[left] {$x=0$} -- (11,-1) node[right] {$x$} ;
			
			\draw (0,1) -- (11,1);
			
			\draw (0,2) -- (3.4,2) -- (3.5,1.2) -- (3.6,2) -- (11,2);
			
			\draw (0,0) -- (3.4,0) -- (3.5,0.8) -- (3.6,0) -- (11,0);
			
			
			\draw (-0,0) -- (-0,2); 
			
			\node[left] at (0,0) {$\tau=0$};
			\node[left] at (0,1) {$\tau=\frac{\beta}{2}$};
			\node[left] at (0,2) {$\tau=\beta$};
			
			\node[blue] at (1.7,1.3) {\Large $M$};
			\node at (5.5,1.3) {\Large $B$};
			\node[blue] at (1.7,2.3) {\Large $N$};
			\node at (5.5,2.3) {\Large $E$};
			\node[blue] at (1.7,-0.3) {\Large $N$};
			\node at (5.5,-0.3) {\Large $E$};
			\node[red] at (8.5,1.3) {\Large $R$};
			\draw[green!50!black, thick, |-|] (3.5,1.2) -- (3.5,1) node[midway, right] {\small $\alpha$};

			\draw[blue, very thick] (0,1) -- (3.5,1); 
			\draw[blue, very thick] (0,2) -- (3.4,2); 
			\draw[blue, very thick] (0,0) -- (3.4,0); 
			
			\draw[red, thick, (-)] (8,1.0) -- (9,1.0);
			
			\node at (1.5,2.8) {TFD$_1$};
			\node at (7.5,2.8) { TFD$_2 $};
			\draw[thick] (9,3) -- (9,2.4) -- (11,2.4);
			\node at (10.3,2.7) {$w = x + i\tau$};
			
			\draw[gray, <->, dashed] (11.1,0) -- (11.1,2);
			\node[gray] at (12,0.1) {\scriptsize $\tau = 0$};
			\node[gray] at (12,1.9) {\scriptsize $\tau = \beta$};
		\end{tikzpicture}
		\caption{Bounded-slit setup: two Euclidean thermal cylinders are glued at $\tau = \beta/2$, with joining point $\alpha$ and identified boundaries $\tau = 0 \sim \beta$.}
		\label{fig:BS-model}
	\end{figure}

	We consider a two-dimensional conformal field theory (CFT) defined on a thermal cylinder with spatial coordinate \( x \in \mathbb{R} \) and Euclidean time \( \tau \sim \tau + \beta \), where the inverse temperature \( \beta \) sets the periodicity of the thermal circle. To model a joining quench that mimics the Hayden-Preskill (HP) information retrieval scenario, we prepare a Euclidean path integral state with one vertical slit on the bounded cylinder. The Euclidean geometry is given by the semi-infinite thermal cylinder
	\be
	\mathcal{C}_{\geq 0}=\{w=x+i \tau \in \mathbb{C} \mid x \in[0, \infty), \tau \in[0,\beta)\} \cong \mathbb{R}_{\geq 0} \times S^1_\beta,
	\ee
	where the only slit $\Sigma_1$ is a line segment of the form:
	\be
	\Sigma_1=\left\{w=X_1+i \tau \left\lvert\, \tau \in\left[0, \frac{\beta}{2}-\alpha\right] \cup\left[\frac{\beta}{2}+\alpha, \beta\right]\right.\right\}.
	\ee
	We partition the half thermal cylinder into the following regions (with analytic continuation to Lorentzian time on the last two lines):
	\begin{equation}
		\begin{aligned}
			\textbf{Region M (Alice)}: \quad & x \in [0,\ X_1], \quad \tau = \tfrac{\beta}{2}, \\
			\textbf{Region B (Black\ Hole)}: \quad & x \in (X_1,\ \infty), \quad \tau = \tfrac{\beta}{2}, \\
			\textbf{Region N (Charlie)}: \quad & x \in [0,\ X_1], \quad \tau = 0, \\
			\textbf{Region E (Early\ Radiation)}: \quad & x \in (X_1,\ \infty), \quad \tau = 0, \\
			\textbf{Region R (Late\ Radiation)}: \quad & x \in [X_2,\ X_3], \quad  \tau= i t+\tfrac{\beta}{2},\\
			\textbf{Region B' (Black\ Hole\ after\ emission)}: \quad & x \in [0,\ X_2) \cup (X_3,\ \infty), \quad \tau= i t+\tfrac{\beta}{2}.
		\end{aligned}
	\end{equation}
	This configuration is illustrated in Fig.\ref{fig:BS-model} ,
	We can map this geometry to an annulus $\mathcal{A}=\left\{\zeta \left\lvert\, \rho \leq \abs{\zeta} \leq 1\right.\right\}$ by a function $\zeta(w)$
		\begin{equation}
	\zeta(w)= f^{-1}(e^{-2\pi w / \beta}),
	\end{equation}
	where the $f$ function is given by
	\begin{equation}\label{eq: map-from-disk-to-annulus}
		\begin{aligned}	
			f(\zeta, a, \rho)&=\frac{ \abs{a} \theta_{3}(\frac{\log(-\frac{\zeta}{\rho a})}{2 \pi i} ; \frac{\log(\rho)}{\pi i}) }{  \theta_{3}(\frac{\log(-\frac{\zeta  a }{\rho})}{2 \pi i} ; \frac{\log(\rho)}{\pi i})}.
		\end{aligned}
	\end{equation}
	The map~\eqref{eq: map-from-disk-to-annulus} can be obtained within the Schottky–Klein prime–function approach developed by Crowdy and Marshall~\cite{crowdy2006conformal, crowdy2011schottky, crowdy2012conformal}; see Appendix~\ref{sec:method-of-maps} for additional details. Here,  $a$ is a real parameter with $a = e^{-2\pi X_1/\beta}$. Moreover, the moduli parameter $\rho$ needs to be determined by $\alpha, \beta$ and $X_1$. 
 
	Although the bounded–slit geometry is initially defined on a half thermal cylinder, the analytic continuation from Euclidean to Lorentzian time naturally extends the conformal map to the complementary half. Consequently, the full thermal cylinder is mapped to a double-up annulus \(\rho<|\zeta|<1/\rho\), i.e. the annulus together with its inversion about the unit circle.

	For the single–slit setup we work with two coordinate charts \((w,\bar w )\) and \((\xi,\bar\xi )\). 
	For the bounded–slit setup we additionally employ \((z,\bar z )\) and \((\zeta,\bar\zeta )\), where
    $z=e^{-2\pi w / \beta}$.
	
\begin{figure}[htbp]
		\centering
		
		\begin{subfigure}[b]{0.3\textwidth}
			\includegraphics[width=\linewidth]{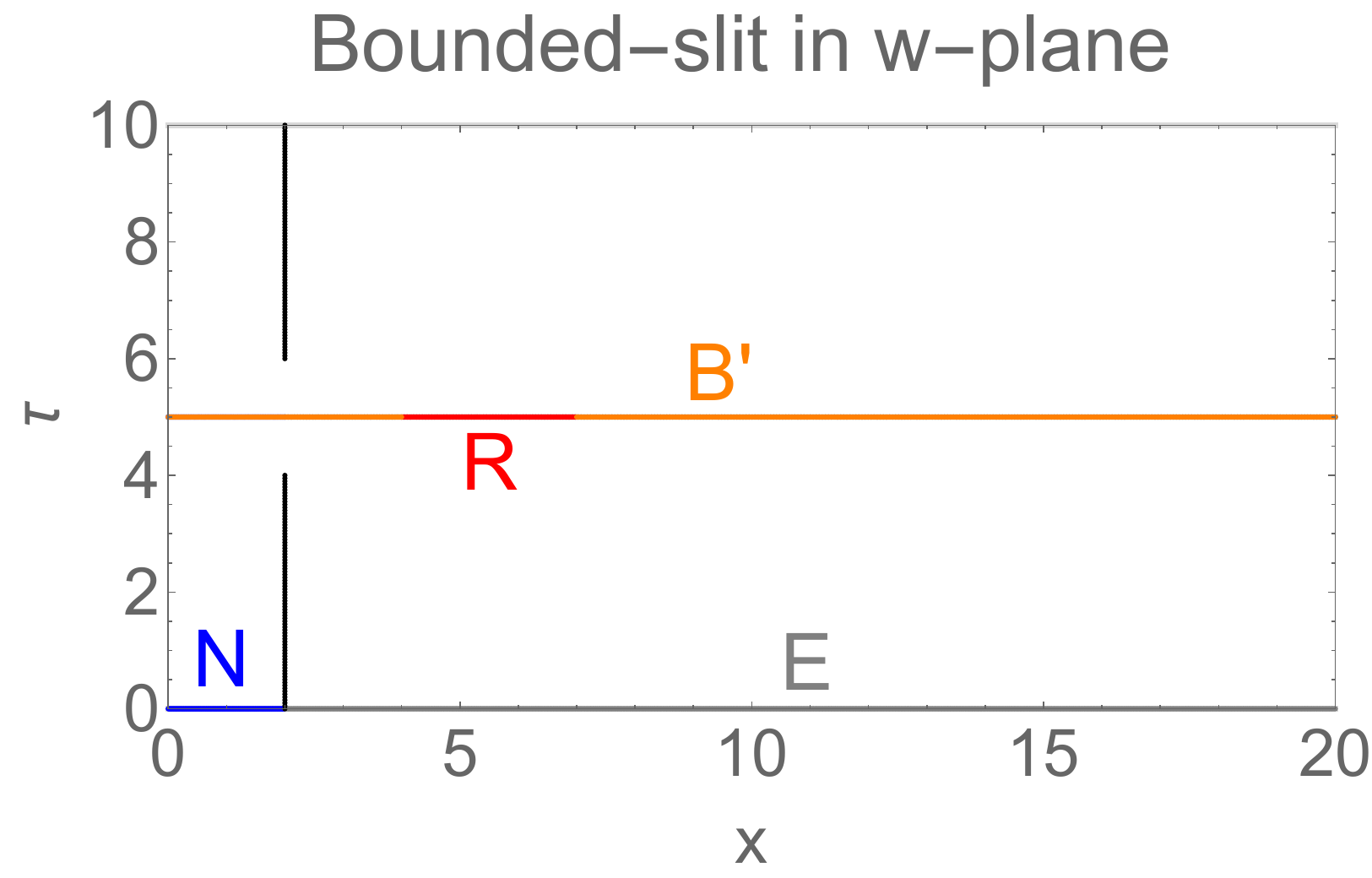}
		\end{subfigure}
		\hfill
		\begin{subfigure}[b]{0.3\textwidth}
			\includegraphics[width=\linewidth]{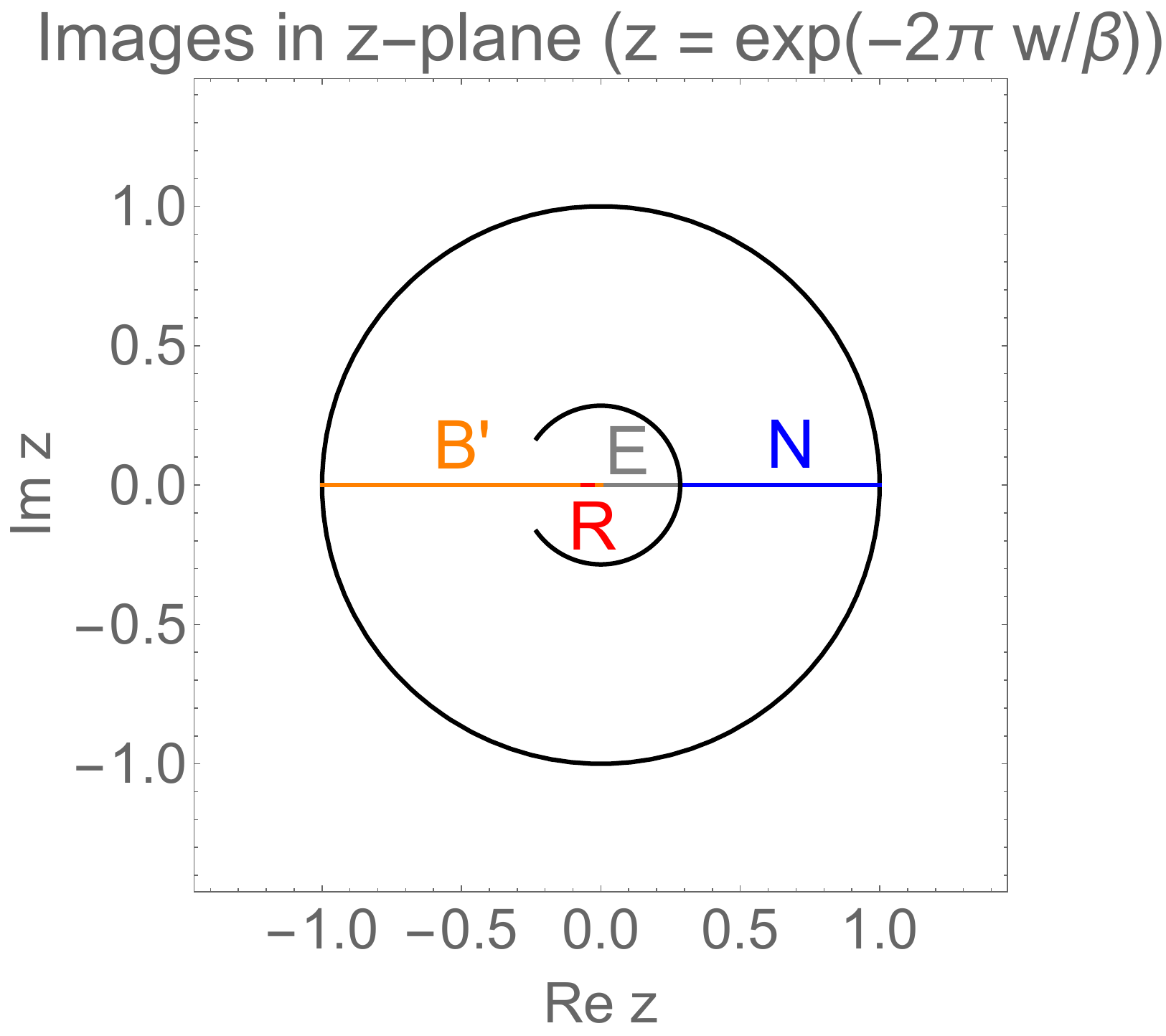}
				\end{subfigure}
				\hfill
			\begin{subfigure}[b]{0.3\textwidth}
				\includegraphics[width=\linewidth]{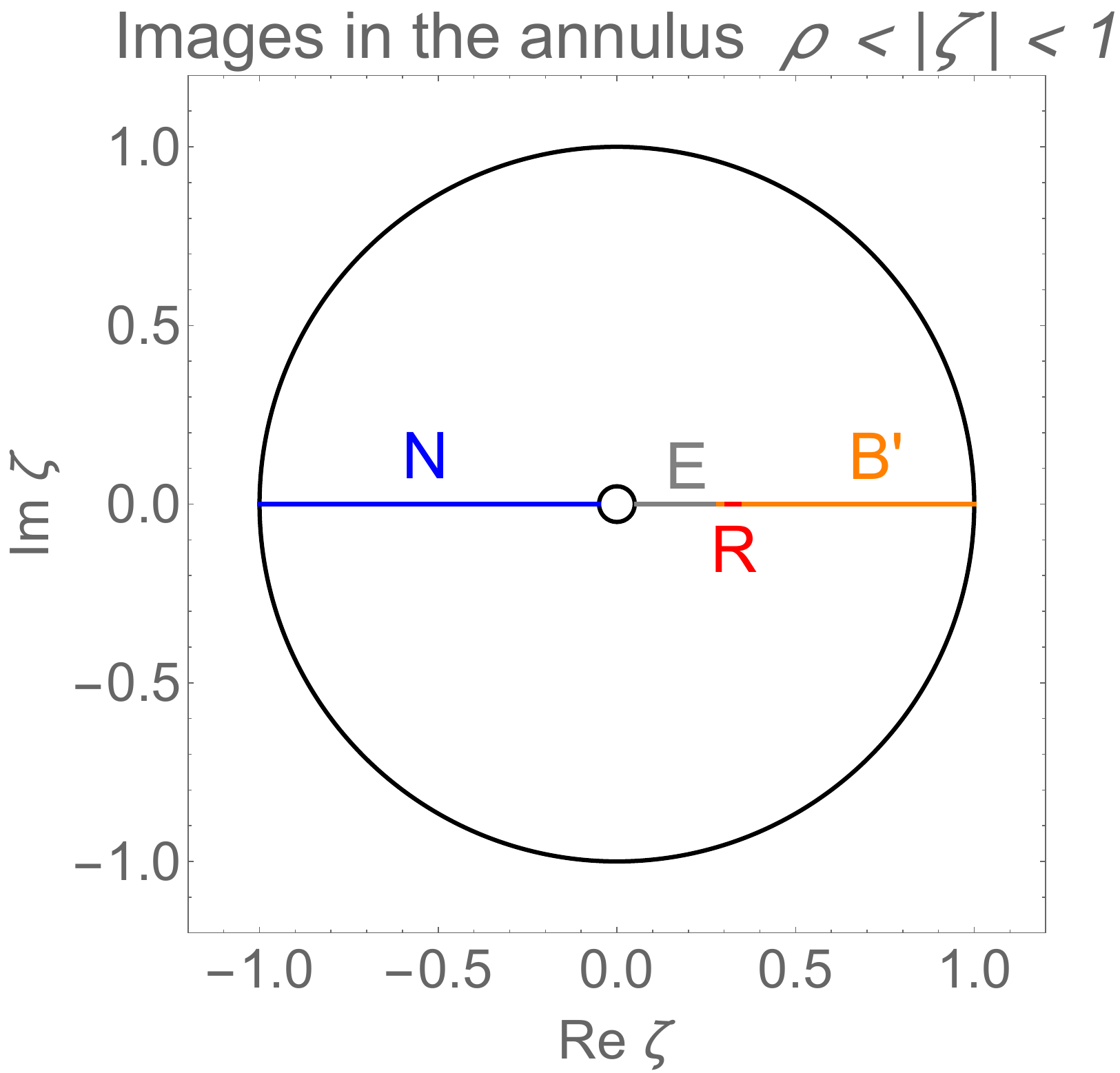}
		\end{subfigure}\\
		
	\caption{%
	The bounded-slit quench model in the $w$ ,$z$ , and $\zeta$ coordinates, with the slit explicitly shown in black. 
	The blue region denotes the $N$ subsystem, the orange double interval corresponds to $B'$, the gray interval corresponds to $E$
	and the red single interval represents $R$. 
	The left panel is plotted in the $w$ coordinate, the middle panel is in the $z$ coordinate, while the right panel is in the $\zeta$ coordinate.
}
		\label{fig:bounded-slit-map}
	\end{figure}

	The mapping chain can be summarized as
	\begin{equation}
	\left[	(w,\bar w )
		\;\xrightarrow{\;z(w)\;}
		(z,\bar z )
		\;\xrightarrow{\;f^{-1}(z)\;}
		(\zeta,\bar\zeta)
		\;\right]=\left[\;
		(w,\bar w )
		\;\xrightarrow{\;\zeta(w)\;}
		(\zeta,\bar\zeta)\right].
	\end{equation}
	Those mappings are depicted in Fig. \ref{fig:bounded-slit-map}.
 
\subsection{Method in this paper}
A key prediction of the Hayden--Preskill protocol is that quantum information thrown into an old black hole can be recovered from the Hawking radiation after a scrambling time. In holographic or field-theoretic realizations, this recovery is expected to manifest as a characteristic pattern in the entanglement structure among subsystems. Mutual information, defined as
\begin{equation}
	I(A : B) = S_A + S_B - S_{A \cup B},
\end{equation}
serves as a powerful diagnostic of correlations and entanglement between distinct regions, and is particularly well-suited for detecting information transfer and recovery.

In our setup, the post-quench geometry prepared by the Euclidean path integral gives rise to a four-party entangled state involving subsystems \(M\), \(B\), \(N\), and \(E\), corresponding respectively to Alice, the black hole interior, the reference observer, and the early radiation.  
We choose \(R\) to be a single interval inside \(B\), representing the late radiation emitted after the Page time, and define \(B'=(B\cup M)\setminus R\) as the post-emission black hole.  
By computing the mutual information \(I(N:B')\), we aim to test whether and when the message initially entangled with subsystem \(N\) becomes recoverable from \(R\), thereby probing signatures of the information mirror behavior~\cite{Hayden:2007cs}. To this end, we compute the time evolution of \(I(N:B')\) under the real time evolutions. 

In a holographic CFT, $I(N\!:\!B')>0$ corresponds to the connected phase of the 
holographic extremal surface (i.e. geodesic in our examples) for $N\cup B'$, while $I(N\!:\!B')=0$ indicates the 
disconnected phase. In the context of the HP information mirror phenomenon, the reduction of $I(N\!:\!B')$ serves as an indicator of the progressive 
decoupling of the information initially entangled with the reference system $N$ 
from $B'$. In this work, we will employ $I(N\!:\!B')$ as the primary diagnostic to analyze the 
transfer of information from $B'$ to $R$.

Since we are specifically interested in the mutual information between \(N\) and \(B'\), we need the explicit coordinates of the interval endpoints on the thermal cylinder.  
We denote them as \(w_{\alpha}(t)\) and \(\bar w_{\alpha}(t)\), where the subscript indicates the corresponding region (\(M\), \(R\), \(B\), \(N\), or \(E\)), and the additional label \(l\) or \(r\) specifies whether it is the left or right endpoint in the spatial coordinate \(x\) on the thermal cylinder slice at the indicated (Euclidean or Lorentzian) time after analytic continuation.
For example, \(w_{Ml}(t)\) is the left endpoint of region \(M\) in the \(w\)-coordinate, while \(w_{Rr}(t)\) is the right endpoint of the radiation interval \(R\).  
Explicitly, they are given by  
\begin{equation}
	\begin{aligned}
		w_{Ml}(t)      &= -t+ \tfrac{i\beta}{2}, &
		\quad \bar w_{Ml}(t) &= t-\tfrac{i\beta}{2}, \\[4pt]
		w_{Rl}(t)      &= X_2 - t + \tfrac{i\beta}{2}, &
		\quad \bar w_{Rl}(t) &= X_2 + t - \tfrac{i\beta}{2}, \\[4pt]
		w_{Rr}(t)      &= X_3 - t + \tfrac{i\beta}{2}, &
		\quad \bar w_{Rr}(t) &= X_3 + t - \tfrac{i\beta}{2}, \\[4pt]
		w_{Br}(t)      &= \lim_{X\to\infty}\!\bigl(X -t + \tfrac{i\beta}{2}\bigr), &
		\quad \bar w_{Br}(t) &= \lim_{X\to\infty}\!\bigl(X+t  - \tfrac{i\beta}{2}\bigr), \\[4pt]
		w_{Nl}(t)      &= 0, &
		\quad \bar w_{Nl}(t) &= 0, \\[4pt]
		w_{Nr}(t)      &= X_1, &
		\quad \bar w_{Nr}(t) &= X_1, \\[4pt]
		w_{Er}(t)      &= \lim_{X\to\infty}\!\bigl(X\bigr), &
		\quad \bar w_{Er}(t) &= \lim_{X\to\infty}\!\bigl(X\bigr).
	\end{aligned}
\end{equation}

In the twist-operator formalism~\cite{Calabrese:2004eu,Calabrese:2009qy}, the \(n\)-th Rényi entropies for \(N\), \(B'\), and \(N\cup B'\), and hence the mutual information, are obtained from twist-operator correlators as
\begin{equation}
	\begin{aligned}
		&S_{N}^{(n)}
		=\frac{1}{1-n}\,
		\log\Bigl\langle
		\sigma_{n}\!\bigl(w_{Nl},\bar w_{Nl}\bigr)\,
		\bar\sigma_{n}\!\bigl(w_{Nr},\bar w_{Nr}\bigr)
		\Bigr\rangle_{\mathrm{SS/BS}}, \\[6pt]
		&S_{B'}^{(n)}
		=\frac{1}{1-n}\,
		\log\Bigl\langle
		\sigma_{n}\!\bigl(w_{Ml},\bar w_{Ml}\bigr)\,
		\bar\sigma_{n}\!\bigl(w_{Rl},\bar w_{Rl}\bigr)\,
		\sigma_{n}\!\bigl(w_{Rr},\bar w_{Rr}\bigr)\,
		\bar\sigma_{n}\!\bigl(w_{Br},\bar w_{Br}\bigr)
		\Bigr\rangle_{\mathrm{SS/BS}}, \\[6pt]
		&S_{N \cup B'}^{(n)}\no
		&=\frac{1}{1-n}\,
		\log\Bigl\langle
		\sigma_{n}\!\bigl(w_{Nl},\bar w_{Nl}\bigr)\,
		\bar\sigma_{n}\!\bigl(w_{Nr},\bar w_{Nr}\bigr)\,
		\sigma_{n}\!\bigl(w_{Ml},\bar w_{Ml}\bigr)\,
		\bar\sigma_{n}\!\bigl(w_{Rl},\bar w_{Rl}\bigr)\,
		\sigma_{n}\!\bigl(w_{Rr},\bar w_{Rr}\bigr)\,
		\bar\sigma_{n}\!\bigl(w_{Br},\bar w_{Br}\bigr)
		\Bigr\rangle_{\mathrm{SS/BS}} .
	\end{aligned}
\end{equation}
Here, \(\sigma_n\) and \(\bar\sigma_n\) denote the twist and anti-twist operators, respectively. 
The subscript \(\mathrm{SS/BS}\) indicates that the expressions apply to both the single-slit and bounded-slit geometries.
\subsubsection{Single-slit model}

In this geometry, the Rényi entropies are obtained directly from twist-operator correlators 
on the upper half-plane (UHP), with coordinates \((\xi,\bar\xi)\). Explicitly, we have
\begin{equation}
	\begin{aligned}\label{eq:SS-mutual-renyi}
		S_N^{(n)} &= \frac{1}{1-n} \log \Bigl[
		|\xi'(w_{Nl})|^{\Delta_n}
		|\xi'(w_{Nr})|^{\Delta_n}
		\left\langle
		\sigma_n(\xi_{Nl}, \bar\xi_{Nl}) \,
		\bar\sigma_n(\xi_{Nr}, \bar\xi_{Nr})
		\right\rangle_{\mathrm{UHP}}
		\Bigr], \\[6pt]
		%
		S_{B'}^{(n)} &= \frac{1}{1-n} \log \Bigl[
		\prod_{\alpha \in \{Ml,Rl,Rr,Br\}} |\xi'(w_\alpha)|^{\Delta_n}
		\left\langle
		\sigma_{n}(\xi_{Ml},\bar \xi_{Ml})\,
		\bar\sigma_{n}(\xi_{Rl},\bar \xi_{Rl})\,
		\sigma_{n}(\xi_{Rr},\bar \xi_{Rr})\,
		\bar\sigma_{n}(\xi_{Br},\bar \xi_{Br})
		\right\rangle_{\mathrm{UHP}}
		\Bigr], \\[6pt]
		%
		S_{N \cup B'}^{(n)} &= \frac{1}{1-n} \log \Bigl[
		\prod_{\alpha \in \{Nl,Nr,Ml,Rl,Rr,Br\}} |\xi'(w_\alpha)|^{\Delta_n}\Bigr.\\& \times
		\Bigl.\left\langle
		\sigma_{n}(\xi_{Nl}, \bar\xi_{Nl})\,
		\bar\sigma_{n}(\xi_{Nr}, \bar\xi_{Nr})\,
		\sigma_{n}(\xi_{Ml}, \bar\xi_{Ml})\,
		\bar\sigma_{n}(\xi_{Rl}, \bar\xi_{Rl})\,
		\sigma_{n}(\xi_{Rr}, \bar\xi_{Rr})\,
		\bar\sigma_{n}(\xi_{Br}, \bar\xi_{Br})
		\right\rangle_{\mathrm{UHP}}
		\Bigr], \\[6pt]
		%
		I_{N:B'}^{(n)} &= S_N^{(n)} + S_{B'}^{(n)} - S_{N \cup B'}^{(n)} .
	\end{aligned}
\end{equation}
Here, the twist operators have conformal dimension 
\(\Delta_n=\tfrac{c}{12}\!\left(n-\tfrac{1}{n}\right)\), and all correlators are
evaluated on the upper half–plane (UHP). 
In the \(\xi\)-coordinate we adopt the shorthand
\be
\xi(w)=\xi,\qquad 
\xi(w_{\alpha})=\xi_{\alpha},\qquad 
\bar\xi(\bar w_{\alpha})=\bar\xi_{\alpha},
\ee
so that each endpoint \(w_{\alpha}\) is mapped to a point \(\xi_{\alpha}\). 

\subsubsection{Bounded-slit model}
For the bounded-slit case, since the complement of \(B' \cup N\) is \(E \cup R\), 
we can use the property of a pure state for entanglement entropy to simplify 
\(S_{N \cup B'}^{(n)}\) as
\begin{equation}
	\begin{aligned}
		S_{N \cup B'}^{(n)}   
		= \frac{1}{1-n} \log \Big\langle\,
		\sigma_n(w_{Rl}, \bar w_{Rl}) \,
		\bar\sigma_n(w_{Rr}, \bar w_{Rr}) \,
		\sigma_n(w_{Er}, \bar w_{Er}) \,
		\bar\sigma_n(w_{Nr}, \bar w_{Nr})
		\,\Big\rangle_{\mathrm{BS}} .
	\end{aligned}
\end{equation}

In the \(\zeta\) coordinate, the Rényi entropies can be expressed more explicitly as
\begin{equation}\label{eq:BS-mutual-renyi}
	\begin{aligned}
		S_N^{(n)} &= \frac{1}{1-n} \log \Bigl[
		|\zeta'(w_{Nl})|^{\Delta_n}
		|\zeta'(w_{Nr})|^{\Delta_n}
		\left\langle \sigma_n(\zeta_{Nl}, \bar\zeta_{Nl}) \,
		\bar\sigma_n(\zeta_{Nr}, \bar\zeta_{Nr})
		\right\rangle_{\mathcal{A}}
		\Bigr], \\[6pt]
		%
		S_{B'}^{(n)} &= \frac{1}{1-n} \log \Bigl[
		\prod_{\alpha \in \{Ml,Rl,Rr,Br\}} |\zeta'(w_\alpha)|^{\Delta_n}
		\left\langle
		\sigma_{n}(\zeta_{Ml},\bar \zeta_{Ml})\,
		\bar\sigma_{n}(\zeta_{Rl},\bar \zeta_{Rl})\,
		\sigma_{n}(\zeta_{Rr},\bar \zeta_{Rr})\,
		\bar\sigma_{n}(\zeta_{Br},\bar \zeta_{Br})
		\right\rangle_{\mathcal{A}}
		\Bigr], \\[6pt]
		%
		S_{N \cup B'}^{(n)} &= \frac{1}{1-n} \log \Bigl[
		\prod_{\alpha \in \{Rl,Rr,Er,Nr\}} |\zeta'(w_\alpha)|^{\Delta_n}
		\left\langle
		\sigma_{n}(\zeta_{Rl},\bar \zeta_{Rl})\,
		\bar\sigma_{n}(\zeta_{Rr},\bar \zeta_{Rr})\,
		\sigma_{n}(\zeta_{Er},\bar \zeta_{Er})\,
		\bar\sigma_{n}(\zeta_{Nr},\bar \zeta_{Nr})
		\right\rangle_{\mathcal{A}}
		\Bigr], \\[6pt]
		%
		I_{N:B'}^{(n)} &= S_N^{(n)} + S_{B'}^{(n)} - S_{N \cup B'}^{(n)} .
	\end{aligned}
\end{equation}
Here, the conformal dimension of the twist operators is 
\(\Delta_n=\tfrac{c}{12}(n-\tfrac{1}{n})\), and \(\mathcal A\) denotes the annulus geometry.  In the \(\zeta\) coordinate we use the shorthand
\be
\zeta(w)=\zeta, \qquad
\zeta(w_{\alpha})=\zeta_{\alpha}, \qquad
\bar \zeta(\bar w_{\alpha})=\bar\zeta_{\alpha},
\ee
thus each endpoint \(w_\alpha\) is mapped to a corresponding point \(\zeta_\alpha\).  
For notational simplicity, we use \(\zeta(w)\) both for the conformal transformation and for the resulting complex coordinate \(\zeta\).

\section{Evolution of Mutual Information in free
	fermion CFT}\label{sec:rational-mutual-information}
		
		We now specialize the general R\'enyi–entropy framework to the two–dimensional free massless Dirac fermion CFT. Our objective is to compute the mutual information between the reference system $N$ and the post–emission black hole $B'$. As a preparation, we briefly recall the twist–field construction that implements the replica method in the fermionic theory\cite{Casini:2005rm,Azeyanagi:2007bj,Casini:2009sr,Takayanagi:2010wp,Numasawa:2016emc}.
		
		Consider the Dirac fermion on an $n$–sheeted cover of the complex plane branched at $z=0$. Unfolding the cover is equivalent to working with $n$ decoupled replicas $\{\psi^{(k)},\bar\psi^{(k)}\}$ on a single copy of $\CC$ subject to the twisted boundary conditions (we assume $n$ is an odd integer):
		\begin{equation}
			\psi_L^{(k)}(e^{2\pi i} z)=\psi_L^{(k+1)}(z),\qquad
			\psi_R^{(k)}(e^{-2\pi i}\bar z)=\psi_R^{(k+1)}(\bar z),
		\end{equation}
		where $\psi_L^{(k)}$ and $\psi_R^{(k)}$ denote the chiral and anti–chiral components, respectively.
		
		Since the model is free, a discrete Fourier transform along the replica index diagonalizes the boundary condition,
		\begin{equation}
			\psi_L^{(p)}(z)=\frac{1}{\sqrt{n}}\sum_{p=0}^{n-1} e^{\frac{2\pi i k p}{n}}\,\tilde\psi_L^{(k)}(z),\qquad
			\psi_R^{(p)}(\bar z)=\frac{1}{\sqrt{n}}\sum_{p=0}^{n-1} e^{\frac{2\pi i k p}{n}}\,\tilde\psi_R^{(k)}(\bar z),
		\end{equation}
		so that each $k$–sector is a single fermion with the $\mathbb{Z}_n$ orbifold boundary condition:
		\begin{equation}
			\tilde\psi_L^{(k)}(e^{2\pi i}z)=e^{\frac{2\pi i k}{n}}\tilde\psi_L^{(k)}(z),\qquad
			\tilde\psi_R^{(k)}(e^{-2\pi i}\bar z)=e^{\frac{2\pi i k}{n}}\tilde\psi_R^{(k)}(\bar z).
		\end{equation}
		
		 Bosonization in each sector with free scalars $H_{L}^{(k)}$, $H_{R}^{(k)}$ leads to 
		\begin{equation}
			\tilde\psi_L^{(k)}(z)=e^{i H_L^{(k)}(z)},\qquad
			\tilde\psi_R^{(k)}(\bar z)=e^{i H_R^{(k)}(\bar z)}.
		\end{equation}
		Then the twist fields become vertex operators carrying charges $\pm k/n$:
		\begin{equation}
			\sigma_n^{(k)}(z)=e^{\,i\frac{k}{n} H_L^{(k)}(z)},\qquad
			\bar\sigma_n^{(k)}(\bar z)=e^{\,i\frac{k}{n} H_R^{(k)}(\bar z)}.
		\end{equation}
		Multiplying the contributions from the non–trivial sectors $k=1,\ldots,n-1$ yields the full twist operator. Writing the left/right gluing at the boundary in a unified form, one has
		\begin{equation}
			\sigma_n(z,\bar z)\;=\;\prod_{k=-\frac{n-1}{2}}^{\frac{n-1}{2}}
			\exp\!\left[i\frac{k}{n}\big(H_L^{(k)}(z)\,\pm\,H_R^{(k)}(\bar z)\big)\right],
		\end{equation}
		with the upper (lower) sign corresponding to Dirichlet (Neumann) boundary conditions \cite{Takayanagi:2010wp}. 
		The $k=0$ mode is untwisted and does not contribute. Each twisted sector carries conformal 
		weights $h_k=\bar h_k=\tfrac12 \tfrac{k}{n}(1-\tfrac{k}{n})$, and summing over 
		$k=1,\dots,n-1$ gives the total scaling dimension 
		$\Delta_n=\tfrac{c}{12}(n-1/n)$ with $c=1$ for the Dirac fermion. 
		
		The full replica twist field is then obtained as the product over all nontrivial sectors,
		\begin{equation}
			\sigma_n(z,\bar z)=\prod_{k=-\frac{n-1}{2}}^{\frac{n-1}{2}}\sigma^{(k)}(z,\bar z).
		\end{equation}
		Correlation functions of these twist fields are reduced to products of vertex–operator correlators. 
		For an untwisted free boson $H=H_L+H_R$, a vertex operator with charges $(\alpha,\bar\alpha)$ is defined by
		\begin{equation}
			V_{\alpha,\bar\alpha}(z,\bar z)
			=:\!\exp\!\big(i\alpha H_L(z)+i\bar\alpha H_R(\bar z)\big)\!:,
		\end{equation}
		with conformal weights $(h,\bar h)=(\alpha^2/2,\bar\alpha^2/2)$. 
		In the $\mathbb{Z}_n$ orbifold twist sector relevant to the replica method, the twisted mode expansion 
		shifts the effective weight to $h_k=\bar h_k=\tfrac12 \tfrac{k}{n}(1-\tfrac{k}{n})$.

\subsection{Single slit model calculation}
	 We begin with the free fermion CFT calculation in the single-slit setup, which illustrates the method used in this paper and yields analytic formulae providing intuition for the information recovery. 
 
	    This case we need to study the 2d free fermion CFT on an upper half plane.
 
		\begin{equation}
		\xi(w)=\sqrt{\frac{\cosh \left(\frac{\pi}{\beta}\left(w-X_1\right)+i \frac{ \pi \alpha}{\beta}\right)}{\cosh \left(\frac{ \pi}{\beta}\left(w-X_1\right)-i \frac{\pi \alpha}{\beta}\right)}} .
		\end{equation}
	
	  We consider the (Rényi) entanglement entropy with one interval. The two point functions of twist operators on the upper half plane read	  
	  \ba
	  \left\langle\sigma_n\left(\xi_1, \bar{\xi}_1\right) \sigma_{-n}\left(\xi_2, \bar{\xi}_2\right)\right\rangle_{UHP}=\tilde{d}_n\left(\frac{a^{\prime 2}\left(\xi_1-\bar{\xi}_2\right)\left(\bar{\xi}_1-\xi_2\right)}{\left|\xi_1-\bar{\xi}_1\right|\left|\xi_2-\bar{\xi}_2\right|\left(\xi_1-\xi_2\right)\left(\bar{\xi}_1-\bar{\xi}_2\right)}\right)^{\frac{1}{12}\left(n-\frac{1}{n}\right)}.
	  \ea
	    By a coordinate transformation from $w$ coordinate to $\xi$ coordinate, we  get
	    \begin{equation}
	    	\begin{aligned}
	    		\left\langle\sigma_n\left(w_1, \bar{w}_1\right) \sigma_{-n}\left(w_2, \bar{w}_2\right)\right\rangle_{SS}= & \tilde{d}_n\left(\sqrt{\left.\left.\left.\left.\frac{d \xi}{d w}\right|_{w=w_1} \frac{d \bar{\xi}}{d \bar{w}}\right|_{\bar{w}=\bar{w}_1} \frac{d \xi}{d w}\right|_{w=w_2} \frac{d \bar{\xi}}{d \bar{w}}\right|_{\bar{w}=\bar{w}_2}}\right. \\
	    		& \left.\frac{a^{\prime 2}\left(\xi_1-\bar{\xi}_2\right)\left(\bar{\xi}_1-\xi_2\right)}{\left|\xi_1-\bar{\xi}_1\right|\left|\xi_2-\bar{\xi}_2\right|\left(\xi_1-\xi_2\right)\left(\bar{\xi}_1-\bar{\xi}_2\right)}\right)^{\frac{1}{12}\left(n-\frac{1}{n}\right)}.
	    	\end{aligned}
	    \end{equation}
	    From the replica trick, we obtain
	    	    \begin{equation}
	    	\begin{aligned}
	    		S_A^{(n)}= &\frac{\frac{1}{12}\left(n-\frac{1}{n}\right)}{1-n} \log \left[ \sqrt{\left.\left.\left.\left.\frac{d \xi}{d w}\right|_{w=w_1} \frac{d \bar{\xi}}{d \bar{w}}\right|_{\bar{w}=\bar{w}_1} \frac{d \xi}{d w}\right|_{w=w_2} \frac{d \bar{\xi}}{d \bar{w}}\right|_{\bar{w}=\bar{w}_2}}\right. \\
	    		& \left.\frac{a^{\prime 2}\left(\xi_1-\bar{\xi}_2\right)\left(\bar{\xi}_1-\xi_2\right)}{\left|\xi_1-\bar{\xi}_1\right|\left|\xi_2-\bar{\xi}_2\right|\left(\xi_1-\xi_2\right)\left(\bar{\xi}_1-\bar{\xi}_2\right)}\right]+\frac{1}{1-n}\log \left[\tilde{d}_n\right]\\
	    		= &-\frac{1}{12}\left(1+\frac{1}{n}\right) \log \left[ \sqrt{\left.\left.\left.\left.\frac{d \xi}{d w}\right|_{w=w_1} \frac{d \bar{\xi}}{d \bar{w}}\right|_{\bar{w}=\bar{w}_1} \frac{d \xi}{d w}\right|_{w=w_2} \frac{d \bar{\xi}}{d \bar{w}}\right|_{\bar{w}=\bar{w}_2}}\right. \\
	    		& \left.\frac{a^{\prime 2}\left(\xi_1-\bar{\xi}_2\right)\left(\bar{\xi}_1-\xi_2\right)}{\left|\xi_1-\bar{\xi}_1\right|\left|\xi_2-\bar{\xi}_2\right|\left(\xi_1-\xi_2\right)\left(\bar{\xi}_1-\bar{\xi}_2\right)}\right] 
	    	\end{aligned}
	    \end{equation}
	    Our goal is to compute the dynamics of the mutual information between the
	    post–emission black hole $B'$ and the reference system $N$. For simplicity, we set the
	    UV cutoff to unity ($a'=1$) and choose a conformal boundary condition with
	    vanishing boundary entropy ($g=1$). In addition, we normalize the twist
	    operators so that the plane two–point normalization constant satisfies
	    $c_n=1$. With these conventions, the UHP normalization reduces to
	    $\tilde d_n=1$ and omitted here.

	   We now recall a general fact crucial for multi-interval configurations in the free massless Dirac fermion \cite{Casini:2005rm}: the entanglement entropy of a $p$-component set $A_1\cup\cdots\cup A_p$ on the complex plane is exactly given by 
       
	  \begin{equation}
	  	\label{eq:multi-interval-free-Dirac}
	  	S\!\left(A_1 \cup \cdots \cup A_p\right)=\frac{1}{3}
	  	\left(\sum_{i,j} \log \big|a_i-b_j\big|
	  	-\sum_{i<j} \log \big|a_i-a_j\big|
	  	-\sum_{i<j} \log \big|b_i-b_j\big| \right),
	  \end{equation}
	  with $a_i$ and $b_i$ the left and right endpoints of $A_i$ (we have set the UV cutoff to $1$). A remarkable consequence of \eqref{eq:multi-interval-free-Dirac} is the extensivity of mutual information:
	  \begin{equation}
	  	I(A,B\cup C)=I(A,B)+I(A,C)\,.
	  \end{equation}
	 Although our setup is defined on the UHP, the method of images can be employed to reduce the computation to that on the full plane. 
	 Following Ref.~\cite{Berthiere:2019lks}, which provides a generalization of Eq.~\eqref{eq:multi-interval-free-Dirac} to boundary geometries, 
	 correlators on the UHP are mapped to multi-interval correlators on the complex plane with mirror endpoints. 
	 Consequently, Eq.~\eqref{eq:multi-interval-free-Dirac} can be applied directly to our boundary setup.

	 By employing the method of images for the calculation of entanglement entropy, the general formula can be extended to the UHP. Specifically, for each interval $A_i$ we define its image interval $A_{-i}$, with endpoints identified as
	 \begin{equation}
	 	a_{-i} \equiv b_i^{*}, \qquad b_{-i} \equiv a_i^{*},
	 \end{equation}
	 where the symbol “$*$” denotes the image point under the method of images.
	 
	 The entanglement entropy of the union of these intervals is then found to be
	 \begin{equation}
	 	S\!\left(A_{-p} \cup \ldots \cup A_{-1} \cup A_1 \cup \ldots \cup A_p\right)
	 	= \frac{1}{6}\left( 
	 	\sum_{i,j} \log |a_i-b_j|
	 	- \sum_{i<j} \log |a_i-a_j|
	 	- \sum_{i<j} \log |b_i-b_j|
	 	\right).
	 \end{equation}
	 
	 As an illustration, we can apply this formula to compute the entanglement entropy for a single interval on the UHP.
	  
	     \begin{equation}
	     	\begin{split}\label{eq:ss-2pt-rcft}
	    S_A&=-\left.\partial_n \log \left\langle\sigma_n\left(\xi_1, \bar{\xi}_1\right) \sigma_{-n}\left(\xi_2, \bar{\xi}_2\right)\right\rangle_{U H P}\right|_{n=1}\\
	    &=\frac{1}{6}\left(\log \left|a_{-1}-b_{-1}\right|+ \log \left|a_{-1}-b_{1}\right|+\log \left|a_{1}-b_{-1}\right|+ \log \left|a_{1}-b_{1}\right|- \log \left|a_{-1}-a_{1}\right|- \log \left|b_{-1}-b_{1}\right| \right)\\
	    &=\frac{1}{6}\left(\log \sqrt{\left|(\bar{\xi_2}-\bar{\xi_1}) (\xi_2-\xi_1)\right|}+ \log \left|\xi_2-\bar{\xi_2}\right|+\log \left|\xi_1-\bar{\xi_1}\right| \right. \\ & \left.+ \log \sqrt{\left|(\bar{\xi_2}-\bar{\xi_1}) (\xi_2-\xi_1)\right|}- \log \sqrt{\left|(\xi_1-\bar{\xi_2})(\xi_2-\bar{\xi_1})\right|}- \log \sqrt{\left|(\xi_2-\bar{\xi_1})(\xi_1-\bar{\xi_2})\right|} \right)\\ &=-\frac{1}{6}\log\left(\frac{\left(\xi_1-\bar{\xi}_2\right)\left(\bar{\xi}_1-\xi_2\right)}{\left|\xi_1-\bar{\xi}_1\right|\left|\xi_2-\bar{\xi}_2\right|\left(\xi_1-\xi_2\right)\left(\bar{\xi}_1-\bar{\xi}_2\right)}\right).
	     	\end{split}
    \end{equation}
	    
	  In principle, one needs to evaluate the entanglement entropies 
	  $S_N$, $S_B'$, and $S_{B'\cup N}$ in order to obtain the dynamics of 
	  the mutual information $I(B':N)$. However, thanks to the extensivity of mutual information in this case, the calculation can be simplified. 
	  Instead of treating the full region $B'$, it suffices to compute the 
	  contributions from its two disconnected components, namely 
	  $I(B'_{\text{left}}:N)$ and $I(B'_{\text{right}}:N)$. 
	  
	  For the clarity of the derivation, we introduce the explicit definitions of 
	  these components. The left part is taken as 
	  \begin{equation}
	  	B'_{\text{left}} \equiv 
	  	\Bigl\{\, w = x+i\tau \;\Big|\; 
	  	x \in [\Re(w_{Ml}),\,\Re(w_{Rl})),\; 
	  	\tau = \tfrac{\beta}{2}\,\Bigr\},
	  \end{equation}
	  while the right part is defined as 
	  \begin{equation}
	  	B'_{\text{right}} \equiv 
	  	\Bigl\{\, w = x+i\tau \;\Big|\; 
	  	x \in (\Re(w_{Rr}),\,\Re(w_{Br})),\; 
	  	\tau = \tfrac{\beta}{2}\,\Bigr\}.
	  \end{equation}
	  These intervals correspond to the left and right extensions of $B'$ 
	  along the Euclidean time slice at $\tau=\beta/2$, bounded respectively 
	  by the points $(w_{Bl},w_{Rl})$ and $(w_{Rr},w_{Br})$.
	  
	  Thus, the mutual information can be decomposed into five separate contributions,  
	  \begin{equation}
	  	\begin{split}
	  		I\!\left(B^{\prime}: N\right)
	  		&= I\!\left(B_{\text{left}}^{\prime}: N\right) + I\!\left(B_{\text{right}}^{\prime}: N\right) \\
	  		&= S(B_{\text{left}}^{\prime}) + S(B_{\text{right}}^{\prime}) 
	  		+ 2\,S(N) - S(B_{\text{left}}^{\prime} \cup N) 
	  		- S(B_{\text{right}}^{\prime} \cup N) \, .
	  	\end{split}
	  \end{equation}
	  In the following, we will evaluate these five different terms one by one. 
	  By previous calculation, we have already obtained the form of single interval entanglement entropy. By taking the conformal mapping into account we find
	  \begin{equation}
	  	\begin{split}
	  		S(A)&=-\frac{1}{6}\log\left( 
	  		\sqrt{\left.\left.\left.\left.\frac{d \xi}{d w}\right|_{w=w_1} \frac{d \bar{\xi}}{d \bar{w}}\right|_{\bar{w}=\bar{w}_1} \frac{d \xi}{d w}\right|_{w=w_2} \frac{d \bar{\xi}}{d \bar{w}}\right|_{\bar{w}=\bar{w}_2}}
	  		 \frac{\left(\xi_1-\bar{\xi}_2\right)\left(\bar{\xi}_1-\xi_2\right)}{\left|\xi_1-\bar{\xi}_1\right|\left|\xi_2-\bar{\xi}_2\right|\left(\xi_1-\xi_2\right)\left(\bar{\xi}_1-\bar{\xi}_2\right)}\right)
	  	\end{split}
	  \end{equation}

	  
	  Consider two disjoint intervals on the upper half-plane (UHP), 
	  \(A_1=[a_1,b_1]\) and \(A_2=[a_2,b_2]\), together with their image intervals
	  \(A_{-1}=[a_{-1},b_{-1}]=[\bar{b}_1,\bar{a}_1]\) and \(A_{-2}=[a_{-2},b_{-2}]=[\bar{b}_2,\bar{a}_2]\). We use only the
	  index “\(-i\)” to denote images (no conjugation symbol). We start from the
	  UHP master formula with \(p=2\):
	  \begin{equation}
	  	\begin{split}\label{eq:master-UHP}
	  	&S\!\left(A_{-2}\cup A_{-1}\cup A_1\cup A_2\right)\\
	  	&=\frac{1}{6}\left(
	  	\sum_{i,j\in\{-2,-1,1,2\}}\log|a_i-b_j|
	  	-\sum_{\substack{i<j\\ i,j\in\{-2,-1,1,2\}}}\log|a_i-a_j|
	  	-\sum_{\substack{i<j\\ i,j\in\{-2,-1,1,2\}}}\log|b_i-b_j|
	  	\right).
	  		\end{split}
	  \end{equation}
	  
	  By construction of image points across the boundary, we use the explicit coordinates in method of image
	\begin{equation}
		\begin{split}
			a_1=(\xi_1, \bar{\xi}_1), b_1=(\xi_2, \bar{\xi}_2),
			a_2=(\xi_3, \bar{\xi}_3), b_2=(\xi_4, \bar{\xi}_4),\\
			a_{-1}=(\bar{\xi}_2, \xi_2), b_{-1}=(\bar{\xi}_1 ,\xi_1),
			a_{-2}=(\bar{\xi}_4, \xi_4), b_{-2}=(\bar{\xi}_3 ,\xi_3),\\
		\end{split}
	\end{equation}
	\eqref{eq:master-UHP} can be further simplified to
\begin{equation}
	\begin{split}
		&S\!\left(A_{-2}\cup A_{-1}\cup A_1\cup A_2\right)\\
		&=\frac{1}{6}\,
		\log\!\Bigg(
		\sqrt{\left.\left.\left.\left.\left.\left.\left.\left.\frac{d \xi}{d w}\right|_{w=w_1} \frac{d \bar{\xi}}{d \bar{w}}\right|_{\bar{w}=\bar{w}_1} \frac{d \xi}{d w}\right|_{w=w_2} \frac{d \bar{\xi}}{d \bar{w}}\right|_{\bar{w}=\bar{w}_2}\frac{d \xi}{d w}\right|_{w=w_3} \frac{d \bar{\xi}}{d \bar{w}}\right|_{\bar{w}=\bar{w}_3} \frac{d \xi}{d w}\right|_{w=w_4} \frac{d \bar{\xi}}{d \bar{w}}\right|_{\bar{w}=\bar{w}_4}} \\[6pt]
		&\times
			|\xi_1-\bar{\xi}_1|\,
			|\xi_2-\bar{\xi}_2|\,
			|\xi_3-\bar{\xi}_3|\,
			|\xi_4-\bar{\xi}_4| \\[6pt]
		&\times
		\frac{
			|(\xi_1-\xi_2)(\bar{\xi}_1-\bar{\xi}_2)|\,
			|(\bar{\xi}_1-\xi_3)(\xi_1-\bar{\xi}_3)|\,
			|(\xi_2-\xi_3)(\bar{\xi}_2-\bar{\xi}_3)|}{
			|(\bar{\xi}_1-\xi_2)(\xi_1-\bar{\xi}_2)|\,
			|(\xi_1-\xi_3)(\bar{\xi}_1-\bar{\xi}_3)|\,
			|(\bar{\xi}_2-\xi_3)(\xi_2-\bar{\xi}_3)|} \\[6pt]
		&\times
		\frac{
			|(\xi_1-\xi_4)(\bar{\xi}_1-\bar{\xi}_4)|\,
			|(\bar{\xi}_2-\xi_4)(\xi_2-\bar{\xi}_4)|\,
			|(\xi_3-\xi_4)(\bar{\xi}_3-\bar{\xi}_4)|}{
			|(\bar{\xi}_1-\xi_4)(\xi_1-\bar{\xi}_4)|\,
			|(\xi_2-\xi_4)(\bar{\xi}_2-\bar{\xi}_4)|\,
			|(\bar{\xi}_3-\xi_4)(\xi_3-\bar{\xi}_4)|} 
		\Bigg),
	\end{split}
\end{equation}
which is consistent with \eqref{eq:ss-2pt-rcft}.

\begin{figure}[htbp]
	\centering		
	\begin{subfigure}{0.45\linewidth}
		\centering
		\includegraphics[width=\linewidth]{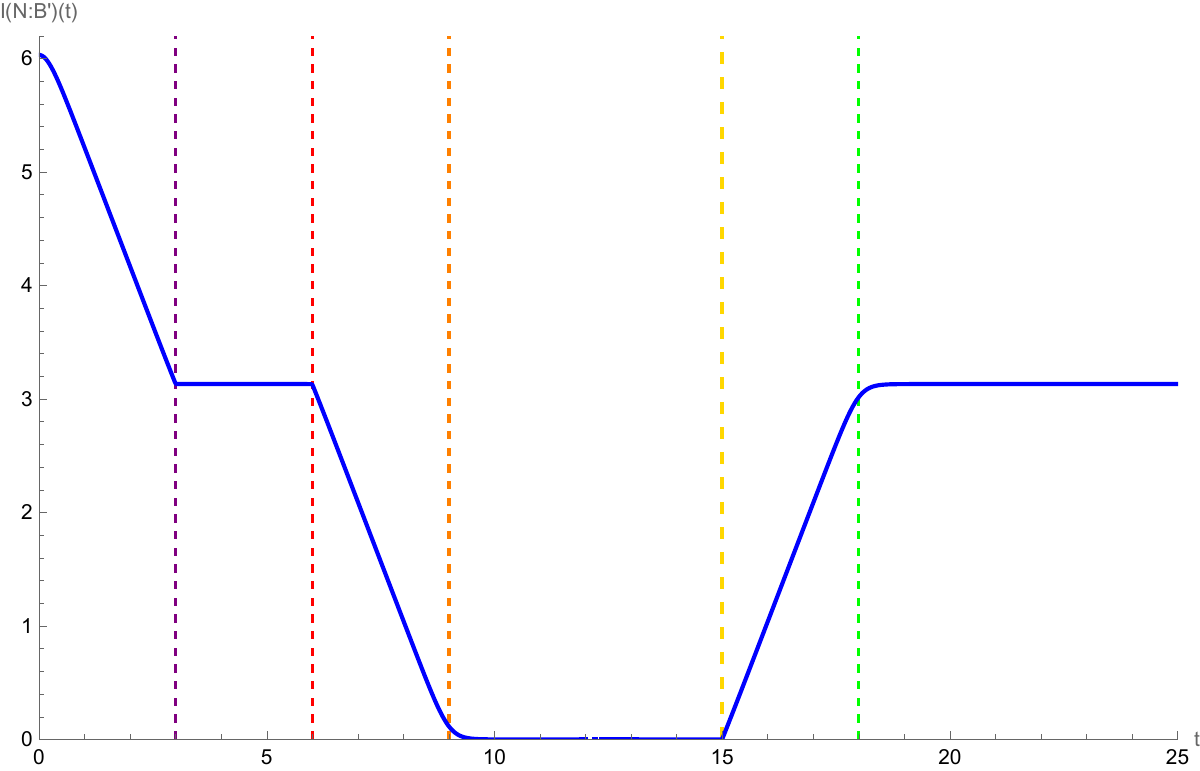}
		\caption{Mutual information $I(N\!:\!B')$ for the radiation length larger than the reference system ($|R|=3 L$).}
		
	\end{subfigure}
	\hfill
	\begin{subfigure}{0.45\linewidth}
		\centering
		\includegraphics[width=\linewidth]{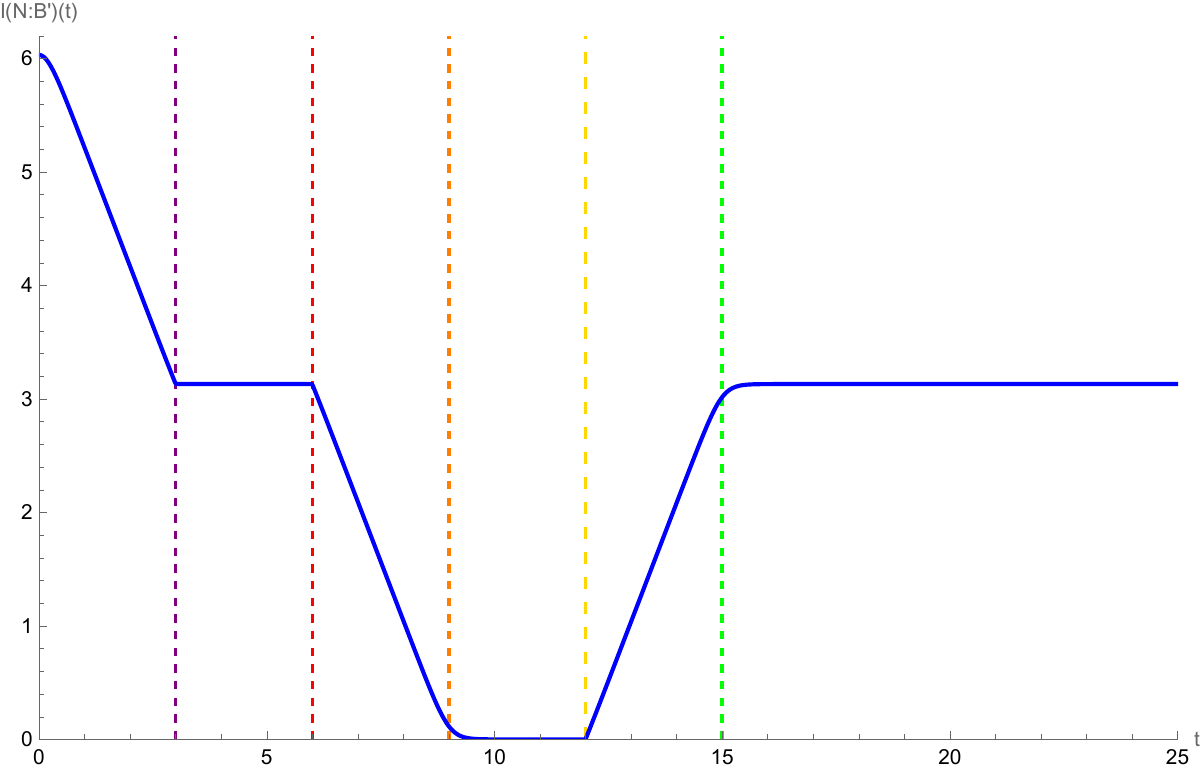}
		\caption{Mutual information $I(N\!:\!B')$ for the radiation length larger than the reference system ($|R|=2 L$).}
	\end{subfigure}
	\\	\begin{subfigure}{0.45\linewidth}
		\centering
		\includegraphics[width=\linewidth]{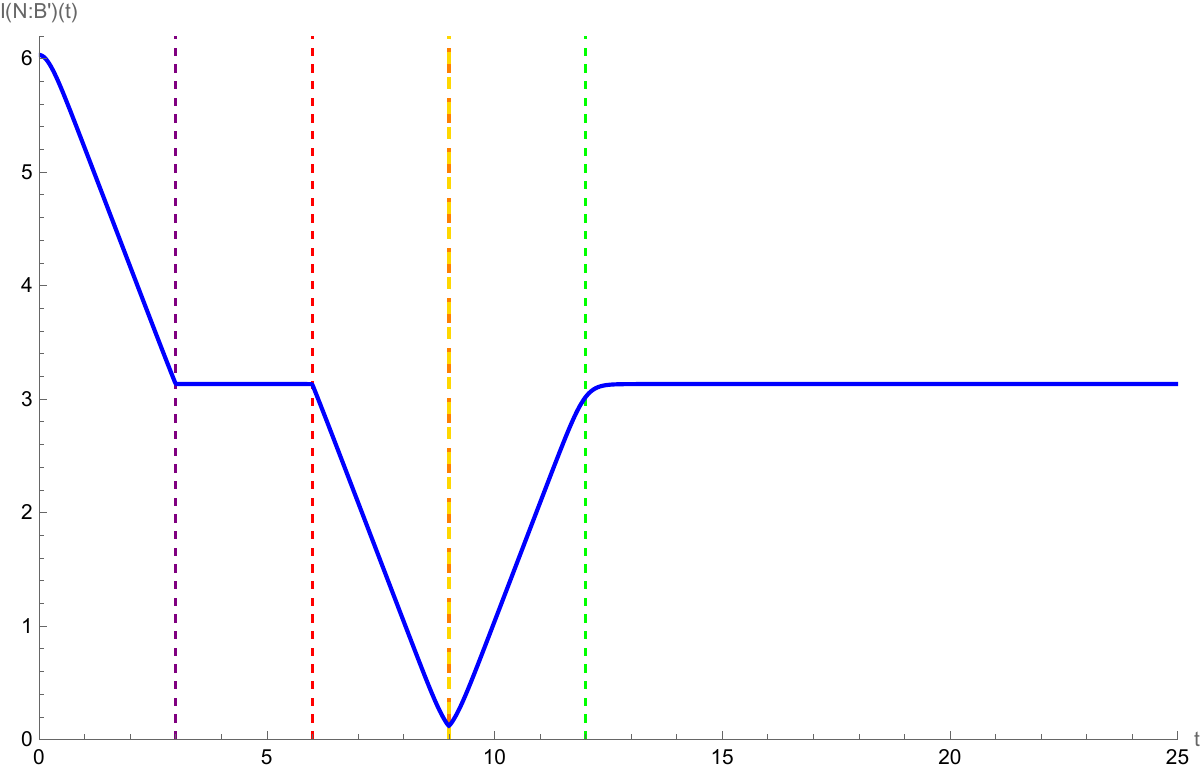}
		\caption{Mutual information $I(N\!:\!B')$ for the radiation length equal to the reference system ($|R|=L$).}
		
	\end{subfigure}
	\hfill
	\begin{subfigure}{0.45\linewidth}
		\centering
		\includegraphics[width=\linewidth]{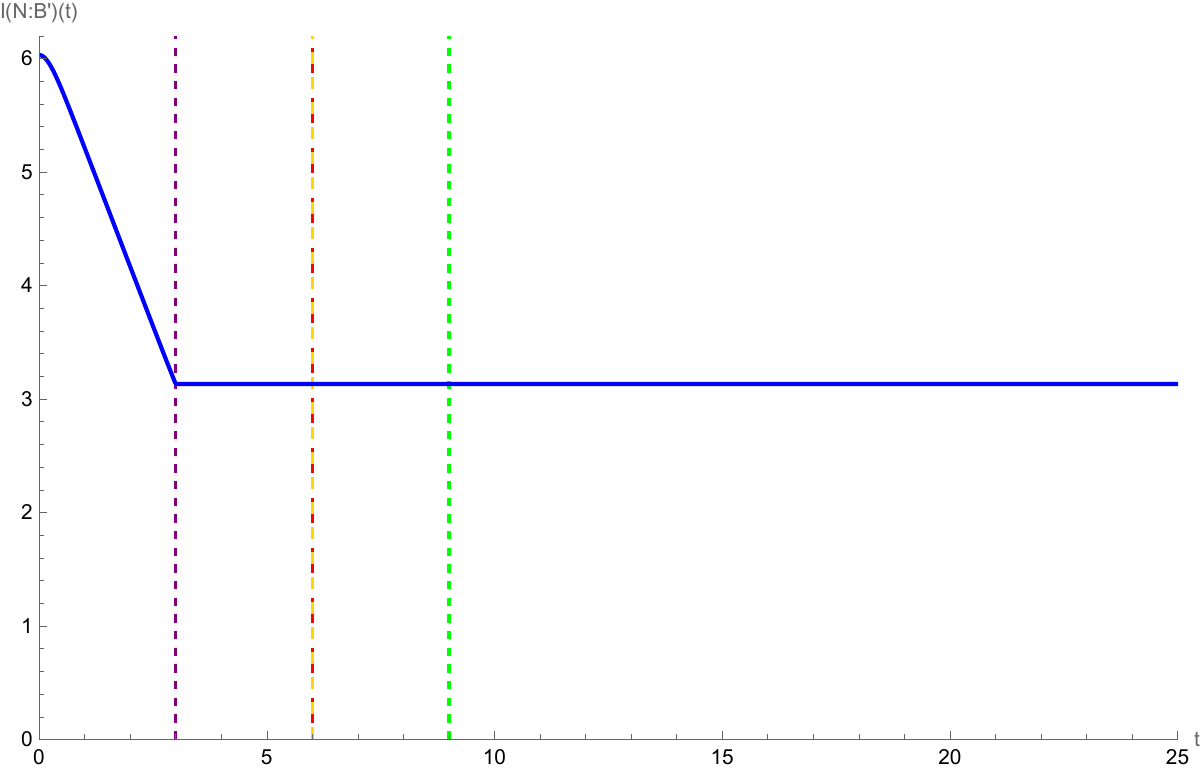}
		\caption{Mutual information $I(N\!:\!B')$ for the radiation length smaller than the reference system ($|R|=0L$).}
	\end{subfigure}
\caption{Plots of $I(N:B')$ as a function of the time for different radiation lengths in the free fermion CFT.  
	In free fermion theories, integrability forbids scrambling and conserves quasi-particle number.  
	Entangled quasi-particles created in $M$ propagate ballistically to the left and right.  
	As a result, the mutual information $I(N:B')$ displays a non-monotonic behavior: it initially decreases when the left-moving partners leave $M$, and later partially recovers once the right-moving partners, after entering $R$, return to the $B'$ subsystem. The parameters we used in this plot are $\alpha=0.01,   \beta = 1 , \epsilon=1, c= 1, \delta= 0.01, X_1=L=3, X_2= 2 L=6 , X_3= 2 L+|R|$ }
\label{fig:SS-Mutual-RCFT}
\end{figure}

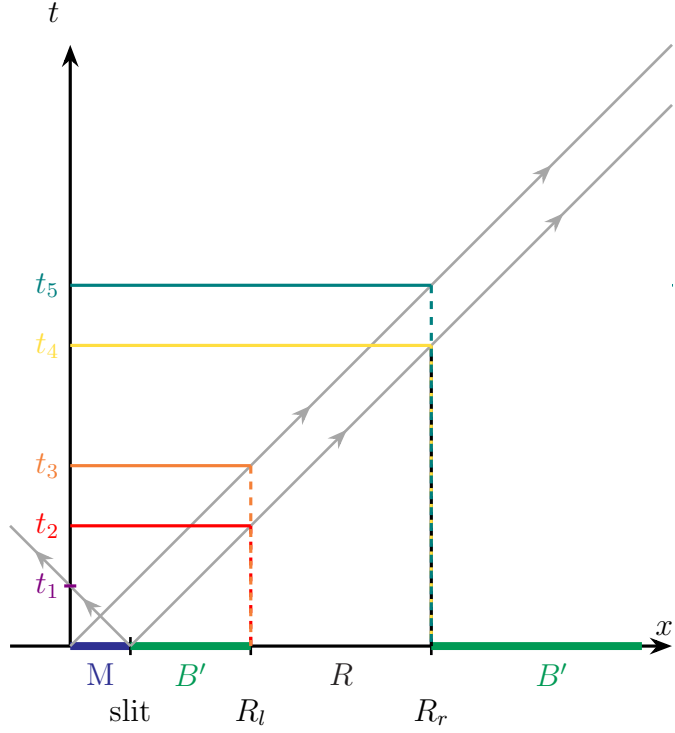
\begin{figure}[htbp]
	\centering
	\begin{tikzpicture}[scale=0.8,>=Stealth]
		
		\draw[->, very thick] (0,0) -- (0,10);         
		\draw[->, very thick] (-1,0) -- (10,0);        
		
		\node[anchor=south east] at (0,10.2) {$t$};
		\node[anchor=south east] at (10.2,0) {$x$};
		
		\draw[line width=3pt, Blue] (0,0) -- (1.0,0);
		\node[Blue, below] at (0.5,-0.1) {M};
		
		\draw[line width=1pt] (1,-0.15) -- (1,0.15);
		\node[below] at (1,-0.7) {slit};
		\draw[line width=1pt] (3,-0.15) -- (3,0.15);
		\node[below] at (3,-0.7) {$R_l$};
		\draw[line width=1pt] (6,-0.15) -- (6,0.15);
		\node[below] at (6,-0.7) {$R_r$};
		
		\draw[line width=3pt, ForestGreen] (1,0) -- (3,0);
		\draw[line width=3pt, ForestGreen] (6,0) -- (9.5,0);
		\node[ForestGreen, below] at (2,-0.1) {$B'$};
		\node[ForestGreen, below] at (8,-0.1) {$B'$};
		\node[Black, below] at (4.5,-0.1) {$R$};
		
		\draw[line width=1.2pt] (6,0) -- (6,5);
		
		\draw[line width=1pt, gray!70,
		postaction={decorate},
		decoration={markings,
			mark=at position 0.4 with {\arrow{>}},
			mark=at position 0.8 with {\arrow{>}}}] (0,0) -- (10,10);
		\draw[line width=1pt, gray!70,
		postaction={decorate},
		decoration={markings,
			mark=at position 0.4 with {\arrow{>}},
			mark=at position 0.8 with {\arrow{>}}}] (1,0) -- (-1,2);	
		
		\draw[line width=1pt, gray!70,
		postaction={decorate},
		decoration={markings,
			mark=at position 0.4 with {\arrow{>}},
			mark=at position 0.8 with {\arrow{>}}}] (1,0) -- (10,9);

		\def\yone{1}
		\def\ytwo{2}
		\def\ythr{3}
		\def\yfor{5}
		\def\yfive{6}
		
		\draw[very thick, violet]      (-0.1,\yone) -- (0.1,\yone);
		\draw[very thick, red]         (0,\ytwo) -- (3,\ytwo);
		\draw[very thick, Orange]      (0,\ythr) -- (3,\ythr);
		\draw[very thick, Goldenrod]   (0,\yfor) -- (6,\yfor);
		\draw[very thick, teal]        (0,\yfive) -- (6,\yfive);
		
		\draw[line width=1pt, teal]      (10,\yfive) -- (10,\yfive); 
		
		\draw[dashed, line width=1.2pt, red] (3,\ytwo) -- (3,0);   
		\draw[dashed, line width=1.2pt, Orange] (3,\ythr) -- (3,0);
		\draw[dashed, line width=1.2pt, Goldenrod] (6,\yfor) -- (6,0);
		\draw[dashed, line width=1.2pt, teal] (6,\yfive) -- (6,0); 
		
		\node[left, violet]    at (0,\yone) {$t_1$};
		\node[left, red]       at (0,\ytwo) {$t_2$};
		\node[left, Orange]    at (0,\ythr) {$t_3$};
		\node[left, Goldenrod] at (0,\yfor) {$t_4$};
		\node[left, teal]      at (0,\yfive) {$t_5$};
		
	\end{tikzpicture}\caption{Characteristic times for the single-slit (no-boundary) quench. Two lightlike fronts (speed $v=1$ ) are emitted from the endpoints of $M$ at $x=0$ and $x= X_1$. With no reflecting boundary, the left-moving front from $x=X_1$ escapes to $x \rightarrow-\infty$. The five marked times are:
	$t_1=X_1$ : this left-moving front reaches the left endpoint of $M$;
	$t_2=R_l-X_1$ : first entry of a right-moving front into the radiation interval
	$R=\left[R_l, R_r\right]$;
	$t_3=R_l-0$ :  entry time of right-moving front from left endpoint of $M$;
	$t_4=R_r-X_1$ : first exit time from $R$;
	$t_5=R_r-0$ : last exit time.
	With $X_1=1, R_l=3, R_r=6$, we have $t_1=1, t_2=2, t_3=3, t_4=5, t_5=$ 6.}\label{fig:SS-time}
\end{figure}

	In this way, we obtain the mutual information between the reference system \(N\) and the post–emission
	black hole \(B'\). This is explicitly plotted in Fig.~\ref{fig:SS-Mutual-RCFT}.
	Its time evolutions can be completely explained by the quasi-particle picture, whose characteristic times 
    $t_1,t_2,t_3,t_4$ and $t_5$ are defined in  Fig.~\ref{fig:SS-time}. The initial decay of mutual information started from $t=0$ and ends at $t=t_1$, due to the left-moving modes which escape out of $M$.
Between \(t_2\) and \(t_5\),
	the subsequent piecewise–linear decrease, saturation, and partial revival arise from the
	ballistic propagation of the right–moving partners across the radiation segment \(R\):
crossing \(R\) progressively depletes their correlations with \(B'\) (linear decay), until all relevant quasi-particles have crossed (saturation), after which the mutual information partially revives. Eventually, at late time $t>t_5$, the value approaches one half of the initial mutual information because the left–moving partners initially entangled with \(N\) have left the
    \(B'\) region. 

\subsection{Bounded slit model calculation}

 In order to analyze the bounded slit model, we now consider a conformal map from the complex plane with coordinate \(w\)  to a cylinder coordinate \(y\). By introducing the cylindrical coordinate $y$ through
	$\zeta=\rho e^y, \quad \bar{\zeta}=\rho e^{\bar{y}},
	$ after mapping we have $0 \leq \operatorname{Re} y \leq$ $-\log \rho$ and $0 \leq \operatorname{Im} y \bmod 2\pi  \leq 2 \pi$. 
	Under this transformation, the correlation function of vertex operators 
	\(\sigma^{(k)}(w_1)\) and \(\sigma^{(-k)}(w_2)\)
	on the complex plane is related to the correlation function on the cylinder by
	\begin{equation}
		\left\langle\sigma^{(k)}\left(w_1, \bar{w}_1\right) \sigma^{(-k)}\left(w_2, \bar{w}_2\right)\right\rangle_{\mathrm{BS}}=\prod_{i=1}^2\left(\frac{d y}{d w}\left(w_i\right)\right)^{h_k}\left(\frac{d \bar{y}}{d \bar{w}}\left(\bar{w}_i\right)\right)^{\bar{h}_k}\left\langle\sigma^{(k)}\left(y_1, \bar{y}_1\right) \sigma^{(-k)}\left(y_2, \bar{y}_2\right)\right\rangle_{\text {cylinder }}.
	\end{equation}
	Here, let us list each $y_i$ we will use below:
	\begin{equation}
		\begin{array}{ll}
			y_{M l}(t)=\log\!\bigl[\tfrac{\zeta_{M l}(t)}{\rho}\bigr], &\qquad
			\bar y_{M l}(t)=\log\!\bigl[\tfrac{\bar{\zeta}_{M l}(t)}{\rho}\bigr], \\[6pt]
			y_{R l}(t)=\log\!\bigl[\tfrac{\zeta_{R l}(t)}{\rho}\bigr], &\qquad
			\bar y_{R l}(t)=\log\!\bigl[\tfrac{\bar \zeta_{R l}(t)}{\rho}\bigr], \\[6pt]
			y_{R r}(t)=\log\!\bigl[\tfrac{\zeta_{R r}(t)}{\rho}\bigr], &\qquad
			\bar y_{R r}(t)=\log\!\bigl[\tfrac{\bar \zeta_{R r}(t)}{\rho}\bigr], \\[6pt]
			y_{B r}(t)=\log\!\bigl[\tfrac{a}{\rho}\bigr], &
			\bar y_{B r}(t)=\log\!\bigl[\tfrac{a}{\rho}\bigr], \\[6pt]
			y_{E r}(t)=\log\!\bigl[\tfrac{a}{\rho}\bigr], &
			\bar y_{E r}(t)=\log\!\bigl[\tfrac{a}{\rho}\bigr], \\[6pt]
			y_{N l}(t)=\log\!\bigl[\tfrac{1}{\rho}\bigr]+\pi i, &
			\bar y_{N l}(t)=\log\!\bigl[\tfrac{1}{\rho}\bigr]-\pi i, \\[6pt]
			y_{N r}(t)=\pi i, &
			\bar y_{N r}(t)=-\pi i.
		\end{array}
	\end{equation}
	The correlation function of vertex operators on cylinder with the Dirichlet boundary condition is given by
		\begin{equation}
	\begin{aligned}
		& \left\langle V_{\left(k_R, k_L\right)}\left(y_1, \bar{y}_1\right) V_{\left(-k_R,-k_L\right)}\left(y_2, \bar{y}_2\right)\right\rangle \\
		= & \frac{\langle B| e^{-2 \pi s H} V_{\left(k_L, k_R\right)}\left(y_1, \bar{y}_1\right) V_{\left(-k_L,-k_R\right)}\left(y_2, \bar{y}_2\right)|B\rangle_N}{\langle B| e^{-2 \pi s H}|B\rangle_N} \\
		= & \frac{\theta_3\left(\left.\frac{ k}{2 \pi n}\left[\left(y_1-y_2\right)-\left(\bar{y}_1-\bar{y}_2\right)\right] \right\rvert\, \frac{i}{\pi s }\right)}{\theta_3\left(0 \left\lvert\, \frac{i}{\pi s }\right.\right)}\\
		& \cdot\left(\frac{\eta(2 i s)^3}{\theta_1\left(\left.\frac{y_2-y_1}{2 \pi i} \right\rvert\, 2 i s\right)}\right)^{\frac{k^2}{n^2}} \cdot\left(\frac{\eta(2 i s)^3}{\theta_1\left(\left.\frac{\bar{y}_2-\bar{y}_1}{2 \pi i} \right\rvert\, 2 i s\right)}\right)^{\frac{k^2}{n^2}} \cdot\left(
		\frac{
			\theta_1\!\Bigl(\dfrac{y_1+\bar y_2}{2\pi i}\,\Big|\,2is\Bigr)\,
			\theta_1\!\Bigl(\dfrac{y_2+\bar y_1}{2\pi i}\,\Big|\,2is\Bigr)}
		{\theta_1\!\Bigl(\dfrac{y_1+\bar y_1}{2\pi i}\,\Big|\,2is\Bigr)\,
			\theta_1\!\Bigl(\dfrac{y_2+\bar y_2}{2\pi i}\,\Big|\,2is\Bigr)}
		\right)^{\frac{k^2}{n^2}}.
	\end{aligned}
	\end{equation}
	Here we introduce $s=-\frac{1}{2 \pi} \log \rho$;  $k_L$ and $k_R$ denote the left- and right-moving charge of the vertex operator, for the Dirichlet boundary condition we have $k_L=k_R =\frac{k}{n}$.  In our case, as a simple check, one can find the correlation function in $w$ plane can be represent as 
	\begin{equation}
		\begin{aligned}
			&\bigl\langle
			\sigma_n(w_1,\bar w_1)\,
			\sigma_{-n}(w_2,\bar w_2)
			\bigr\rangle_{\text{BS}}\\
			&=\Biggl[
			\sqrt{
				\left.\frac{\mathrm{d}y}{\mathrm{d}w}\right|_{w=w_1}\!
				\left.\frac{\mathrm{d}\bar y}{\mathrm{d}\bar w}\right|_{\bar w=w_1}\!
				\left.\frac{\mathrm{d}y}{\mathrm{d}w}\right|_{w=w_2}\!
				\left.\frac{\mathrm{d}\bar y}{\mathrm{d}\bar w}\right|_{\bar w=w_2}}
			\\[8pt]
			&\cdot
			\frac{\eta(2is)^{3}}
			{\theta_1\!\Bigl(\tfrac{y_2-y_1}{2\pi i}\,\Big|\,2is\Bigr)}
			\;
			\frac{\eta(2is)^{3}}
			{\theta_1\!\Bigl(\tfrac{\bar y_2-\bar y_1}{2\pi i}\,\Big|\,2is\Bigr)}
		 \cdot
			\frac{
				\theta_1\!\Bigl(\dfrac{y_1+\bar y_1}{2\pi i}\,\Big|\,2is\Bigr)\,
				\theta_1\!\Bigl(\dfrac{y_2+\bar y_2}{2\pi i}\,\Big|\,2is\Bigr)}
			{\theta_1\!\Bigl(\dfrac{y_1+\bar y_2}{2\pi i}\,\Big|\,2is\Bigr)\,
				\theta_1\!\Bigl(\dfrac{y_2+\bar y_1}{2\pi i}\,\Big|\,2is\Bigr)}
			\Biggr]^{\frac{1}{12}\!\left(n-\frac1n\right)}  
			\\& \cdot \prod^{\frac{n-1}{2}}_{-\frac{n-1}{2}} \frac{\theta_3\left(\left.\frac{k}{2 \pi n}\left[\left(y_1-y_2\right)-\left(\bar{y}_1-\bar{y}_2\right)\right] \right\rvert\, \frac{i}{\pi s}\right)}{\theta_3\left(0 \left\lvert\, \frac{i}{\pi s}\right.\right)}.
		\end{aligned}
	\end{equation}
	 Finally, we can calculate the entanglement entropy from the formula by definition, assuming $w_1$ and $w_2$ are the end points of subsystem $A$,
	\begin{equation}
 S_A^{(n)}=\frac{1}{1-n} \log \left\langle\sigma_n\left(w_1, \bar{w}_1\right) \sigma_{-n}\left(w_2, \bar{w}_2\right)\right\rangle_{\mathrm{BS}},
 	\end{equation}
which is evaluated as follows:
\begin{equation}
	\begin{aligned}
		S_A^{(n)} 
		&=\frac{1}{1-n}\Biggl\{
		\frac{1}{12}\!\left(n-\frac{1}{n}\right)\,
		\log\!\Biggl[
		\sqrt{
			\left.\frac{dy}{dw}\right|_{w_1}
			\left.\frac{d\bar y}{d\bar w}\right|_{\bar w_1}
			\left.\frac{dy}{dw}\right|_{w_2}
			\left.\frac{d\bar y}{d\bar w}\right|_{\bar w_2}}
		\\[-2pt]
		&\qquad\qquad\qquad\qquad\cdot
		\frac{\eta(2is)^{3}}{\theta_1\!\left(\frac{y_2-y_1}{2\pi i}\,\Big|\,2is\right)}\;
		\frac{\eta(2is)^{3}}{\theta_1\!\left(\frac{\bar y_2-\bar y_1}{2\pi i}\,\Big|\,2is\right)}\;
		\frac{\theta_1\!\left(\frac{y_1+\bar y_1}{2\pi i}\,\Big|\,2is\right)\,
			\theta_1\!\left(\frac{y_2+\bar y_2}{2\pi i}\,\Big|\,2is\right)}
		{\theta_1\!\left(\frac{y_1+\bar y_2}{2\pi i}\,\Big|\,2is\right)\,
			\theta_1\!\left(\frac{y_2+\bar y_1}{2\pi i}\,\Big|\,2is\right)}
		\Biggr]
		\\[4pt]
		&\quad
		+\;\log
		\prod_{\substack{k=-(n-1)/2}}^{(n-1)/2}
		\frac{\theta_3\!\left(\frac{k}{2\pi n}\big[(y_1-y_2)-(\bar y_1-\bar y_2)\big]\,\Big|\,\frac{i}{\pi s}\right)}
		{\theta_3\!\left(0\,\Big|\,\frac{i}{\pi s}\right)}
		\Biggr\}.
	\end{aligned}
\end{equation}

To evaluate the mutual information we must generalize the computation from the 
two–point functions of vertex operators to the four–point function. 
Fortunately, this was carried out in \cite{Nozaki:2023fkx}. 
For the Dirichlet boundary conditions, by using the method developed in \cite{Takayanagi:2010wp,Numasawa:2016emc,Takayanagi:2022xpv,Nozaki:2023fkx}, the double–interval entanglement entropy as 
   \begin{equation} 
   	\begin{aligned} 
   S_{A\cup B}^{(n)} 
   =\frac{1}{1-n}\,\log
   \Biggl\{
   \Biggl[
   \sqrt{
   	\left.\frac{dy}{dw}\right|_{w_1}
   	\left.\frac{d\bar y}{d\bar w}\right|_{\bar w_1}
   	\left.\frac{dy}{dw}\right|_{w_2}
   	\left.\frac{d\bar y}{d\bar w}\right|_{\bar w_2}
   	\left.\frac{dy}{dw}\right|_{w_3}
   	\left.\frac{d\bar y}{d\bar w}\right|_{\bar w_3}
   	\left.\frac{dy}{dw}\right|_{w_4}
   	\left.\frac{d\bar y}{d\bar w}\right|_{\bar w_4}
   }
   \\[-2pt]
   \qquad\qquad\cdot\;
   \frac{\eta(2is)^{12}\;
   	\theta_1\!\left(\frac{y_4-y_2}{2\pi i}\,\Big|\,2is\right)
   	\theta_1\!\left(\frac{y_3-y_1}{2\pi i}\,\Big|\,2is\right)}
   {\theta_1\!\left(\frac{y_4-y_3}{2\pi i}\,\Big|\,2is\right)
   	\theta_1\!\left(\frac{y_4-y_1}{2\pi i}\,\Big|\,2is\right)
   	\theta_1\!\left(\frac{y_3-y_2}{2\pi i}\,\Big|\,2is\right)
   	\theta_1\!\left(\frac{y_2-y_1}{2\pi i}\,\Big|\,2is\right)}
   \\
   \qquad\qquad\cdot\;
   \frac{
   	\theta_1\!\left(\frac{\bar y_4-\bar y_2}{2\pi i}\,\Big|\,2is\right)
   	\theta_1\!\left(\frac{\bar y_3-\bar y_1}{2\pi i}\,\Big|\,2is\right)}
   {\theta_1\!\left(\frac{\bar y_4-\bar y_3}{2\pi i}\,\Big|\,2is\right)
   	\theta_1\!\left(\frac{\bar y_4-\bar y_1}{2\pi i}\,\Big|\,2is\right)
   	\theta_1\!\left(\frac{\bar y_3-\bar y_2}{2\pi i}\,\Big|\,2is\right)
   	\theta_1\!\left(\frac{\bar y_2-\bar y_1}{2\pi i}\,\Big|\,2is\right)}
   \\
   \qquad\qquad\cdot\;
   \frac{
   	\theta_1\!\left(\frac{y_1+\bar y_2}{2\pi i}\,\Big|\,2is\right)
   	\theta_1\!\left(\frac{y_1+\bar y_4}{2\pi i}\,\Big|\,2is\right)
   	\theta_1\!\left(\frac{y_2+\bar y_1}{2\pi i}\,\Big|\,2is\right)
   	\theta_1\!\left(\frac{y_2+\bar y_3}{2\pi i}\,\Big|\,2is\right)}
   {\theta_1\!\left(\frac{y_1+\bar y_1}{2\pi i}\,\Big|\,2is\right)
   	\theta_1\!\left(\frac{y_1+\bar y_3}{2\pi i}\,\Big|\,2is\right)
   	\theta_1\!\left(\frac{y_2+\bar y_2}{2\pi i}\,\Big|\,2is\right)
   	\theta_1\!\left(\frac{y_2+\bar y_4}{2\pi i}\,\Big|\,2is\right)}
   \\
   \qquad\qquad\cdot\;
   \frac{
   	\theta_1\!\left(\frac{y_3+\bar y_2}{2\pi i}\,\Big|\,2is\right)
   	\theta_1\!\left(\frac{y_3+\bar y_4}{2\pi i}\,\Big|\,2is\right)
   	\theta_1\!\left(\frac{y_4+\bar y_1}{2\pi i}\,\Big|\,2is\right)
   	\theta_1\!\left(\frac{y_4+\bar y_3}{2\pi i}\,\Big|\,2is\right)}
   {\theta_1\!\left(\frac{y_3+\bar y_1}{2\pi i}\,\Big|\,2is\right)
   	\theta_1\!\left(\frac{y_3+\bar y_3}{2\pi i}\,\Big|\,2is\right)
   	\theta_1\!\left(\frac{y_4+\bar y_2}{2\pi i}\,\Big|\,2is\right)
   	\theta_1\!\left(\frac{y_4+\bar y_4}{2\pi i}\,\Big|\,2is\right)}
   \Biggr]^{\frac{1}{12}\left(n-\frac{1}{n}\right)}
   \\[4pt]
   \qquad\cdot\;
   \prod_{k=-(n-1)/2}^{(n-1)/2}
   \frac{
   	\theta_3\!\left(
   	\frac{k}{2\pi n}\Big[
   	(y_1-y_2+y_3-y_4)-(\bar y_1-\bar y_2+\bar y_3-\bar y_4)
   	\Big]\;\Big|\;\frac{i}{\pi s}\right)}
   {\theta_3\!\left(0\;\Big|\;\frac{i}{\pi s}\right)}
   \Biggr\}.
   \end{aligned}
\end{equation}
Here we define subsystem $A=[w_1,w_2]$ and subsystem $B=[w_3,w_4]$, 
with twist operators $\sigma_n$ inserted at their respective endpoints. 
This allows us to simplify the mutual information formula in \eqref{eq:BS-mutual-renyi}. 
In what follows, we focus on the third Rényi entropy and the corresponding mutual information, instead of the genuine entanglement entropy (or von-Neumann entropy) at $n=1$, because the analytical continuation to $n=1$ is technically complicated in this example:
\begin{equation}
	S_A^{(3)}(t)=-\tfrac{1}{2}\,
	\log \operatorname{Tr}\!\left[\rho_A(t)^3\right].
\end{equation}

\begin{figure}[htbp]
		\centering		
		\begin{subfigure}{0.45\linewidth}
			\centering
			\includegraphics[width=\linewidth]{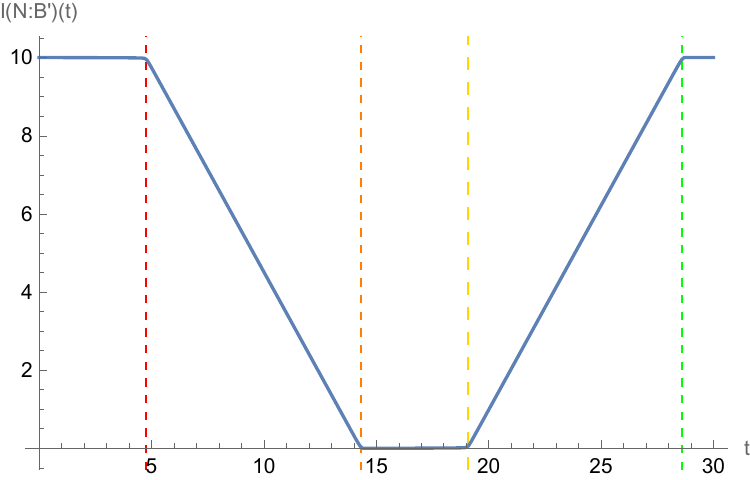}
			\caption{Mutual information $I(N\!:\!B')$ for the radiation length larger than the reference system ($|R|=3L$).}
			
		\end{subfigure}
		\hfill
		\begin{subfigure}{0.45\linewidth}
			\centering
			\includegraphics[width=\linewidth]{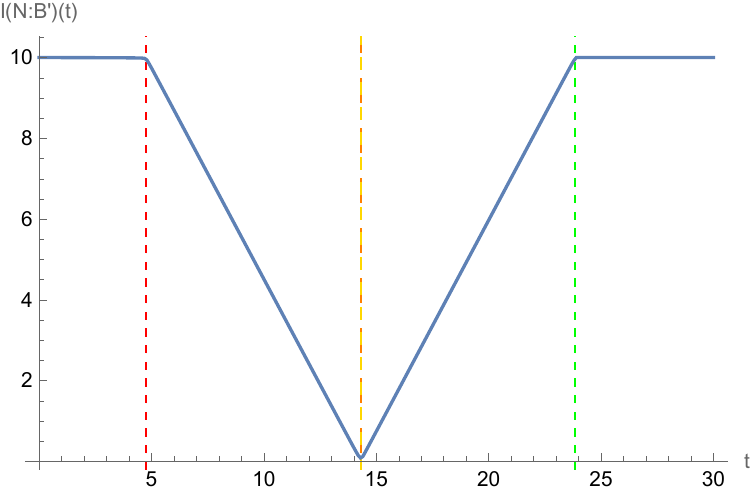}
			\caption{Mutual information $I(N\!:\!B')$ for the radiation length larger than the reference system ($|R|=2L$).}
		\end{subfigure}
		\\	\begin{subfigure}{0.45\linewidth}
			\centering
			\includegraphics[width=\linewidth]{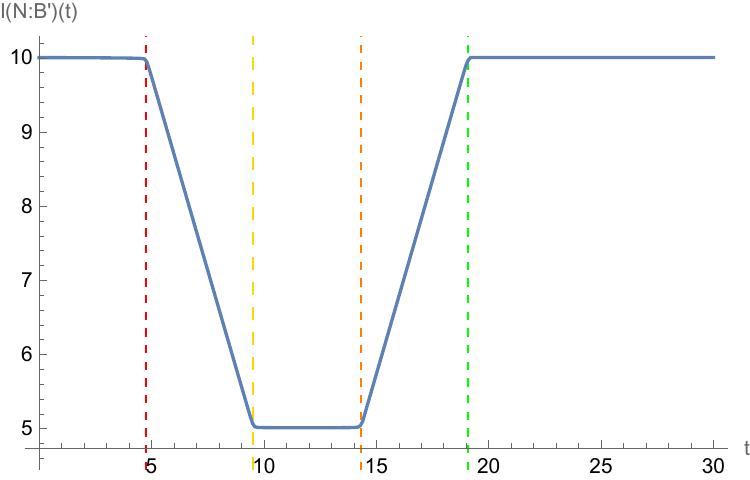}
			\caption{Mutual information $I(N\!:\!B')$ for the radiation length equal to the reference system ($|R|=L$).}
			
		\end{subfigure}
		\hfill
		\begin{subfigure}{0.45\linewidth}
			\centering
			\includegraphics[width=\linewidth]{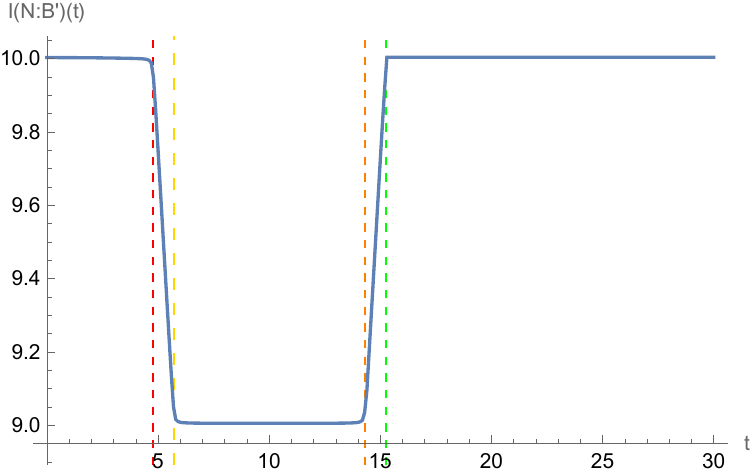}
			\caption{Mutual information $I(N\!:\!B')$ for the radiation length smaller than the reference system ($|R|=0.3L$).}
		\end{subfigure}
		
		\caption{Plots of $I(N\!:\!B')$ as a function of time for different radiation lengths in the free fermion CFT.   
	In the free fermion CFT, where integrability prevents scrambling and particle number is conserved, these quasi-particles reflect back and forth within the system. As a result, the mutual information $I\left(N: B^{\prime}\right)$ exhibits non-monotonic behavior: it initially decreases as entangled quasi-particles exit to R , but eventually recovers as they return. This stands in contrast to holographic CFT, where such return is suppressed due to strong chaos and scrambling. Note that we set $L$ as $L=|M|=|N|$.} \label{fig:BS-Mutual-RCFT}
	\end{figure}

Finally, after some calculation (see Appendix~\ref{sec:detail-of-calculation} for details), 
		we obtain the third Rényi mutual information between the reference system \(N\) and the post–emission black hole \(B'\), 
		as shown in Fig.~\ref{fig:BS-Mutual-RCFT}. 
		The characteristic times $t_1,t_2,t_3$ and $t_4$ of the curves coincide with those based on the quasi-particle picture indicated in Fig.~\ref{fig:BS-time}.
		At late times the mutual information returns to its initial value, as the left–moving partners 
		originally entangled with \(N\) are reflected at the boundary and recombine with their counterparts. 
		In the intermediate regime between \(t_1\) and \(t_4\), the mutual information displays a sequence of decay, plateau, and revival in an obvious way. 
		This pattern reflects the ballistic motion of the right–moving partners: as they enter and leave the radiation interval \(R\), 
		the correlations with \(N\) are first reduced, then temporarily saturated, and eventually restored through boundary reflections, 
		leading to full revivals. Note also that we can have $I(N:B')=0$  only for $|R|\geq 2L$ as only in that case we can accommodate the all modes propagated from $M$ inside $R$, where $L$ is the size of $M$.

	\begin{figure}[H]
		\centering
		\begin{tikzpicture}[scale=0.8,>=Stealth]
			
			\draw[->, very thick] (0,0) -- (0,10);         
			\draw[->, very thick] (0,0) -- (10,0);         
			
			\node[anchor=south east] at (0,10.2) {$t$};
			\node[anchor=south east] at (10.2,0) {$x$};
			
			\draw[line width=3pt, Blue] (0,0) -- (1.0,0);  
			\node[Blue, below] at (0.5,-0.1) {M};
			
			\draw[line width=1pt] (1,-0.15) -- (1,0.15);
			\node[below] at (1,-0.7) {slit};
			\draw[line width=1pt] (3,-0.15) -- (3,0.15);
			\node[below] at (3,-0.7) {$R_l$};
			\draw[line width=1pt] (6,-0.15) -- (6,0.15);
			\node[below] at (6,-0.7) {$R_r$};
			\draw[line width=3pt, ForestGreen] (1,0) -- (3,0);
			\draw[line width=3pt, ForestGreen] (6,0) -- (9.5,0);
			\node[ForestGreen, below] at (2,-0.1) {$B'$};
			\node[ForestGreen, below] at (8,-0.1) {$B'$};
			\node[Black, below] at (4.5,-0.1) {$R$};
			\draw[line width=1.2pt] (6,0) -- (6,5);
			
			\draw[line width=1pt, gray!70,
			postaction={decorate},
			decoration={markings,
				mark=at position 0.06 with {\arrow{>}},
				mark=at position 0.40 with {\arrow{>}},
				mark=at position 0.80 with {\arrow{>}}}]
			(1,0) -- (0,1) -- (9,10);
			
			\draw[line width=1pt, gray!70,
			postaction={decorate},
			decoration={markings,
				mark=at position 0.40 with {\arrow{>}},
				mark=at position 0.80 with {\arrow{>}}}]
			(1,0) -- (10,9);
			\draw[line width=1pt, gray!70,
			postaction={decorate},
			decoration={markings,
				mark=at position 0.40 with {\arrow{>}},
				mark=at position 0.80 with {\arrow{>}}}]
			(0,0) -- (10,10);
			
			\def\yone{2}
			\def\ytwo{4}
			\def\ythr{5}
			\def\yfor{7}
			
			\draw[very thick, red]      (0,\yone) -- (3,\yone);
			\draw[very thick, Orange]         (0,\ytwo) -- (3,\ytwo);
			\draw[very thick, Goldenrod]      (0,\ythr) -- (6,\ythr);
			\draw[very thick, teal]   (0,\yfor) -- (6,\yfor);
			
			\draw[dashed, line width=1.2pt, red]    (3,\yone) -- (3,0);
			\draw[dashed, line width=1.2pt, Orange]       (3,\ytwo) -- (3,0);
			\draw[dashed, line width=1.2pt, Goldenrod]    (6,\ythr) -- (6,0);
			\draw[dashed, line width=1.2pt, teal] (6,\yfor) -- (6,0);
			
			\node[left, red]    at (0,\yone) {$t_1$};
			\node[left, Orange]       at (0,\ytwo) {$t_2$};
			\node[left, Goldenrod]    at (0,\ythr) {$t_3$};
			\node[left, teal] at (0,\yfor) {$t_4$};
			
		\end{tikzpicture}\caption{Bounded–slit causal structure and characteristic times. Space is on the horizontal axis ($x$) and time on the vertical axis ($t$). A local joining at the slit position $x=X_1$ emits two lightlike fronts (gray lines): one moves right immediately; the other moves left, reflects at the boundary $x=0$, and then travels right. The late–radiation interval is $R=[R_l,R_r]$ (dashed verticals). The four colored time slices mark:
		(i) $t_1$—the first arrival of an $M$–sourced front at $R_l$;
		(ii) $t_2$—the last (reflected) front to enter $R$;
		(iii) $t_3$—the first exit at $R_r$;
		(iv) $t_4$—the exit of the reflected front from $R$.
		With light speed set to unity, the arrival/exit times are
		$t_1=R_l-X_1$, $t_2=R_l+X_1$, $t_3=R_r-X_1$, and $t_4=R_r+X_1$.
		For the parameters shown ($X_1=1$, $R_l=3$, $R_r=6$) one has $(t_1,t_2,t_3,t_4)=(2,4,5,7)$.}
		\label{fig:BS-time}
	\end{figure}

\section{Holographic mutual information dynamics}
\label{sec:holographic-mutual-information}

In 2d holographic CFTs, the entanglement entropy of a boundary region \(A\) is computed by the length of space-like geodesic \(\gamma_A\) homologous to \(A\) in \(\mathrm{AdS}_3\) \cite{Ryu:2006bv,Ryu:2006ef,Hubeny:2007xt}:
\begin{equation}
	S_A \;=\; \frac{\mathcal L(\gamma_A)}{4G_N}\,,
	\qquad
	c \;=\; \frac{3 \ell}{2G_N}\, ,
\end{equation}
where \(G_N\) is the bulk Newton constant, \(\ell\) the \(\mathrm{AdS}\) radius  ,
and \(c\) the central charge. If there are multiple geodesics 
$\gamma_A$, we choose the smallest one among them.

For a boundary conformal field theory (BCFT), the bulk terminates on an end-of-world (EOW)  brane \(Q\) obeying
\(K_{ab}-K\,h_{ab}=T\,h_{ab}\), where \(T\) is the tension of brane \cite{Takayanagi:2011zk,Fujita:2011fp}. Refer also to
\cite{Ugajin:2013xxa,Shimaji:2018czt,Caputa:2019avh} for examples of calculations similar to those in the present paper. The geodesic $\gamma_A$ is allowed to end on \(Q\) so that it is orthogonal to $Q$. When $\gamma_A$ is an interval, there are then two generic candidates: a  connected  geodesic joining the two endpoints on \(\partial\mathrm{AdS}\), and a  disconnected  configuration formed by two segments, each stretching from one endpoint to \(Q\). The correct entanglement entropy is given by the minimum of the candidate lengths, divided by \(4G_N\).

We work in the \(\mathrm{AdS}_3\) Poincar\'e coordinate \((\kappa,\bar\kappa,\lambda)\) whose metric reads
\begin{equation}
	ds^2 \;=\; \ell^2 \frac{d\lambda^2+d\kappa\,d\bar\kappa}{\lambda^2}.
\end{equation}
The geodesic length between two bulk points
\(X=(\kappa_1,\bar\kappa_1,\lambda_1)\) and \(Y=(\kappa_2,\bar\kappa_2,\lambda_2)\) is computed as
\begin{equation}
	\label{eq:L-XY}
	\mathcal L(\Gamma_X^Y)
	\;=\;\operatorname{arccosh}\!\left[
	\frac{(\kappa_1-\kappa_2)(\bar\kappa_1-\bar\kappa_2)+\lambda_1^{2}+\lambda_2^{2}}
	{2\,\lambda_1\lambda_2}\right],
\end{equation}
where \(\Gamma_{X}^{Y}\) denotes the bulk geodesic between \(X\) and \(Y\). 

Conformal transformations in the dual CFT can be lifted to the bulk diffeomorphism in the AdS$_3$. Writing \(\kappa=f(w)\) for a holomorphic map in the 2d CFT (with \(w=x+i\tau\) the thermal cylinder coordinate), a convenient bulk extension \cite{Roberts:2012aq} reads
\begin{equation}
	\begin{aligned}
		\kappa \,&=\, f(w)\;-\;\frac{2 \nu^2\,(f')^2\,\bar f''}{4|f'|^2+\nu^2|f''|^2},\qquad
		\bar\kappa \,=\, \bar f(\bar w)\;-\;\frac{2 \nu^2\,(\bar f')^2\,f''}{4|f'|^2+\nu^2|f''|^2},\\[4pt]
		\lambda \,&=\, \frac{4 \nu\,(f'\bar f')^{3/2}}{4|f'|^2+\nu^2|f''|^2},
	\end{aligned}
\end{equation}
where \(\nu\) is the bulk radial coordinate. In the coordinate system \((w,\bar w,\nu)\), the metric becomes
\begin{equation}
	ds^2=\frac{d\nu^2}{\nu^2}+T(w)\,dw^2+\bar T(\bar w)\,d\bar w^2
	+\Bigl(\frac{1}{\nu^2}+\nu^2 T(w)\bar T(\bar w)\Bigr)\,dw\,d\bar w,
\end{equation}
with
\begin{equation}
	T(w)=\frac{3(f'')^2-2f'f'''}{4\,f'^2},\qquad
	\bar T(\bar w)=\frac{3(\bar f'')^2-2\bar f'\bar f'''}{4\,\bar f'^2},
\end{equation}
where \(T(w)\) and \(\bar T(\bar w)\) are the holomorphic and anti–holomorphic stress tensors in the \(w\) coordinate.

\subsection{Single slit model calculation}
First, we present the dynamical behavior of mutual information in the single--slit model. 
Unlike the case of free fermions, here we cannot rely on the additivity property of mutual information. 
Instead, we need to evaluate each contribution separately by employing the holographic entanglement entropy formula. That is, $S(N)$, $S (B') $, and $S (B'\cup N)$.

In the single–slit geometry the boundary is the upper half–plane (UHP), with coordinate \(\xi=f(w)\) obtained from the thermal cylinder (Sec.~\ref{subsec:setup1}). For a boundary interval \(A=[\xi_1,\xi_2]\) with \(\Im \xi_{1,2}>0\), there are two candidates of contributions to the holographic entanglement entropy on the UHP: the connected geodesic and disconnected ones:
\begin{align}
	\label{eq:S-conn-UHP}
	S_{\mathrm{con}}(A)
	&= \frac{c}{6 \ell}\,\log\!\frac{|\xi_{1}-\xi_{2}|^{2}}{\varepsilon_{1}\varepsilon_{2}},\\[4pt]
	\label{eq:S-dis-UHP}
	S_{\mathrm{dis}}(A)
	&= \frac{c}{6 \ell}\,\log\!\frac{|\xi_{1}-\bar\xi_{1}|}{\varepsilon_{1}}
	\;+\; \frac{c}{6}\,\log\!\frac{|\xi_{2}-\bar\xi_{2}|}{\varepsilon_{2}}\,,
\end{align}
Here \(\varepsilon_i\) are the UV cutoffs in the \(\xi\) coordinate. Since we work with a tensionless EOW brane for simplicity, the boundary entropy \(S_{\mathrm{bdy}}=\log g\) vanishes. The entanglement entropy of \(A\) reads
\begin{equation}
	\label{eq:min-conn-dis}
	S(A)=\min\{S_{\mathrm{con}}(A),\,S_{\mathrm{dis}}(A)\},
\end{equation}
and the phase transition occurs when
\begin{equation}
	\label{eq:transition-criterion}
	\frac{|\xi_{1}-\xi_{2}|^{2}}{|\xi_{1}-\bar\xi_{1}|\,|\xi_{2}-\bar\xi_{2}|}=1\,.
\end{equation}

\begin{figure}[htbp]
		\centering		
		\begin{subfigure}{0.45\linewidth}
			\centering
			\includegraphics[width=\linewidth]{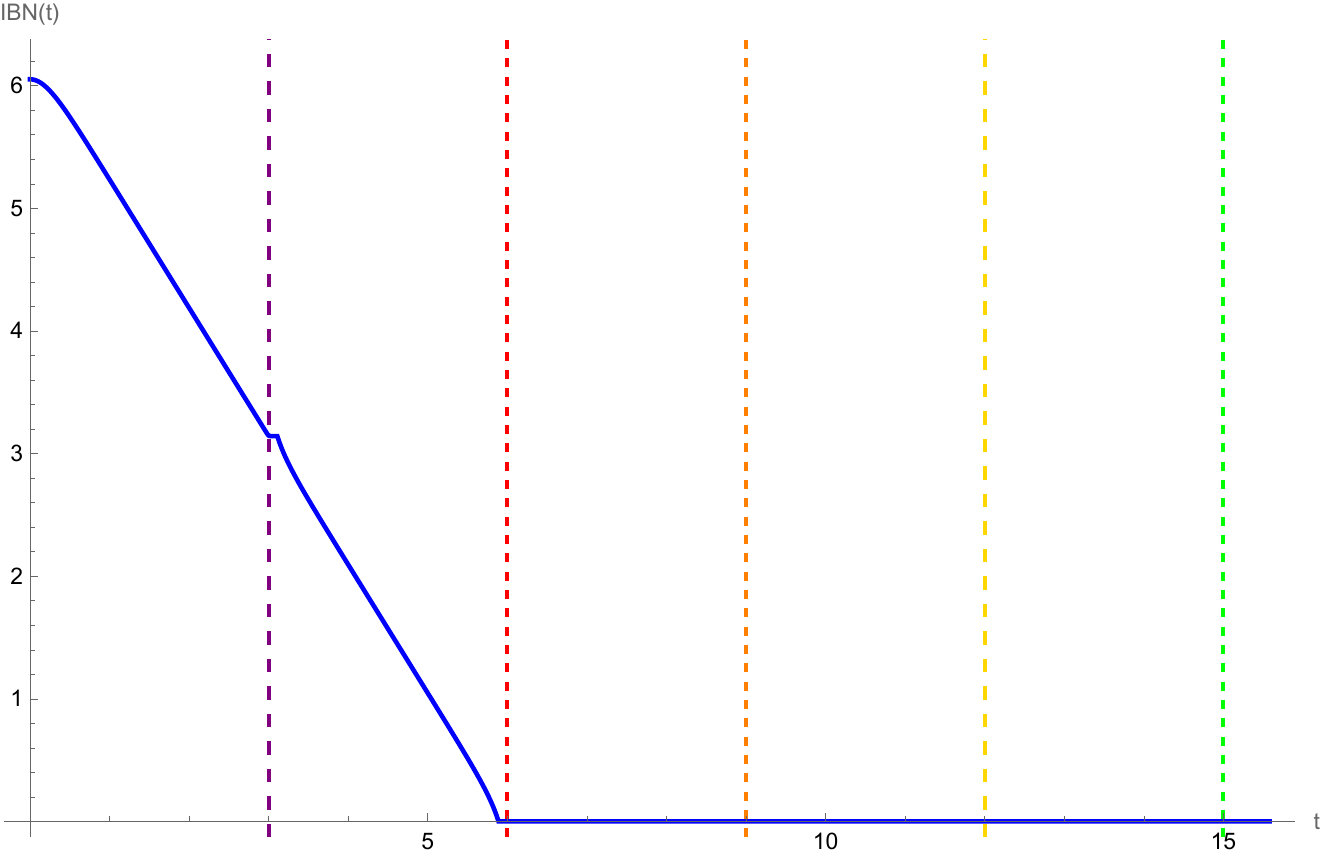}
			\caption{Mutual information $I(N\!:\!B')$ for the radiation length larger than the reference system ($|R|=2L$).}
		\end{subfigure}
		\hfill
		\begin{subfigure}{0.45\linewidth}
			\centering
			\includegraphics[width=\linewidth]{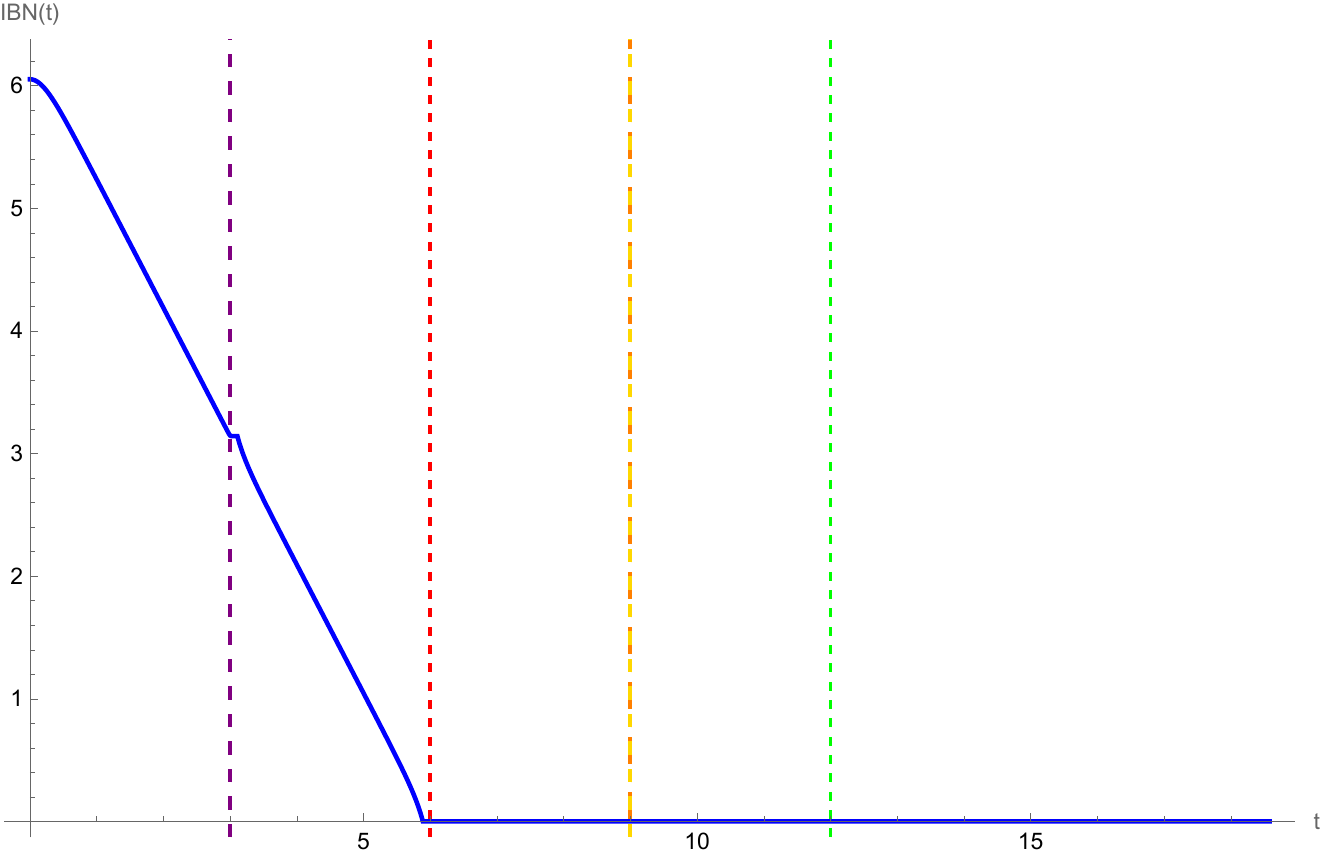}
			\caption{Mutual information $I(N\!:\!B')$ for the radiation length equal to the reference system ($|R|=L$).}
		\end{subfigure}
		\\	\begin{subfigure}{0.45\linewidth}
			\centering
			\includegraphics[width=\linewidth]{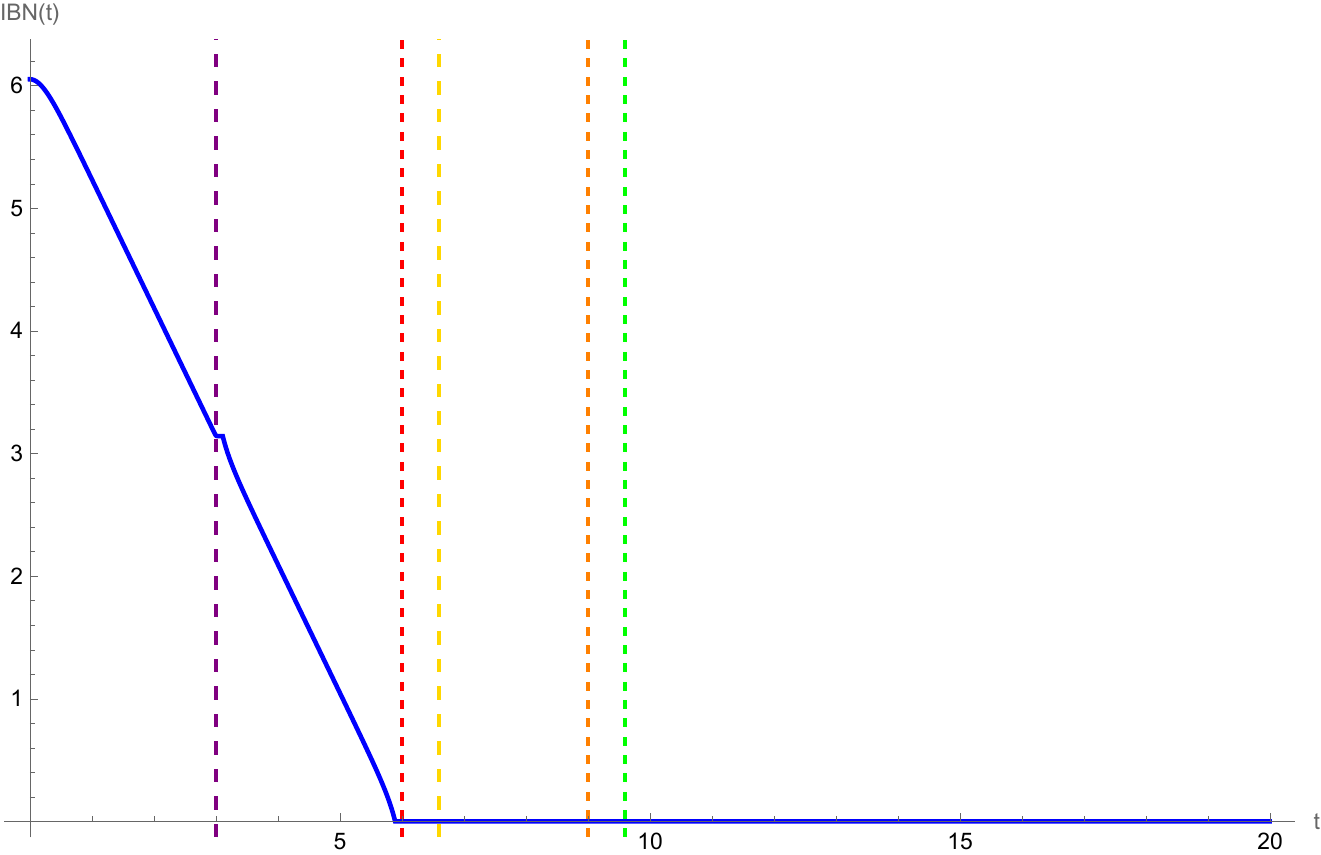}
			\caption{Mutual information $I(N\!:\!B')$ for the radiation length smaller than the reference system ($|R|=0.2L$).}
			
		\end{subfigure}
		\hfill
		\begin{subfigure}{0.45\linewidth}
			\centering
			\includegraphics[width=\linewidth]{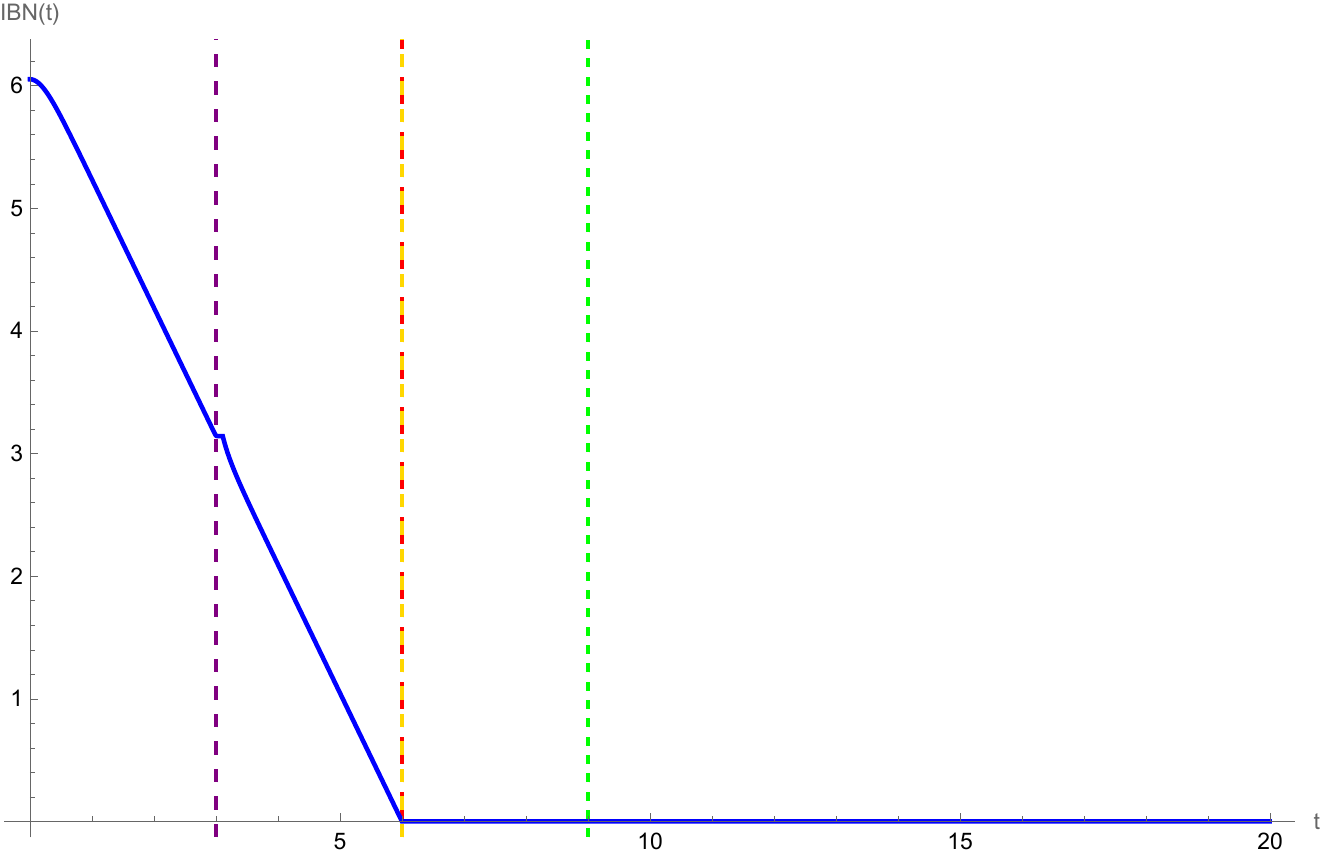}
			\caption{Mutual information $I(N\!:\!B')$ for the radiation length smaller than the reference system ($|R|=0L$).}
		\end{subfigure}
		\caption{Plots of $I(N\!:\!B')$ as a function of time for different radiation lengths in holographic CFT. Here we take $\alpha=1,   \beta = 1 , \epsilon=1, c= 1, \delta= 0.01, X_1=L=3, X_2= 2 L=6 , X_3= 2 L+|R|$. }\label{fig:SS-Mutual-HCFT}
	\end{figure}

If the endpoints are specified on the \(w\)–cylinder and mapped by \(\xi=\xi(w)\), the cutoffs transform as \(\varepsilon_{i}=\varepsilon_{i}^{(w)}|\xi'(w_i)|\). The two contributions then read
\begin{align}
	S_{\mathrm{con}}(A)
	&=\frac{c}{6 \ell}\,\log\!\left[
	\frac{|\xi_{1}-\xi_{2}|^{2}}{\varepsilon_{1}^{(w)}\varepsilon_{2}^{(w)}|\xi'(w_{1})\xi'(w_{2})|}
	\right],\\[4pt]
	S_{\mathrm{dis}}(A)
	&=\frac{c}{6 \ell}\sum_{i=1,2}
	\log\!\left[
\frac{|\xi_{i}-\bar\xi_{i}|}{\varepsilon_{i}^{(w)}|\xi'(w_{i})|}
	\right].
\end{align}
We can similarly compute the holographic entanglement entropy 
$S(A)$ when $A$ consists of multiple intervals. We plotted the time evolution of  the mutual information $I(N:B')$ in  Fig.\ref{fig:SS-Mutual-HCFT}.  At early times, the minimal configuration for $S_{B'}$ is 
	\(\mathcal{L}(\Gamma_{R_r}^{B_r})+\mathcal{L}(\Gamma_{Q}^{R_l})+\mathcal{L}(\Gamma_{Q}^{M_l})\);
	after the first transition time $t_1$ in Fig.\ref{fig:SS-time}, it switches to 
	\(\mathcal{L}(\Gamma_{M_l}^{R_l})+\mathcal{L}(\Gamma_{R_r}^{B_r})\).
	Here $t_1$ is defined with respect to the TFD pairing between $M$ and $N$: it is the first time when the left moving
	rightmost member of the quasi-particle pair that initially entangles $M$ with $N$ reaches the left boundary of $M$.
	Similarly, the geodesic for $S_{B'\cup N}$ is initially 
	\(\mathcal{L}(\Gamma_{R_r}^{E_r})+\mathcal{L}(\Gamma_{M_l}^{N_l})+\mathcal{L}(\Gamma_{Q}^{R_l})\);
	around the later transition time $t_2$ in Fig.\ref{fig:SS-time}, it changes to 
	\(\mathcal{L}(\Gamma_{M_l}^{R_l})+\mathcal{L}(\Gamma_{R_r}^{E_r})+\mathcal{L}(\Gamma_{Q}^{N_l})\).
	The time $t_2$ thus marks the entry of the same rightmost $M$–$N$ entangled front into the radiation interval. 
	
 The qualitative difference between Fig.~\ref{fig:SS-Mutual-HCFT} and Fig.~\ref{fig:SS-Mutual-RCFT}
 is already visible in the single–slit quench. In free fermion CFT case  the mutual information retains a clear
light–cone structure and is sensitive to the size of the radiation interval \(R\).
 By contrast, in holographic CFTs, strong scrambling reorganizes entanglement in a nonlocal way:
 the geodesic that control \(I(N\!:\!B')\) in this geometry are largely
 insensitive to \(|R|\). In fact, even in the limit \(|R|\to 0\) our holographic computation still
 exhibits information transfer out of the black hole region \(B\).
 
 This insensitivity is an artifact of the single–slit construction rather than a feature of the
 HP experiment. The entanglement between $M$ and $N$ can escape in the left direction as the system is extended infinitely for $x<0$.
 To faithfully test the HP recovery one must analyze the bounded–slit geometry, where the Euclidean path integral prepares two maximally
 entangled pairs and the complement relation \((B'\!\cup N)^c=E\cup R\) holds. In that setup
 the mutual information develops a genuine, nontrivial dependence on the size and position of \(R\),
 allowing us to probe the \(R\)-dependence of information recovery as we will see soon later.

\subsection{Bounded slit model calculation}
	For two arbitrary points 
\(X=\bigl(\zeta_1,\bar{\zeta}_1,\eta_1\bigr)\) and 
\(Y=\bigl(\zeta_2,\bar{\zeta}_2,\eta_2\bigr)\) in \(\mathrm{AdS}_3\), 
the geodesic length is

\begin{equation}
	\mathcal{L}(\Gamma_X^Y)=\operatorname{arccosh}\!\left[
	\frac{(\zeta_1-\zeta_2)(\bar\zeta_1-\bar\zeta_2)+\eta_1^{2}+\eta_2^{2}}
	{2\eta_1\eta_2}\right].
\end{equation}

Write the boundary complex coordinates in polar form
\begin{equation}
	\zeta_i=r_i e^{i\theta_i},\qquad 
	\bar\zeta_i=r_i e^{-i\theta_i}\quad(i=1,2).
\end{equation}

The invariant inner product becomes
\begin{equation}
	-X\!\cdot\!Y
	=\frac{|\zeta_1-\zeta_2|^{2}+\eta_1^{2}+\eta_2^{2}}
	{2\eta_1\eta_2},\qquad
	|\zeta_1-\zeta_2|^{2}=r_1^{2}+r_2^{2}-2r_1r_2\cos(\theta_1-\theta_2).
\end{equation}

Accordingly, the geodesic distance in polar coordinates 
\((r,\theta,\eta)\) is
\begin{equation}
	\mathcal{L}(\Gamma_X^Y)=\operatorname{arccosh}\!\left[
	\frac{r_1^{2}+r_2^{2}-2r_1r_2\cos(\theta_1-\theta_2)
		+\eta_1^{2}+\eta_2^{2}}
	{2\eta_1\eta_2}\right].
\end{equation} 

The bounded–slit Euclidean path integral prepares a BCFT state on an annulus
\(\mathcal A=\{\rho\le|\zeta|\le 1\}\). In the gravity dual of the annulus,
the holographic geometry depends on the modulus \(\rho\) through the competition
of two saddle topologies~\cite{Fujita:2011fp,Hawking:1982dh}. When
\(\pi< -\log \rho\), the dual contains two disconnected end–of–the–world
branes; when \(\pi> -\log \rho\), the branes connect and form a single surface.
In the connected case, the mutual information between \(B'\) and \(N\) is
trivially zero at all times. Therefore we focus on the former phase (BTZ phase) below.

In our setup, since we focus on the BTZ phase and require the ratio
\(\tfrac{X_1}{\beta}=\tfrac{\log a}{-2\pi}\) to be relatively large in order
to enhance the initial entanglement between \(B\) and \(E\), we necessarily
take \(a=e^{-2\pi X_1/\beta}\ll 1\). This corresponds to small \(\rho\), so the dominant saddle is the disconnected–brane phase. This choice
ensures that the system models a strongly scrambling black hole. In this regime,
the competition of the geodesic for \(S_{N\cup B'}\) governs the
time evolution of the mutual information. 

\begin{figure}[htbp]
		\centering
		\begin{tikzpicture}[scale=1.0,>=stealth]
			\def\R{4.0}          
			\def\rho{0.1}       
			\def\Ri{\rho*\R}     
			\def\xmin{-4.8}
			\def\xmax{4.8}
			\def\ymax{4.8}
			
			\draw[->,  thick,gray] (\xmin,0)--(\xmax,0) node[below right] {$r$};
			\draw[->, thick,gray] (0,0)--(0,\ymax) node[left] {$\eta$};
			
			\draw[very thick,black]  ( -\R,0) arc[start angle=180, end angle=0, radius=\R] node[midway,anchor=north east] {$Q_1$};
			\draw[very thick,black]  ( -\Ri,0) arc[start angle=180, end angle=0, radius=\Ri] node[midway,xshift=-4mm,yshift=2mm] {$Q_2$};
			
			\draw[very thick,blue] (-\R,0) -- (-\Ri,0);
			\coordinate (x0) at (0,0);
			\coordinate (xE1) at (\Ri,0);
			\coordinate (xE2) at (1.0,0);     
			\coordinate (xB1) at (1.5,0);     
			\coordinate (xR2) at (2.0,0);     
			\coordinate (xB2) at (\R,0);        
			
			\draw[very thick,gray!70] (xE1)--(xE2);             
			\draw[very thick,orange]  (xE2)--(xB1);             
			\draw[very thick,red]     (xB1)--(xR2);             
			\draw[very thick,orange]  (xR2)--(xB2);             
			
			\node[blue,below=3pt]  at (-2.2,0) {$N$};
			\node[gray!70,below=3pt] at ($(xE1)!0.5!(xE2)$) {$E$};
			\node[orange,below=3pt]  at ($(xE2)!0.5!(xB1)$) {$B'$};
			\node[red,below=3pt]     at ($(xB1)!0.5!(xR2)$) {$R$};
			\node[orange,below=3pt]  at ($(xR2)!0.5!(xB2)$) {$B'$}; 
		
			\def\angC{30}    
			\path (\angC:\R) coordinate (QC); 
			\draw[purple,very thick]
			(xE2) .. controls +(0,0.4) and +(0,0.4) .. (xB1);
			\draw[purple,very thick]
			(xR2) .. controls +(90:1.2) and +(\angC-180:1.0) .. (QC);

			\begin{scope}[shift={(5.3,2.8)}]
				\draw[rounded corners,thick] (0,-0.4) rectangle (4.4,2.4);
				\draw[very thick,blue]   (0.3,2.0)--(1.0,2.0) node[right,black] {~$N$};
				\draw[very thick,gray!70](0.3,1.6)--(1.0,1.6) node[right,black] {~$E$};
				\draw[very thick,orange] (0.3,1.2)--(1.0,1.2) node[right,black] {~$B'$};
				\draw[very thick,red]    (0.3,0.8)--(1.0,0.8) node[right,black] {~$R$};
				\draw[very thick,black]  (0.3,0.3)--(1.0,0.3)  node[right,black] {~EoW brane $Q_{1,2}$};
				\draw[very thick,purple] (0.3,-0.2)--(1.0,-0.2) node[right,black] { HEE surface($B'$) };
			\end{scope}
			
		\end{tikzpicture}
		\caption{Schematic illustration of two disconnected tensionless branes 
			\(Q_1:\ \eta=\sqrt{1-r^2}\) and \(Q_2:\ \eta=\sqrt{\rho^2-r^2}\) 
			in the \(\eta\!-\!r\) plane. The boundary segment between \(r=\rho\) and 
			\(r=1\) is partitioned into subsystems at Lorentzian time \(t=0\): 
			\(N\) denotes the external reference, \(E\) the early radiation, 
			\(B'\) the post-emission black hole, \(R\) the late radiation, 
			and \(M\) the infalling message. 
			The purple curves represent one candidate holographic extremal geodesic for $B'$.}
		\label{fig:brane-profile}
	\end{figure}
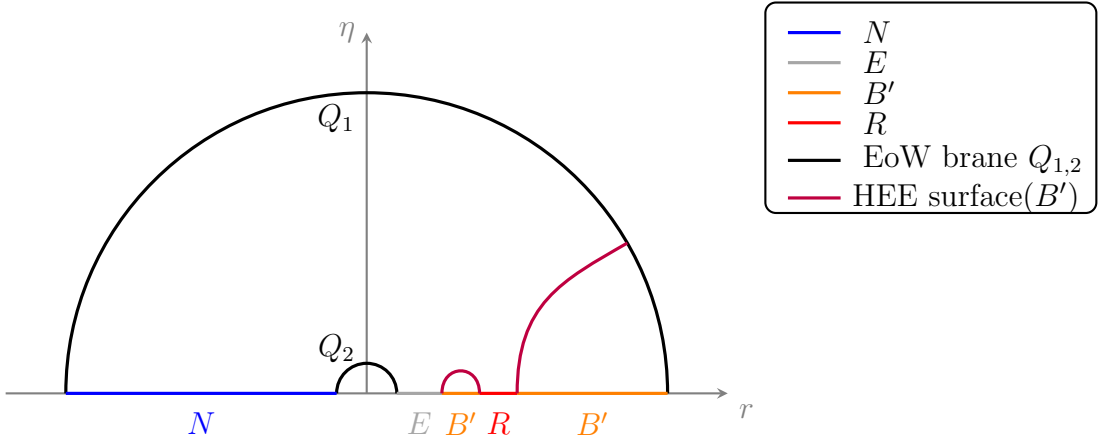

There are two candidates of disconnected tension-less branes as follows from the general rule of holographic entanglement entropy in AdS/BCFT.
  Brane \(Q_1\) is given by 
	\(\eta_1(r)=\sqrt{1-r^{2}}\), while brane \(Q_2\) is described by 
	\(\eta_2(r)=\sqrt{\rho^{2}-r^{2}}\). 
	Their profiles are illustrated in Fig.~\ref{fig:brane-profile} for intuition.

	We mainly employ the Poincaré coordinate
	\begin{equation}
		ds^{2}=\frac{d\eta^{2}+d\zeta\,d\bar\zeta}{\eta^{2}},
	\end{equation}
	with the boundary coordinate 
	\(\zeta=re^{i\theta},\;\bar\zeta=re^{-i\theta}\).
	
	A generic boundary point is described by
	\begin{equation}\label{eq:BoundaryPoint}
		P=(\zeta_1,\bar\zeta_1,\eta_1)
		=(r e^{i\theta},\,r e^{-i\theta},\,\varepsilon),
		\qquad 0<\varepsilon\ll1,
	\end{equation}
	where \(\varepsilon\) is the UV cut-off.
	
	The \(i\)-th Q-brane is anchored on the boundary circle of radius 
	\(r_{Q^{i}}\) and extends into the bulk as
	\begin{equation}\label{eq:QbraneProfile}
		\eta_2(r')=\sqrt{\,r_{Q^{i}}^{\,2}-r'^{\,2}},\qquad 0\le r'\le r_{Q^{i}}.
	\end{equation}
	A generic point on the brane reads
	\begin{equation}
		P'=(\zeta_2,\bar\zeta_2,\eta_2)
		=(r' e^{i\theta'},\,r' e^{-i\theta'},\,\eta_2(r')).
	\end{equation}
	
	The geodesic distance between \(P\) and \(P'\) is given by
	\begin{equation}\label{eq:DistanceBeforeMin}
		\mathcal L(\Gamma_P^{P'})
		=\operatorname{arccosh}\!\left[
		\frac{r^{2}+r'^{2}-2rr'\cos(\theta-\theta')
			+\varepsilon^{2}+\eta_2^{2}(r')}
		{2\varepsilon\,\eta_2(r')}\right].
	\end{equation}
	
	Since \eqref{eq:DistanceBeforeMin} depends on \(\theta'\) 
	only through \(\cos(\theta-\theta')\), the distance is minimized by
	\begin{equation}\label{eq:AngleMin}
		\theta'_{\!*}=\theta.
	\end{equation}
	
	Substituting \eqref{eq:AngleMin} and \eqref{eq:QbraneProfile} into
	\eqref{eq:DistanceBeforeMin} and extremizing with respect to \(r'\) gives
	\begin{equation}\label{eq:OptimalRprime}
		r'_{\!*}
		=\frac{2\,r_{Q^{i}}^{\,2}\,r}{r^{2}+r_{Q^{i}}^{\,2}+\varepsilon^{2}}
		\;\xrightarrow[\varepsilon\to0]{}\;
		\frac{2\,r_{Q^{i}}^{\,2}\,r}{r^{2}+r_{Q^{i}}^{\,2}}.
	\end{equation}
	
	The corresponding bulk depth is computed as
\begin{equation}\label{eq:OptimalXi}
  \eta_{*}=\eta_{2}(r'_{\!*})
  =\frac{r_{Q_i}}{\,r^{2}+r_{Q_i}^{2}+\varepsilon^{2}\,}
   \sqrt{\bigl(r^{2}+r_{Q_i}^{2}+\varepsilon^{2}\bigr)^{2}-4 r^{2} r_{Q_i}^{2}}
  \;\xrightarrow[\varepsilon\to0]{}\;
  \frac{r_{Q_i}\,\bigl|\,r_{Q_i}^{2}-r^{2}\,\bigr|}{\,r^{2}+r_{Q_i}^{2}\,}.
\end{equation}

	Inserting \eqref{eq:AngleMin}–\eqref{eq:OptimalXi} back into
	\eqref{eq:DistanceBeforeMin} yields the minimal geodesic length
	\begin{equation}\label{eq:MinimalDistance}
		\mathcal L_{\min}(\Gamma^P_Q  )
		=\operatorname{arccosh}\!\left[
		\frac{\sqrt{\bigl(r^{2}+r_{Q^{i}}^{\,2}+\varepsilon^{2}\bigr)^{2}
				-4r^{2}r_{Q^{i}}^{\,2}}}
		{2\varepsilon r_{Q^{i}}}\right].
	\end{equation}
	
	Two limiting cases serve as useful consistency checks.  
	First, when $r \to r_{Q^{i}}$, the radicand in \eqref{eq:MinimalDistance} approaches $(2\varepsilon r_{Q^{i}})^{2}$, so the geodesic length $\mathcal L_{\min}$ vanishes, consistent with the expectation that the distance to the brane anchor goes to zero.  
	Second, in the UV limit with fixed $r<r_{Q^{i}}$ and $\varepsilon\to 0$, \eqref{eq:MinimalDistance} reduces to
	\begin{equation}
		\mathcal L_{\min}
		= \log\!\left(\frac{r_{Q^{i}}^{\,2}-r^{2}}{r_{Q^{i}}\,\varepsilon}\right)
		+\mathcal O(\varepsilon),
	\end{equation}
	which reproduces the standard logarithmic divergence associated with the UV cut-off.

	The geodesics connecting the branes $Q_1, Q_2$ to the boundary point $P=\bigl(r\,e^{i\theta},\,r\,e^{-i\theta},\,\varepsilon\bigr)$ is then given by  
	\begin{equation}
		\begin{split}
			&\mathcal{L}\left(\Gamma_{Q_1}^{P}\right)=\operatorname{arccosh}
			\!\left[
			\frac{%
				\sqrt{\bigl(r^{2}+1+\varepsilon^{2}\bigr)^{2}
					-4 r^{2} }
			}{%
				2\,\varepsilon}%
			\right],\\
			&\mathcal{L}\left(\Gamma_{Q_2}^{P}\right)=\operatorname{arccosh}
			\!\left[
			\frac{%
				\sqrt{\bigl(r^{2}+\rho^{\,2}+\varepsilon^{2}\bigr)^{2}
					-4 r^{2} \rho^{\,2}}
			}{%
				2\,\varepsilon\,\rho}%
			\right].
		\end{split}
	\end{equation}

	According to map \cite{Roberts:2012aq},  the connected entropy of an subregion $R=[X_2, X_1]$ is given by the usual formula,
	\begin{equation}
		\begin{split}	
			S_R^{con}=\frac{\mathcal{L}(\Gamma^{1}_{2})}{4 G_N}&=\frac{c}{6 \ell} \operatorname{arccosh}\left[\frac{\left(\zeta_1-\zeta_2\right)\left(\bar{\zeta}_1-\bar{\zeta}_2\right)+\eta_1^2+\eta_2^2}{2 \eta_1 \eta_2}\right] 
			\\&=\frac{c}{6 \ell} \operatorname{arccosh} \left[\frac{\left|f\left(w_1\right)-f\left(w_2\right)\right|^2+ \epsilon^2 \left|f^{\prime}\left(w_1\right)\right|^2+\epsilon^2 \left|f^{\prime}\left(w_2\right)\right|^2}{2 \epsilon^2\left|f^{\prime}\left(w_1\right)\right|\left|f^{\prime}\left(w_2\right)\right|}\right],
		\end{split}	
	\end{equation}
	$\epsilon$ is a small correction in radial distance of $(w,\bar{w}, u)$ coordinate system. The geodesic length connected one point $w_i$ and the brane $Q_1$ is given by $\mathcal{L}\left(\Gamma_{Q_1}^{w_i}\right)$ and connected one point $w_i$ and the brane $Q_1$ is given by $\mathcal{L}\left(\Gamma_{Q_2}^i\right)$, the distance between two branes is noted as $\mathcal{L}\left(\Gamma_{Q_2}^{Q_1}\right)$, where
	\begin{equation}
		\begin{split}
			&\mathcal{L}(\Gamma^{w_i}_{Q_1})=  \operatorname{arccosh}
			\left[
			\frac{
				\sqrt{\bigl(\left|f(w_i)\right|^{2}+1+\epsilon^{2} \left|f'(w_i)\right|^{2}\bigr)^{2}
					-4 \left|f(w_i)\right|^{2} }
			}
			{2 \epsilon \left|f'(w_i)\right|}
			\right]\\
			&\mathcal{L}(\Gamma^{w_i}_{Q_2})=  \operatorname{arccosh}
			\left[
			\frac{
				\sqrt{\bigl(\left|f(w_i)\right|^{2}+\rho^2+\epsilon^{2} \left|f'(w_i)\right|^{2}\bigr)^{2}
					-4 \left|f(w_i)\right|^{2} \rho^2}
			}
			{2 \rho \epsilon \left|f'(w_i)\right|}
			\right]\\
			&\mathcal{L}(\Gamma^{Q_1}_{Q_2})=  \operatorname{arccosh}\left[\frac{1 +\rho^2 }{ 2 \rho}\right],
		\end{split}
	\end{equation}
	again $ \epsilon$ is a small correction in radial distance. When $\rho \approx 0$, one can arrive at $\mathcal{L}(\Gamma^{Q_1}_{Q_2})=\frac{c}{6 \ell} \log \left[\frac{1 }{  \rho}\right]$. 
	Recall that the mutual information between region $N$ (Charlie) and $B^{\prime}$ (the black hole after emission) is defined as
	\begin{equation}
		I\left(N: B^{\prime}\right)=S_N+S_{B^{\prime}}-S_{N \cup B^{\prime}}.
	\end{equation}

Finally, we have the following results of holographic entanglement entropy: 
	\begin{equation}
	\begin{aligned}
		& \quad S_N=\frac{c}{6 \ell} \mathcal{L}\left(\Gamma_{Q_2}^{Q_1}\right)=\frac{c}{6 \ell} \operatorname{arccosh}\left[\frac{1 +\rho^2 }{ 2 \rho}\right] , \\
		& \quad S_{B'}=\frac{c}{6\ell}\,
		\min\Big\{
		\mathcal{L}(\Gamma_{R_r}^{B_r})+\mathcal{L}(\Gamma_{Q_2}^{R_l})+\mathcal{L}(\Gamma_{Q_1}^{Q_2}),\;
		\mathcal{L}(\Gamma_{R_r}^{B_r})+\mathcal{L}(\Gamma_{Q_1}^{R_l}),\;
		\mathcal{L}(\Gamma_{R_r}^{R_l})+\mathcal{L}(\Gamma_{Q_1}^{B_r}),\;\\ &
		\mathcal{L}(\Gamma_{R_r}^{R_l})+\mathcal{L}(\Gamma_{Q_2}^{B_r})+\mathcal{L}(\Gamma_{Q_1}^{Q_2}),\;
		\mathcal{L}(\Gamma_{Q_1}^{B_r})+\mathcal{L}(\Gamma_{Q_1}^{R_l})+\mathcal{L}(\Gamma_{Q_1}^{R_r}),\;\\&
		\mathcal{L}(\Gamma_{Q_2}^{B_r})+\mathcal{L}(\Gamma_{Q_1}^{R_l})+\mathcal{L}(\Gamma_{Q_1}^{Q_2})+\mathcal{L}(\Gamma_{Q_1}^{R_r}),\;
		\mathcal{L}(\Gamma_{Q_2}^{B_r})+\mathcal{L}(\Gamma_{Q_2}^{R_r})+\mathcal{L}(\Gamma_{Q_1}^{R_l}),\;\\&
		\mathcal{L}(\Gamma_{Q_2}^{B_r})+\mathcal{L}(\Gamma_{Q_2}^{R_r})+\mathcal{L}(\Gamma_{Q_2}^{R_l})+\mathcal{L}(\Gamma_{Q_1}^{Q_2}),\; \mathcal{L}(\Gamma_{Q_2}^{B_r})+\mathcal{L}(\Gamma_{Q_1}^{R_r})+\mathcal{L}(\Gamma_{Q_1}^{R_l})+\mathcal{L}(\Gamma_{Q_1}^{Q_2})
		\Big\} ,\\
		& \quad S_{B'\cup N}=\frac{c}{6\ell}\,
		\min\Big\{
		\mathcal{L}(\Gamma_{R_r}^{R_l})+\mathcal{L}(\Gamma_{Q_2}^{E_r}),\;
		\mathcal{L}(\Gamma_{R_r}^{R_l})+\mathcal{L}(\Gamma_{Q_1}^{E_r})+\mathcal{L}(\Gamma_{Q_1}^{Q_2}),\;\\&
		\mathcal{L}(\Gamma_{R_r}^{E_r})+\mathcal{L}(\Gamma_{Q_1}^{R_l})+\mathcal{L}(\Gamma_{Q_1}^{Q_2}),\;
		\mathcal{L}(\Gamma_{R_r}^{E_r})+\mathcal{L}(\Gamma_{Q_2}^{R_l}),\;
		\mathcal{L}(\Gamma_{Q_1}^{R_r})+\mathcal{L}(\Gamma_{Q_1}^{R_l})+\mathcal{L}(\Gamma_{Q_2}^{E_r}),\;\\&
		\mathcal{L}(\Gamma_{Q_1}^{R_r})+\mathcal{L}(\Gamma_{Q_1}^{R_l})+\mathcal{L}(\Gamma_{Q_1}^{E_r})+\mathcal{L}(\Gamma_{Q_1}^{Q_2}),\;
		\mathcal{L}(\Gamma_{Q_2}^{E_r})+\mathcal{L}(\Gamma_{Q_1}^{R_l})+\mathcal{L}(\Gamma_{Q_1}^{R_r}),\;\\&
		\mathcal{L}(\Gamma_{Q_2}^{E_r})+\mathcal{L}(\Gamma_{Q_1}^{R_l})+\mathcal{L}(\Gamma_{Q_2}^{R_r})+\mathcal{L}(\Gamma_{Q_1}^{Q_2}),\;
		\mathcal{L}(\Gamma_{Q_2}^{E_r})+\mathcal{L}(\Gamma_{Q_2}^{R_l})+\mathcal{L}(\Gamma_{Q_2}^{R_r})
		\Big\}.
	\end{aligned}
	\end{equation}
	Using the holographic entanglement formula, we can calculate the dynamical behavior of mutual information as in \eqref{eq:BS-mutual-renyi}.
	
	 When $|R|$ smaller than the threshold, the minimal configuration for $S_{B'}$ at is 
	 \(\mathcal{L}(\Gamma_{R_r}^{R_l})+\mathcal{L}(\Gamma_{Q_2}^{B_r})+\mathcal{L}(\Gamma_{Q_1}^{Q_2})\), the geodesic for $S_{B'\cup N}$ is always
	 \(\mathcal{L}(\Gamma_{R_r}^{R_l})+\mathcal{L}(\Gamma_{Q_2}^{E_r})\). 
	 
\begin{figure}[htbp]
		\centering	
		\begin{subfigure}{0.45\linewidth}
			\centering
			\includegraphics[width=\linewidth]{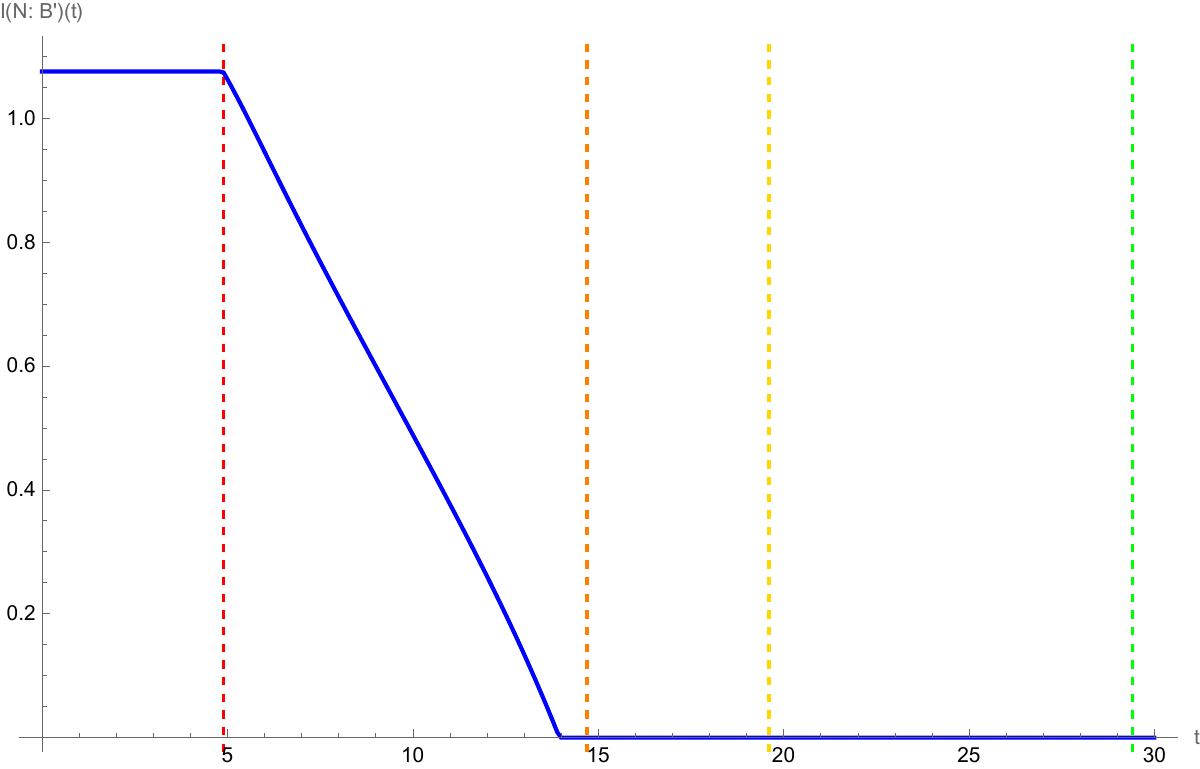}
			\caption{Mutual information $I(N\!:\!B')$ for the radiation length larger than the reference system ($|R|=3L$).}
			
		\end{subfigure}
		\hfill
		\begin{subfigure}{0.45\linewidth}
			\centering
			\includegraphics[width=\linewidth]{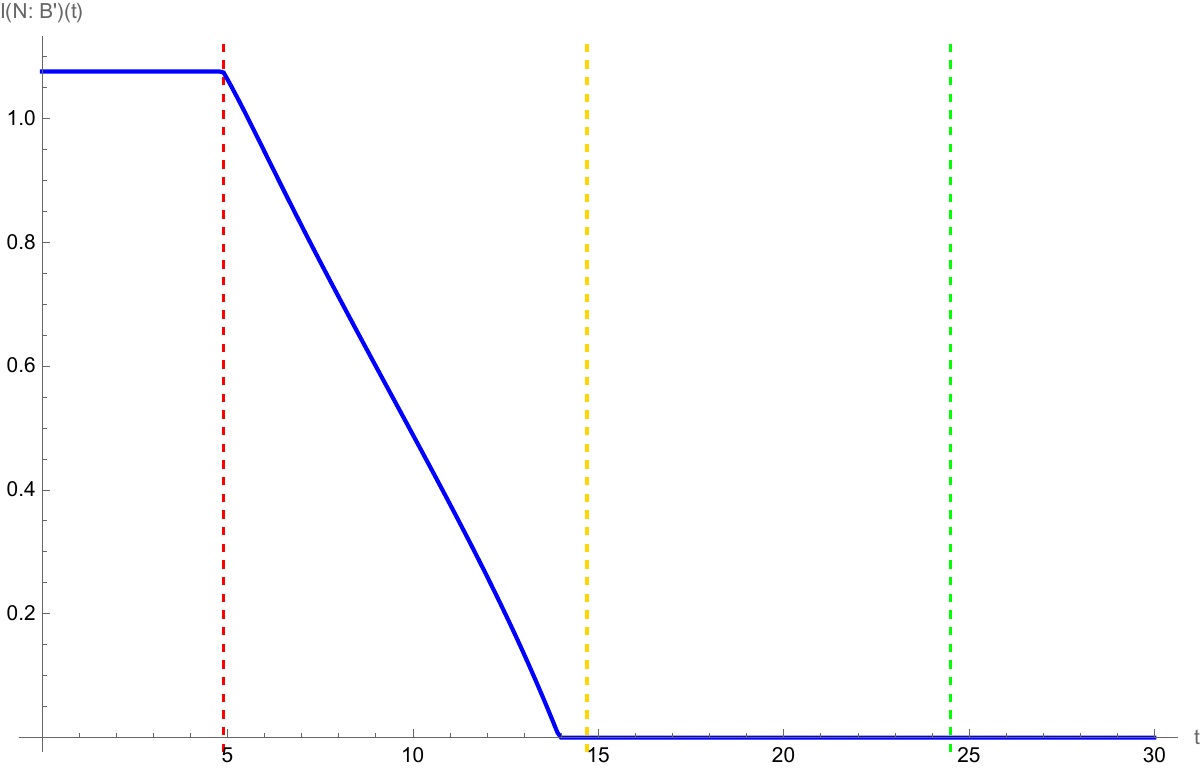}
			\caption{Mutual information $I(N\!:\!B')$ for the radiation length larger than the reference system ($|R|=2L$).}
		\end{subfigure}
		\\	\begin{subfigure}{0.45\linewidth}
			\centering
			\includegraphics[width=\linewidth]{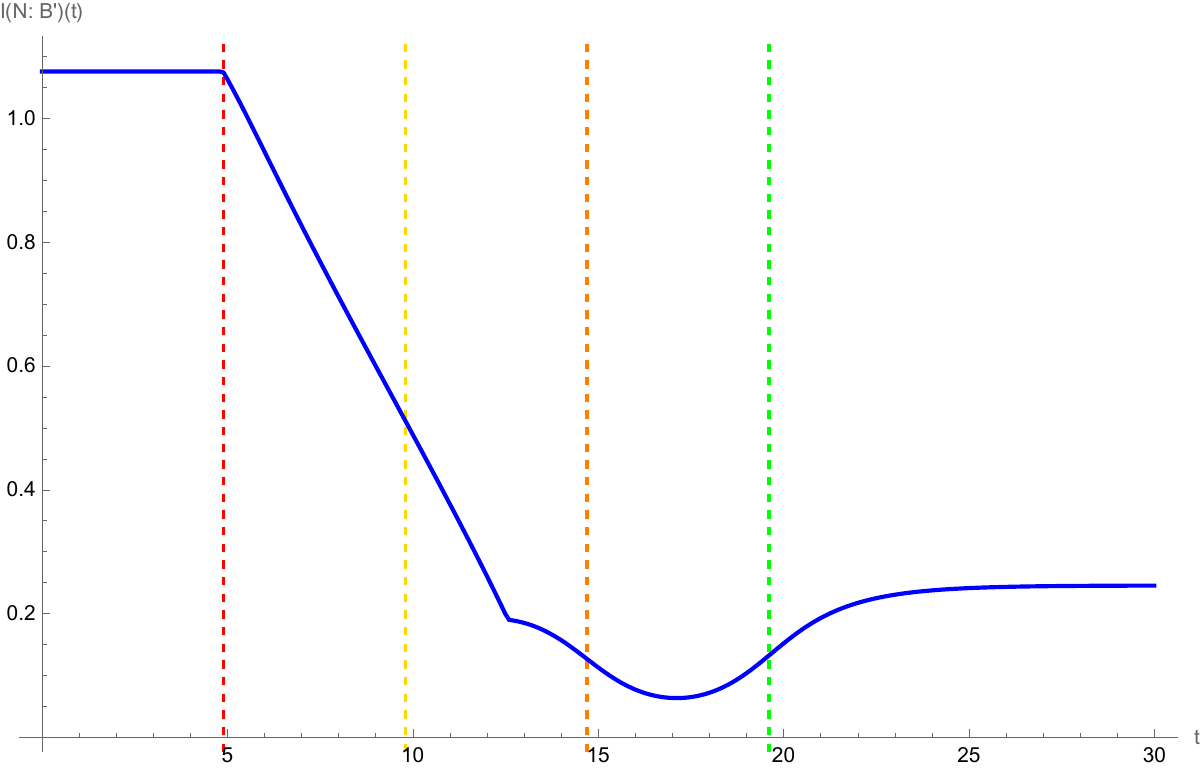}
			\caption{Mutual information $I(N\!:\!B')$ for the radiation length equal to the reference system ($|R|=L$).}
			
		\end{subfigure}
		\hfill
		\begin{subfigure}{0.45\linewidth}
			\centering
			\includegraphics[width=\linewidth]{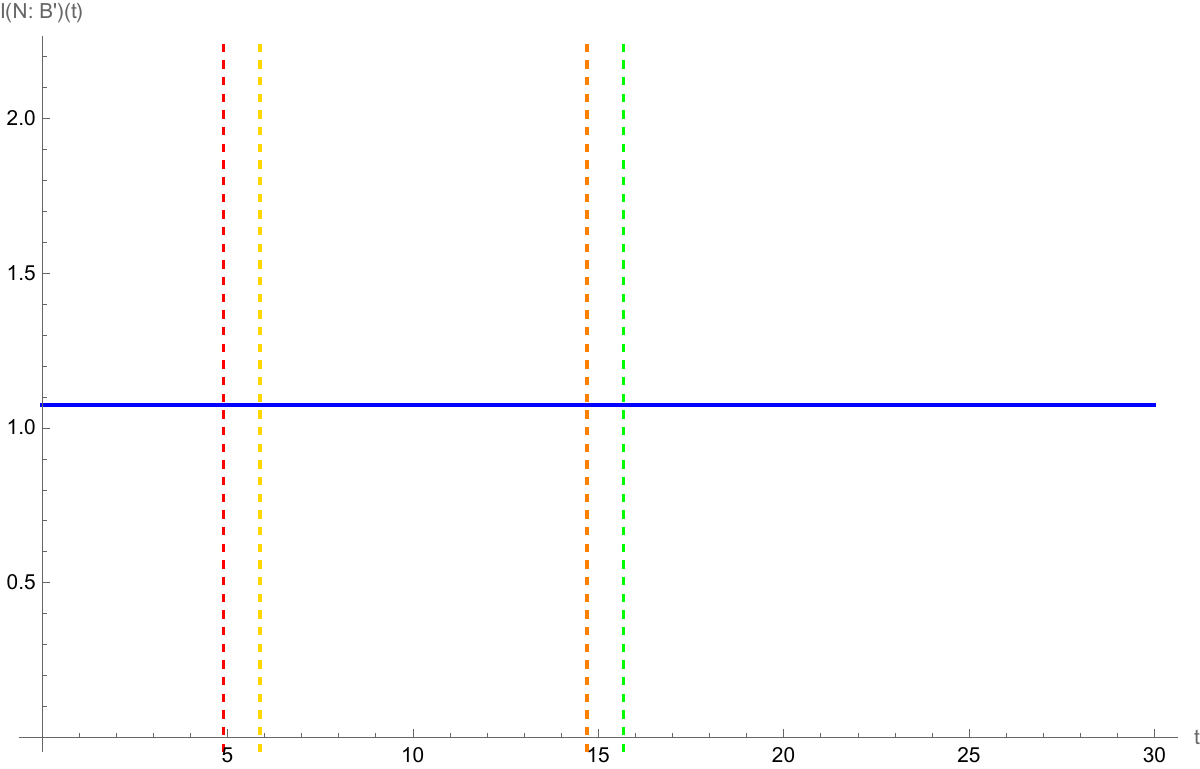}
			\caption{Mutual information $I(N\!:\!B')$ for the radiation length smaller than the reference system ($|R|=0.3L$).}
		\end{subfigure}
		
		\caption{Plots of $I(N\!:\!B')$ as a function of time for different radiation lengths in holographic CFT. Here we take $a=0.046,  \rho=0.0398, \beta = 10, \epsilon=1, \ell=1, c= 1, X_1= L= 2, X_2= 2 L , X_3= 2 L+|R|$. }\label{fig:BS-Mutual-HCFT}
	\end{figure}

	 When $|R|$ larger than the threshold, at early times, the minimal configuration for $S_{B'}$ is 
	 \(\mathcal{L}(\Gamma_{R_r}^{B_r})+\mathcal{L}(\Gamma_{Q_2}^{R_l})+\mathcal{L}(\Gamma_{Q_1}^{Q_2})\);
	 after the first transition time $t_1$ in Fig.\ref{fig:BS-time}, it switches to 
	 \(\mathcal{L}(\Gamma_{R_r}^{B_r})+\mathcal{L}(\Gamma_{Q_1}^{R_l})\).
	 Here $t_1$ is defined with respect to the TFD pairing between $M$ and $N$: it is the first time when the
	 rightmost member of the quasi-particle pair that initially entangles $M$ with $N$ reaches the radiation interval $R$ (equivalently, when the right-moving front on the $M$ side whose partner lies in $N$ first enters $R$).
	 Similarly, the geodesic for $S_{B'\cup N}$ is initially 
	 \(\mathcal{L}(\Gamma_{R_r}^{E_r})+\mathcal{L}(\Gamma_{Q_2}^{R_l})\);
	 after the later transition time $t_3$ in Fig.\ref{fig:BS-time}, it changes to 
	 \(\mathcal{L}(\Gamma_{R_r}^{R_l})+\mathcal{L}(\Gamma_{Q_2}^{E_r})\).
	 The time $t_3$ thus marks the exit of the same rightmost $M$–$N$ entangled front from the radiation interval.
     The time $t_2$ and $t_4$ correspond to similar characteristic times with respect to the initially left-moving modes which are reflected back at the boundary $x=0$.
	 These rearrangements of geodesics govern the time dependence of $I(N:B')$ displayed in Fig.~\ref{fig:BS-Mutual-HCFT}.
	 
	The main result of this calculation is that there exists a critical length for the later radiation subregion. When the radiation subregion’s length exceeds about that of the reference system, the mutual information eventually decays to zero; otherwise, the mutual information remains constant. The Fig.\ref{fig:BS-Mutual-HCFT}  illustrates these two behaviors, which is exactly what we expect to find in the Hayden–Preskill protocol.

	Interestingly, as shown in Fig.\ref{fig:BS-Mutual-HCFT} the behavior of holographic entanglement entropy follows the same causal structure as in the free fermion CFT in the beginning. After the joining quench at $\tau=\beta / 2$, two outgoing wave-fronts form: one propagating toward the late radiation region $\mathrm{R}$, and the other toward the black hole remnant  $\mathrm{B}^{\prime}$ . These fronts effectively carry entangled content from the reference system $N$ into different parts of the spacetime.
	
\begin{figure}[htbp]
		\centering		
		 \includegraphics[width=14cm]{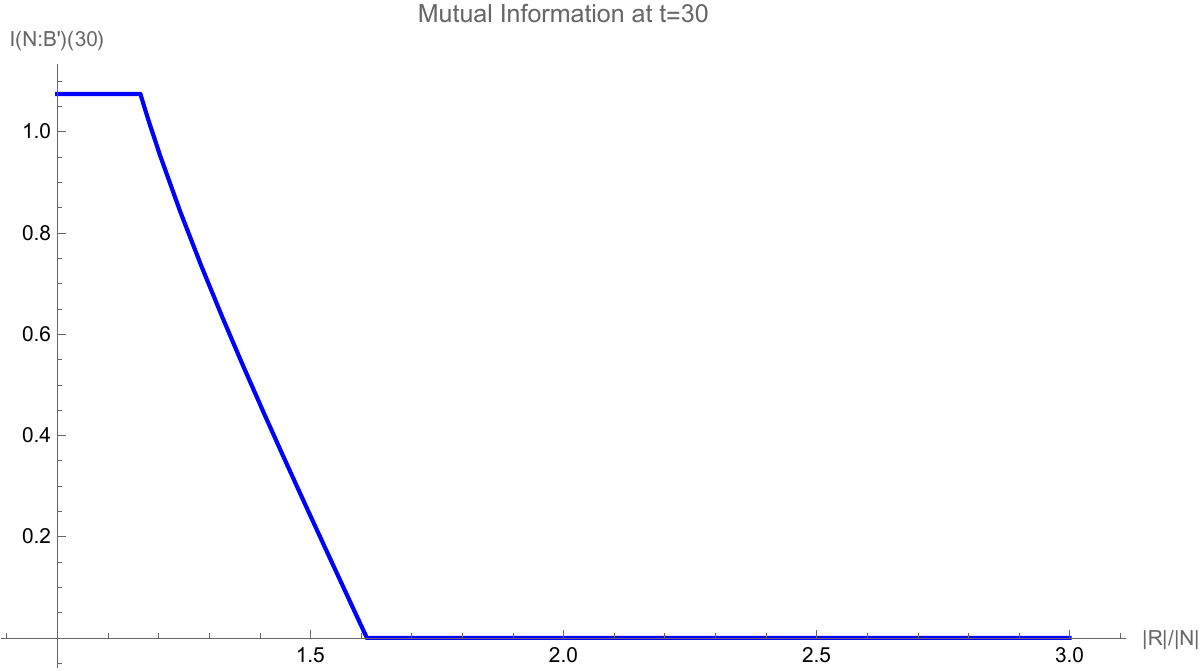}
		\caption{Mutual information $I(N\!:\!B')$ at the late time $t=30$ as a function of $\frac{|R|}{|N|}$ in the holographic CFT. Here we take $a=0.046,  \rho=0.0398, \beta = 10, \epsilon=1,\ell=1, c= 1, X_1= L= 2, X_2= 2 L , X_3= 2 L+|R|$.  The value of $\gamma$ extracted from the plot is in the range 
        $1<\gamma<2$. } \label{fig:BS-Rational-HCFT} 
	\end{figure}

	However, due to the strongly chaotic nature of holographic CFT, governed by the fast scrambling and the growth of operator size, the information carried by these quasi-particle-like excitations is quickly scrambled and absorbed into the black hole or radiated away. As opposed to the free CFT, where owing to its integrability, quasi-particles preserve coherence and return periodically, holographic CFTs suppress such free propagations of excitations and recurrences at a relatively short time scale.
	Consequently, the mutual information $I\left(N: B^{\prime}\right)$ undergoes a sharp transition: it remains to be a positive constant at late time when $|R|< \gamma |N|$ (here $\gamma=O(1)$, numerically $1<\gamma<2$ ). On the other hand, when $|R|> \gamma |N|$, the final value gets vanishing after an indicating irreversible information release to the radiation region. The holographic geodesic which computes $S_{B' \cup N}$ undergoes a geometric reconfiguration, disconnecting $N$ from $B^{\prime}$, and instead associating $N$ with $R$ by a change of the dominant geodesic configuration. This reflects the expected recovery condition peculiar to the Hayden-Preskill model such that the Alice's information $M$ can be recovered from the radiation $R$ 
    (i.e. the mutual information becomes vanishing at late time) when $|R|> \gamma |N|$ as shown in Fig.\ref{fig:BS-Rational-HCFT}. 
\section{Conclusions}\label{sec:conclusions}

We have constructed a continuum realization of the Hayden--Preskill (HP) information–recovery protocol in two–dimensional conformal field theories using local joining quenches. Two Euclidean geometries were central to our analysis: (i) the single slit on the thermal cylinder, conformally mapped to an upper half–plane, and (ii) the bounded slit on the semi–infinite cylinder, conformally mapped to an annulus. The single–slit case offers a full analytical control and serves as a tractable toy model, while the bounded–slit case captures much more complete structures of the HP setup.

In the free Dirac fermion CFT, the mutual information between the reference $N$ and the post–emission black hole $B'$ can be computed exactly. owing to the integrability, the time evolution of mutual information is well described by the quasi-particle picture: the correlations decrease when one partner of an entangled pair leaves $B'$, remain constant while the partner traverses the radiation interval, and revive once the quasi-particles return.

In the holographic CFTs, the behavior is qualitatively different. Geodesics in the AdS$_3/$BCFT$_2$ with tensionless end–of–the–world branes govern the dynamics. The single–slit model already shows the absence of quasi-particle revivals, while the bounded–slit model further reveals a sharp behavior: once the size of the late–radiation interval $R$ is comparable to or larger than that of $N$, the mutual information $I(N\!:\!B')$ drops to zero and does not recover. In our treatment this transition is entirely accounted for by a change of the configurations of geodesics: the connected geodesics that previously minimized the entropy become subdominant to disconnected (or mixed) configurations, which isolate $N$ from $B'$ when $|R|$ is sufficiently large.

The physical picture that emerges is that slit quenches produce entangled fronts propagating ballistically. In integrable theories these fronts behave as stable quasi-particles, preserving correlations and producing non–monotonic entanglement with partial or full revivals. In holographic theories, which are expected to be maximally chaotic \cite{Maldacena:2015waa}, the operator growth and fast scrambling suppress such returns; at the level of our calculations this manifests as an irreversible change in the geodesic configuration and a concomitant decay of $I(N\!:\!B')$ once $R$ becomes sufficiently large. A similar crucial difference of time evolutions of entanglement entropy and mutual information between those in the chaotic CFT and those in integrable ones, has been observed in many examples \cite{Nozaki:2013wia,Asplund:2013zba,Nozaki:2014hna,He:2014mwa,Asplund:2015eha,Caputa:2014eta} in the past. The main point of this paper is that we showed the characteristic features of chaotic CFTs in a setup which faithfully describes the Hayden-Preskill model.

Our analysis was limited to the free Dirac fermion CFT as a representative rational CFT, which is integrable, and to the large–$c$ and strongly coupled CFTs (i.e. the holographic CFTs). Natural extensions include incorporating finite brane tensions and boundary degrees of freedom in the AdS/BCFT, studying quantum $1/c$ corrections, exploring multi–interval or nonuniform radiation patterns, and employing other quantum informational measures. It would also be interesting to interpret the bounded–slit construction from the perspective of quantum error correcting codes and to relate the observed transitions of geodesics to code–subspace reconstructions in a more systematic way.

In summary, the slit–quench framework provides a minimal and controllable continuum setting in which the contrast between quasi-particle revivals and holographic fast scrambling becomes manifest, and the Hayden--Preskill recovery threshold appears geometrically as a reorganization of the dominant configurations of geodesics.

\section*{Acknowledgments}
We thank Hiroki Kanda very much for his collaboration on this work at an early stage.
We are also grateful to Chen Bai, Pengxiang Hao, Akihiro Miyata, Masahiro Nozaki, Farzad Omidi, Maotian Tan, Masataka Watanabe, Xueda Wen, and Yuxuan Zhang for many helpful discussions.
This work is supported by by MEXT KAKENHI Grant-in-Aid for Transformative Research Areas (A) through the ``Extreme Universe'' collaboration: Grant Number 21H05187. TT is also supported by Inamori Research Institute for Science and by JSPS Grant-in-Aid for Scientific Research (B) No.~25K01000.

\appendix
\section{How to deal with the circular slit domain}\label{sec:method-of-maps}

In this appendix we summarize the conformal mapping techniques used to treat circular–slit
domains. Our approach follows the systematic framework developed by
Crowdy~\cite{crowdy2006conformal,crowdy2011schottky,crowdy2012conformal}, which provides
explicit formulas mapping multiply–connected circular domains to canonical slit domains.
Here, “multiply–connected circular domains’’ are regions bounded by disjoint circles, while
“canonical slit domains’’ are domains whose additional boundary components are slits
(straight or curved). A schematic example is shown in Fig.~\ref{fig:bounded-slit-map}.
The key object is the Schottky–Klein prime function, naturally adapted to circular
geometries with multiple boundary components.

This framework has also been used in high–energy theory to analyze path–integral
geometries and entanglement dynamics in 2d CFT
(e.g.~\cite{Numasawa:2016emc,Lap:2024hsy,Lap:2024vwm}). While many applications focus on
parallel–slit domains, here we consider   circular–arc slit domains , where the slit is an
arc of a concentric circle. This setting enlarges the class of tractable geometries and
offers additional control for the constructions used in the main text.

\subsection{Conventions for $\theta$–functions}

Let $q=e^{2\pi i\tau}$. We use
\begin{equation}
	\begin{aligned}
		\eta(\tau) &= q^{\tfrac{1}{24}}\prod_{n=1}^\infty(1-q^n),\\[3pt]
		\theta_1(\nu,\tau) &= 2q^{\tfrac18}\sin(\pi\nu)\prod_{n=1}^\infty(1-q^n)(1-e^{2\pi i\nu}q^n)(1-e^{-2\pi i\nu}q^n),\\[3pt]
		\theta_2(\nu,\tau) &= 2q^{\tfrac18}\cos(\pi\nu)\prod_{n=1}^\infty(1-q^n)(1+e^{2\pi i\nu}q^n)(1+e^{-2\pi i\nu}q^n),\\[3pt]
		\theta_3(\nu,\tau) &= \prod_{n=1}^\infty(1-q^n)(1+e^{2\pi i\nu}q^{n-\tfrac12})(1+e^{-2\pi i\nu}q^{\,n-\tfrac12}),\\[3pt]
		\theta_4(\nu,\tau) &= \prod_{n=1}^\infty(1-q^n)(1-e^{2\pi i\nu}q^{n-\tfrac12})(1-e^{-2\pi i\nu}q^{\,n-\tfrac12}).
	\end{aligned}
\end{equation}

\subsection{From a circular–arc slit to an annulus}

After the exponential map from the bounded–slit quench geometry on the thermal cylinder,
the image is a unit disk with an interior circular–arc slit. Treating the two sides of the
slit as distinct boundary components, the domain is   doubly connected  (two boundary
components), and hence conformally equivalent to an annulus.

Concretely, consider the annulus
\begin{equation}\label{df:anulus}
	D_\zeta=\{\zeta\mid \rho<|\zeta|<1\},\qquad 0<\rho<1,
\end{equation}
with outer ($C_0:|\zeta|=1$) and inner ($C_1:|\zeta|=\rho$) circles. Following
\cite{crowdy2006conformal}, define
\begin{equation}
	\mathcal{G}_0(\zeta,a)=\frac{1}{2\pi i}\,
	\log\!\left(\frac{\omega(\zeta,a)}{|a|\,\omega(\zeta,1/\bar a)}\right),
\end{equation}
where $\omega(\zeta,a)$ is the Schottky–Klein prime function. Then
\begin{equation}
	z=\exp\!\big(2\pi i\,\mathcal{G}_0(\zeta,a)\big)
\end{equation}
maps the annulus to a disk with a circular–arc slit determined by the parameter $a$
(with $\rho<|a|<1$; its argument fixes the slit’s angular position).

Using identities relating prime functions and $\theta$–functions, one obtains an explicit
bounded map $f(\zeta;a,\rho)$ from the annulus to the unit disk with an arc–shaped slit
coaxial with $C_0$ and $C_1$:
\begin{equation}
	\label{eq:f-explicit}
	f(\zeta;a,\rho)=
	\frac{|a|\;\theta_3\!\left(\dfrac{\log\!\left(-\dfrac{\zeta}{\rho a}\right)}{2\pi i}\,;\,\dfrac{\log\rho}{\pi i}\right)}
	{\theta_3\!\left(\dfrac{\log\!\left(-\dfrac{\zeta a}{\rho}\right)}{2\pi i}\,;\,\dfrac{\log\rho}{\pi i}\right)}.
\end{equation}

\section{Energy Density}\label{sec:energy-density}

\subsection{Single slit case}
We first place the 2d CFT on the upper half-plane
\begin{equation}
	\mathbb H = \{ z \in \mathbb C \;|\; \Im z > 0 \}, \qquad \partial\mathbb H=\mathbb R,
\end{equation}
with Cardy boundary condition
\begin{equation}
	T(x)=\overline{T}(x), \qquad x\in\mathbb R .
	\label{eq:CardyBC-UHP}
\end{equation}
For any analytic vector field $\eta(z)$ preserving $\mathbb R$, the Ward identity gives
\begin{equation}
	\oint_{\mathcal C}\frac{dz}{2\pi i}\;\eta(z)\,\langle T(z)\rangle_{UHP}
	+\oint_{\mathcal C}\frac{d\bar z}{2\pi i}\;\bar\eta(\bar z)\,\langle \overline{T}(\bar z)\rangle_{UHP}
	=0, \qquad \mathcal C\subset\mathbb H .
\end{equation}

Invariance under translations and dilations implies $\langle T(z)\rangle_{UHP}$ must be a constant, while Möbius covariance forces this constant to vanish. Hence
\begin{equation}
	\langle T(z)\rangle_{UHP}=0=\langle \overline{T}(\bar z)\rangle_{UHP}.
	\label{eq:vanish-UHP}
\end{equation}

\begin{figure}[htbp]
	\centering
	\includegraphics[width=0.8\linewidth]{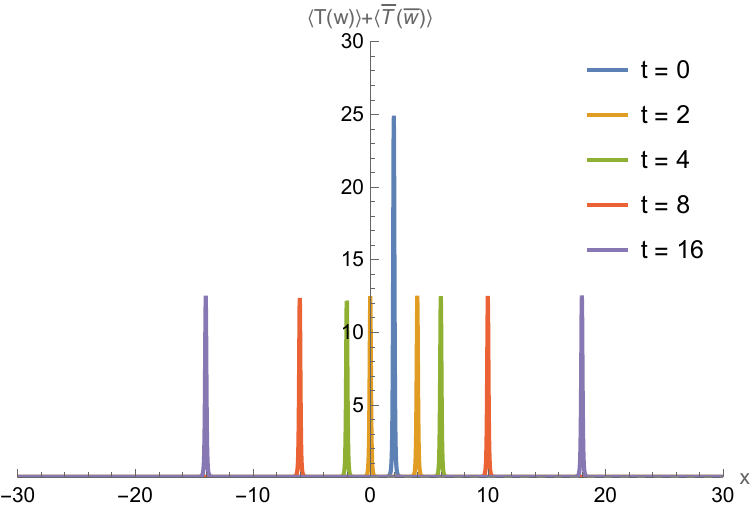}
	\caption{Energy pulses generated by the slit quench: one propagates left and the other right. 
		Apart from the initial separation of two coincident peaks, the pulse height stays nearly constant. 
		Parameters: $\alpha=0.1,\;\beta=10,\;\epsilon=1,\;c=1,\;X_1=2$.}
	\label{fig:SS-energy-pulses}
\end{figure}

Mapping back to the single-slit cylinder by
\begin{equation}
	\xi(w)=\sqrt{\frac{\sinh \!\left(\tfrac{\pi}{\beta}(w-X_1)- i \tfrac{\pi}{\beta}(\tfrac{\beta}{2}+\alpha)\right)}
		{\sinh \!\left(\tfrac{\pi}{\beta}(w-X_1)- i \tfrac{\pi}{\beta}(\tfrac{\beta}{2}-\alpha)\right)}} ,
\end{equation}
and analytically continuing $\tau$ to Lorentzian time, one obtains the pulse-like energy density shown in Fig.~\ref{fig:SS-energy-pulses}.

\subsection{Bounded slit case}
Now consider the CFT on the annulus
\begin{equation}
	\Annulus=\{\rho<|\zeta|<1\},
\end{equation}
with Cardy boundary conditions $T(\zeta)=\overline T(\bar\zeta)$ on $|\zeta|=1,\rho$.  
The Ward identity together with rotational invariance restricts
\(\langle T(\zeta)\rangle_{\Annulus}=A/\zeta^2\).  
Vanishing of the angular momentum operator then enforces $A=0$, so
\begin{equation}
	\langle T(\zeta)\rangle_{\Annulus}=0=\langle \overline T(\bar\zeta)\rangle_{\Annulus}.
\end{equation}

Mapping the annulus to the bounded-slit cylinder $w(\zeta)$ and using the conformal transformation law
\[
\langle T(w)\rangle=\left(\tfrac{d\zeta}{dw}\right)^2\langle T(\zeta)\rangle+\tfrac{c}{12}\{\zeta,w\},
\]
one finds $\langle T(w)\rangle=\tfrac{c}{12}\{\zeta,w\}$, determined solely by the Schwarzian derivative.  

\begin{figure}[htbp]
	\centering
	\includegraphics[width=0.8\linewidth]{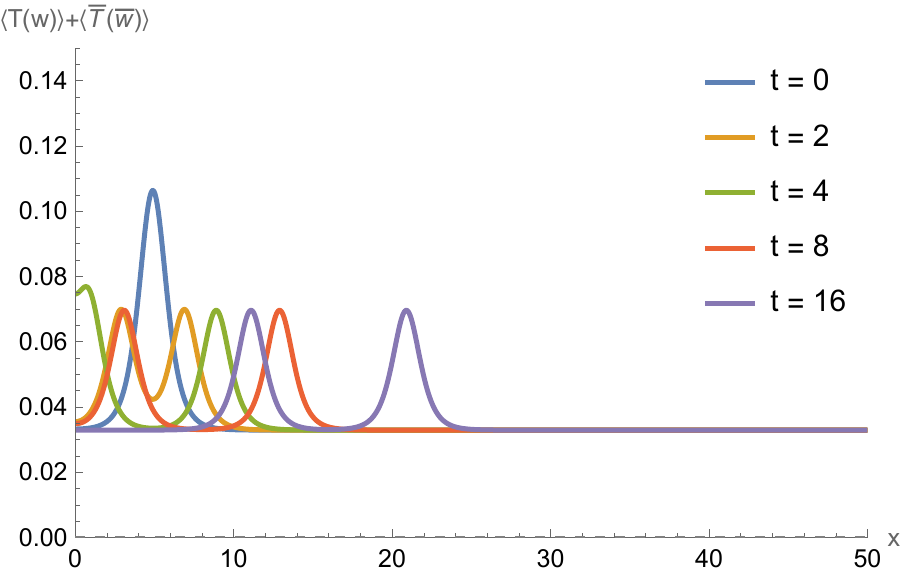}
	\caption{Energy pulses in the bounded-slit quench: one travels left, reflects off the boundary, and returns, while the other travels right. 
		Aside from the initial splitting of coincident peaks, the pulse height remains nearly constant. 
		Parameters: $a=0.1,\;\rho=0.097,\;\beta=10,\;\epsilon=1,\;c=1,\;X_1=2$.}
	\label{fig:BS-energy-pulses}
\end{figure}

At $\tau=\beta/2$ the relevant contour is 
\(
M\cup B=\{\,w=x+i\beta/2\mid x\ge0\,\},
\)
which we then continue to Lorentzian time.  
The resulting chiral+anti-chiral energy density exhibits the pulses shown in Fig.~\ref{fig:BS-energy-pulses}.

\section{Calculation for bounded slit case $I_{N:B'}$ for free fermion }\label{sec:detail-of-calculation} 
 
	Let the annulus modulus be $\tau=2is$ with $s=-\frac{1}{2\pi}\log\rho$.
	We use the cylinder coordinates $y_\alpha$ listed in the main text
	\[
	\begin{array}{l}
		y_{Ml}=\log(1/\rho),\quad y_{Br}=y_{Er}=\log(a/\rho),\\
		y_{Nl}=\log(1/\rho)+\pi i,\quad y_{Nr}=\pi i,\\
		y_{Rl}(t)=\log\!\bigl[w_{Rl}(t)/\rho\bigr],\quad
		y_{Rr}(t)=\log\!\bigl[w_{Rr}(t)/\rho\bigr],
	\end{array}
	\]
	and analogously for $\bar y_\alpha$.
	Introduce the compact notations
	\begin{equation}
		\Delta_{ij}:=\theta_1\!\Bigl(\frac{y_i-y_j}{2\pi i}\,\Big|\,\tau\Bigr),\qquad
		\bar\Delta_{ij}:=\theta_1\!\Bigl(\frac{\bar y_i-\bar y_j}{2\pi i}\,\Big|\,\tau\Bigr),
	\end{equation}
	\begin{equation}
		\Sigma_{ij}:=\theta_1\!\Bigl(\frac{y_i+\bar y_j}{2\pi i}\,\Big|\,\tau\Bigr).
	\end{equation}
	We also factor out the universal Jacobians; they produce only $t$-independent
	constants and drop from mutual information.
	 
	For the free Dirac fermion ($c=1$), using the bosonized vertex building block
	and Dirichlet boundary condition, the  two-point block on the cylinder is 
	\begin{equation}
		\label{eq:D-two-point}
		\mathcal D(i,j)=
		\frac{\eta(\tau)^6}{\Delta_{ij}\,\bar\Delta_{ij}}\,
		\frac{\Sigma_{ii}\,\Sigma_{jj}}{\Sigma_{ij}\,\Sigma_{ji}}.
	\end{equation}
	The four-point block is
	\begin{equation}
		\label{eq:U-four-point}
		\mathcal U_4(1,2,3,4)
		=\frac{\eta(\tau)^{12}\,\Delta_{42}\,\Delta_{31}}{
			\Delta_{43}\,\Delta_{41}\,\Delta_{32}\,\Delta_{21} }
		\cdot
		\frac{\bar\Delta_{42}\,\bar\Delta_{31}}{
			\bar\Delta_{43}\,\bar\Delta_{41}\,\bar\Delta_{32}\,\bar\Delta_{21} }
		\cdot
		\frac{\Sigma_{1 2}\,\Sigma_{1 4}\,\Sigma_{2 1}\,\Sigma_{2 3}\,
			\Sigma_{3 2}\,\Sigma_{3 4}\,\Sigma_{4 1}\,\Sigma_{4 3}}
		{\Sigma_{1 1}\,\Sigma_{1 3}\,\Sigma_{2 2}\,\Sigma_{2 4}\,
			\Sigma_{3 1}\,\Sigma_{3 3}\,\Sigma_{4 2}\,\Sigma_{4 4}}\,.
	\end{equation}
 
  With $\Delta_n=\frac{1}{12}(n-\frac{1}{n})$ and the standard replica prefactor,
	\begin{equation}
		S_{N}^{(n)}=\frac{1}{1-n}\log\left\{\bigl[\mathcal D(Nl,Nr)\bigr]^{\Delta_n}
		\prod_{k=-(n-1)/2}^{(n-1)/2}
		\frac{\theta_3\!\left(\tfrac{k}{2\pi n}\big[(y_{Nl}-y_{Nr})-(\bar y_{Nl}-\bar y_{Nr})\big]\,\big|\,\tfrac{i}{\pi s}\right)}
		{\theta_3\!\left(0\,\big|\,\tfrac{i}{\pi s}\right)}\right\},
	\end{equation}
	
	\begin{equation}
		\begin{split}
		S_{B'}^{(n)}&=\frac{1}{1-n}\log \{\bigl[\mathcal U_4(Ml,Rl,Rr,Br)\bigr]^{\Delta_n} \\
		& \left.\cdot\prod_{k=-(n-1)/2}^{(n-1)/2}
		\frac{\theta_3\!\left(\tfrac{k}{2\pi n}\big[(y_{Ml}-y_{Rl}+y_{Rr}-y_{Br})-(\bar y_{Ml}-\bar y_{Rl}+\bar y_{Rr}-\bar y_{Br})\big]\,\big|\,\tfrac{i}{\pi s}\right)}
		{\theta_3\!\left(0\,\big|\,\tfrac{i}{\pi s}\right)} \right\},  
		\end{split}
	\end{equation}
	
	\begin{equation}
		\begin{split}
			S_{N\cup B'}^{(n)}&=\frac{1}{1-n}\log \{\bigl[\mathcal U_4(Rl,Rr,Er,Nr)\bigr]^{\Delta_n}
			\\
			& \left.
			\prod_{k=-(n-1)/2}^{(n-1)/2}
			\frac{\theta_3\!\left(\tfrac{k}{2\pi n}\big[(y_{Rl}-y_{Rr}+y_{Er}-y_{Nr})-(\bar y_{Rl}-\bar y_{Rr}+\bar y_{Er}-\bar y_{Nr})\big]\,\big|\,\tfrac{i}{\pi s}\right)}
			{\theta_3\!\left(0\,\big|\,\tfrac{i}{\pi s}\right)}\right\}.
		\end{split}
	\end{equation}
	
	Therefore
	\begin{equation}
		\begin{split}
		I_{N:B'}^{(n)}&=\frac{1}{1-n}\,\Delta_n\,
		\log\!\left[
	\sqrt{
		\left.\frac{dy}{dw}\right|_{w_{N_l}}
		\left.\frac{d\bar y}{d\bar w}\right|_{\bar w_{N_l}}
		\left.\frac{dy}{dw}\right|_{w_{M_l}}
		\left.\frac{d\bar y}{d\bar w}\right|_{\bar w_{M_l}}
	}
		 \frac{\mathcal D(Nl,Nr)\,\mathcal U_4(Ml,Rl,Rr,Br)}{\mathcal U_4(Rl,Rr,Er,Nr)}\right]\\
		 &\quad+\;\frac{1}{1-n}\;
		 \log\!\Bigg[
		 \frac{\prod_{k=-(n-1)/2}^{(n-1)/2}
		 	\theta_3\!\left(\tfrac{k}{2\pi n}\big[(y_{Nl}-y_{Nr})-(\bar y_{Nl}-\bar y_{Nr})\big]\;\Big|\;\tfrac{i}{\pi s}\right)}
		 { } \\[6pt]
		 &\qquad\qquad\times\;
		 \prod_{k=-(n-1)/2}^{(n-1)/2}
		 \theta_3\!\left(\tfrac{k}{2\pi n}\big[(y_{Ml}-y_{Rl}+y_{Rr}-y_{Br})-(\bar y_{Ml}-\bar y_{Rl}+\bar y_{Rr}-\bar y_{Br})\big]\;\Big|\;\tfrac{i}{\pi s}\right) \\[6pt]
		 &\qquad\qquad\Bigg/
		 \Bigg(
		 \prod_{k=-(n-1)/2}^{(n-1)/2}
		 \theta_3\!\left(\tfrac{k}{2\pi n}\big[(y_{Rl}-y_{Rr}+y_{Er}-y_{Nr})-(\bar y_{Rl}-\bar y_{Rr}+\bar y_{Er}-\bar y_{Nr})\big]\;\Big|\;\tfrac{i}{\pi s}\right)
		 \theta_3\!\left(0\;\Big|\;\tfrac{i}{\pi s}\right)
		 \Bigg)\Bigg].\\
		&=-\frac{1}{12}\!\Bigl(1+\frac{1}{n}\Bigr)\log\mathcal R\\
		&\quad+\;\frac{1}{1-n}\;
		\log\!\Bigg[
		\frac{\prod_{k=-(n-1)/2}^{(n-1)/2}
			\theta_3\!\left(\tfrac{k}{2\pi n}\big[(y_{Nl}-y_{Nr})-(\bar y_{Nl}-\bar y_{Nr})\big]\;\Big|\;\tfrac{i}{\pi s}\right)}
		{ } \\[6pt]
		&\qquad\qquad\times\;
		\prod_{k=-(n-1)/2}^{(n-1)/2}
		\theta_3\!\left(\tfrac{k}{2\pi n}\big[(y_{Ml}-y_{Rl}+y_{Rr}-y_{Br})-(\bar y_{Ml}-\bar y_{Rl}+\bar y_{Rr}-\bar y_{Br})\big]\;\Big|\;\tfrac{i}{\pi s}\right) \\[6pt]
		&\qquad\qquad\Bigg/
		\Bigg(
		\prod_{k=-(n-1)/2}^{(n-1)/2}
		\theta_3\!\left(\tfrac{k}{2\pi n}\big[(y_{Rl}-y_{Rr}+y_{Er}-y_{Nr})-(\bar y_{Rl}-\bar y_{Rr}+\bar y_{Er}-\bar y_{Nr})\big]\;\Big|\;\tfrac{i}{\pi s}\right)
		\theta_3\!\left(0\;\Big|\;\tfrac{i}{\pi s}\right)
		\Bigg)\Bigg],
		\end{split}
	\end{equation}

	with
	\begin{equation}
		\label{eq:R-master}
		\mathcal R:=
	\sqrt{
		\left.\frac{dy}{dw}\right|_{w_{N_l}}
		\left.\frac{d\bar y}{d\bar w}\right|_{\bar w_{N_l}}
		\left.\frac{dy}{dw}\right|_{w_{M_l}}
		\left.\frac{d\bar y}{d\bar w}\right|_{\bar w_{M_l}}
	}
		\frac{\mathcal D(Nl,Nr)\,\mathcal U_4(Ml,Rl,Rr,Br)}{\mathcal U_4(Rl,Rr,Er,Nr)}.
	\end{equation}
	By instead \eqref{eq:D-two-point} and \eqref{eq:U-four-point} in to \eqref{eq:R-master}, we get simplified form: 
	
	\begin{equation}
		\begin{split}\label{eq:R-master-simplified}
	\mathcal R:=\sqrt{
		\left.\frac{dy}{dw}\right|_{w_{N_l}}
		\left.\frac{d\bar y}{d\bar w}\right|_{\bar w_{N_l}}
		\left.\frac{dy}{dw}\right|_{w_{M_l}}
		\left.\frac{d\bar y}{d\bar w}\right|_{\bar w_{M_l}}
	} &\frac{\eta(\tau)^6}{\Delta_{N_{l}N_{r}}\,\bar\Delta_{N_{l}N_{r}}}\,
	\frac{\Sigma_{N_{l}N_{l}}\,\Sigma_{N_{r}N_{r}}}{\Sigma_{N_{l}N_{r}}\,\Sigma_{N_{r}N_{l}}} \\
	& \frac{\Delta_{R_r M_l}}{
		\Delta_{B_r M_l}\,\Delta_{R_l M_l}}
	\cdot
	\frac{\bar\Delta_{R_r M_l}}{
		\bar\Delta_{B_r M_l}\,\bar\Delta_{R_l M_l}}
	\cdot
	\frac{\Sigma_{M_l R_l}\,\Sigma_{M_l B_r}\,\Sigma_{R_l M_l}\,\Sigma_{B_r M_l}}
	{\Sigma_{M_l M_l}\,\Sigma_{M_l R_r}\,\Sigma_{R_r M_l}}\\
	&\left[ \frac{\Delta_{N_r R_r}}{
		\Delta_{N_r B_r}\,\Delta_{N_r R_l}}
	\cdot
	\frac{\bar\Delta_{N_r R_r}}{
		\bar\Delta_{N_r B_r}\,\bar\Delta_{N_r R_l}}
	\cdot
	\frac{\Sigma_{R_l N_r}\,\Sigma_{N_r B_r}\,\Sigma_{N_r R_l}\,\Sigma_{B_r N_r}}
	{\Sigma_{R_r N_r}\,\Sigma_{N_r R_r}\,\Sigma_{N_r N_r}}\right]^{-1}.
		\end{split}
	\end{equation}

	
		\bibliographystyle{JHEP}
		\bibliography{REF}
		
	\end{document}